\documentclass[aps,prb,10pt,twocolumn,notitlepage,floatfix,bibnotes,nofootinbib]{revtex4-2}

\usepackage{amsmath,amsfonts,amssymb,bm}
\usepackage{mathtools}
\usepackage[colorlinks=true,linkcolor=blue,citecolor=blue]{hyperref}
\usepackage{graphicx}
\usepackage{algorithm}
\usepackage{algorithmicx}
\usepackage[noend]{algpseudocode}
\usepackage{xcolor,xspace}
\usepackage[normalem]{ulem}

\usepackage{siunitx}

\usepackage{tabularx}
\usepackage{booktabs}
\usepackage{multirow}

\newcolumntype{x}[1]{>{\centering\arraybackslash\hspace{0pt}}p{#1}}

\newcommand{\ket}[1]{\vert{#1}\rangle}
\newcommand{\E}{\mathcal{E}}
\renewcommand{\S}{\mathcal{S}}
\newcommand{\F}{\mathbb{F}}
\newcommand{\I}{\mathbb{I}}
\newcommand{\R}{\mathbb{R}}
\newcommand{\Z}{\mathbb{Z}}
\newcommand{\bare}{\mathrm{bare}}
\newcommand{\q}{\mathrm{q}}
\renewcommand{\O}[1]{\ensuremath{O(#1)}\xspace}

\DeclareMathOperator{\supp}{\mathrm{supp}}

\usepackage{cleveref}
\crefname{equation}{Eq.}{Eqs.}
\crefname{section}{Sec.}{Secs.}
\crefname{subsection}{Sec.}{Secs.}
\crefname{appendix}{Appendix}{Appendices}
\crefname{figure}{Figure}{Figures}
\crefname{table}{Table}{Tables}
\crefname{result}{Result}{Results}
\crefname{algorithm}{Algorithm}{Algorithms}

\usepackage[caption=false]{subfig}
\newcommand{\phantomsubfloat}[1]{
    {%
        \captionsetup[subfigure]{labelformat=empty}
        \subfloat[][]{#1}
    }%
}

\appdef \turnpage {%
  \AddToHookNext{shipout/after}{%
    \global\pdfpageattr\expandafter{\the\pdfpageattr/Rotate 90}%
    \AddToHookNext{shipout/after}{%
      \global\pdfpageattr\expandafter{\the\pdfpageattr/Rotate 0}%
    }%
  }%
}

\renewcommand{\algorithmicrequire}{\textbf{Input:}}
\renewcommand{\algorithmicensure}{\textbf{Output:}}

\newcommand{\algrule}[1][0.5pt]{\par\vskip.2\baselineskip\hrule height #1\par\vskip.2\baselineskip}

\usepackage[export]{adjustbox}

\usepackage{capt-of}

\begin{document}

\title{Cored product codes for quantum self-correction in three dimensions}

\date{\today}

\author{Brenden Roberts}
\affiliation{Department of Physics, Harvard University, Cambridge, MA 02138, USA}

\author{Jin Ming Koh}
\affiliation{Department of Physics, Harvard University, Cambridge, MA 02138, USA}

\author{Yi Tan}
\affiliation{Department of Physics, Harvard University, Cambridge, MA 02138, USA}

\author{Norman Y.~Yao}
\affiliation{Department of Physics, Harvard University, Cambridge, MA 02138, USA}
  
\begin{abstract}
The existence of self-correcting quantum memories in three dimensions is a long-standing open question at the interface between quantum computing and many-body physics.
We take the perspective that large contributions to the entropy arising from fine-tuned spatial symmetries, including the assumption of an underlying regular lattice, are responsible for
fundamental challenges to realizing self-correction.
Accordingly, we introduce a class of disordered quantum codes, which we call ``cored product codes''.
These codes are derived from classical factors via the hypergraph product but undergo a coring procedure which allows them to be embedded in a lower number of spatial dimensions while preserving code properties.
As a specific example, we focus on a fractal code based on the aperiodic pinwheel tiling as the classical factor and perform finite temperature numerical simulations on the resulting three-dimensional quantum memory.
We provide evidence that, below a critical temperature, the memory lifetime increases with system size for codes up to $60\,000$ qubits.
\end{abstract}
\maketitle

\section{Introduction}

\begin{figure*}[t]
    \centering
\includegraphics[width=0.8\textwidth]{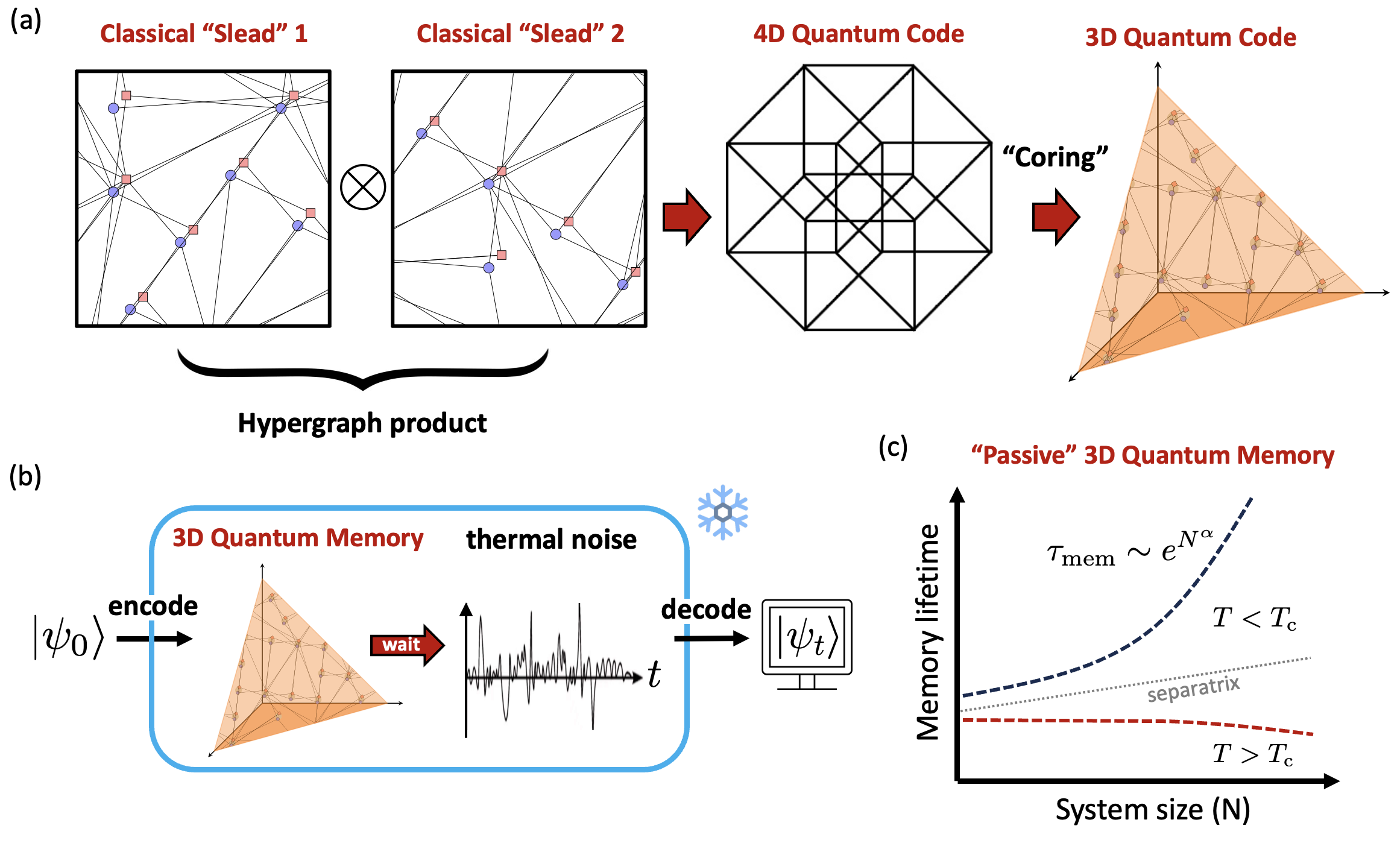}
    \caption{\textbf{Three dimensional self-correcting quantum memory via cored product codes.} \textbf{(a)} Our proposed code construction procedure. A four-dimensional quantum code is produced from the hypergraph product of a pair of two-dimensional  classical ``slead'' codes.
    Our main construction utilizes a pair of so-called classical pinwheel codes~\cite{Tan2024Fracton}.
    The four-dimensional quantum code is then reduced to three dimensions via a coring operation.
    \textbf{(b)} Depicts the schematic setting for a passive quantum memory. 
    In particular, the passive operation of a self-correcting quantum memory entails storing logical information in a quantum code at a sufficiently low temperature. 
    The code is subject to thermal noise at this temperature and decoding is performed once at readout to retrieve the logical information.
    \textbf{(c)} A self-correcting quantum memory ideally exhibits stretched exponential growth (power $\alpha > 0$) in memory lifetime with code size at sufficiently low temperatures.}
\label{fig:introduction_banner}
\end{figure*}

One of the most intriguing open questions at the intersection of quantum error correction and condensed matter physics is the existence of self-correcting quantum memories in few dimensions.
Dating back a quarter-century to the initiation of the field of quantum memory, it was realized that in four or higher dimensions, ``passive'' error correction may provide a lower-overhead alternative to traditional active schemes for protecting quantum information~\cite{Dennis2002,alicki2010thermal}.
It was not clear then whether the lower critical dimension for self-correction is truly four, and although many proposals and results have sharpened the question considerably, its status remains unsettled in three dimensions~\cite{Brown2016RMP}.

At a very high level, self-correction, or equivalently passive error correction, is a paradigm in which particular memory states of a system are stabilized by the action of its own Hamiltonian, without the usual pattern of syndrome measurement, decoding, and recovery.
Typically such properties are modeled by coupling the system to a bath at finite temperature and allowing it to undergo dissipative open-system dynamics.
To be self-correcting requires that there be a fixed (with respect to the memory size) finite temperature below which the system, once initialized in a memory state, remains in this state with finite probability for a time which grows super-polynomially in the memory size.

Various factors contribute to the lifetime of the memory state, which we classify broadly as being either thermodynamic or dynamic effects.
The thermodynamic properties are nicely summarized by the Arrhenius equation $t_\mathrm{mem} \sim e^{\beta \E}$, where $\beta$ is the inverse temperature and $\E$ is a quantity known as the energy barrier, equivalent to the activation energy for processes taking the system between different memory states.
The condition that $\E$ scales algebraically with system size therefore suggests a suitable lifetime.
Unfortunately, dynamical effects tend to spoil this analysis.
One must consider, for example, the entropy of those processes moving between memory states, and if this factor is large enough the memory lifetime may be parametrically shorter than the simple prediction of the Arrhenius equation.

In an early no-go result for passive quantum memories, it was shown that a class of commuting Hamiltonians called stabilizer models cannot be self-correcting in two dimensions due to the generality of constant-height energy barriers~\cite{Bravyi2009NoGo}.
Subsequently, a large class of models, namely translation invariant stabilizer Hamiltonians with topological order, were proved to be incapable of self-correction in two and three dimensions~\cite{Yoshida2011Feasibility}.
Subsequently, many and varied proposals have been put forward which avoid the conditions of these no-go theorems, and the status of some remains outstanding.
We provide a limited overview of this body of prior work in Sec.~\ref{sec:intro_background}.

Perhaps the most well-known attempt at passive error correction in three dimensions consists of Haah's families of cubic codes, including the canonical example known as ``Haah's code'', whose low-energy states have a detailed structure designed to be amenable to self-correction.
Haah's code exhibits a type of glassy thermodynamics driven by an intricate fractal symmetry, spontaneously breaking translation invariance to scale invariance in a hallmark of the exotic phase of matter known as type-II fracton order.
Haah's codes avoid no-go results via this generalized form of ``topological order'' which allows the ground state degeneracy to grow with the system size.

Using the framework described above, we consider both the thermodynamics and dynamics of Haah's code coupled to a thermal bath.
The primary features in the former category are its lack of a thermal phase transition, and also an energy barrier that grows with the logarithm of the system size~\cite{Bravyi2011Energy}.
Thus while the energy required to implement a logical error does grow as a function of the system size, this growth is relatively slow and there is no finite-temperature ordered phase which would guarantee self-correction in the ground states.
For the latter category, one must consider the entropy of the time series of local quantum jumps by which the bath may implement a logical error.
Evidently a source of multiplicity is translation invariance, which permits logical representatives throughout the lattice, so adds an extensive contribution to entropy.
More important, however, is the scale invariance of the logical operators: notice that any self-similar (more precisely, self-affine) error may be implemented by various degenerate single-qubit error sequences which differ by ordering.
The freedom of ordering at each scale leads to an exponential multiplicity of degenerate sequences.
This deleterious feature is closely tied to the translation invariance of the stabilizer Hamiltonian.
The combined effect of the above properties is that Haah's code has not been shown to be self-correcting, instead numerically exhibiting ``partial self-correction'', meaning that the memory time does not grow with the size of the code asymptotically but instead reaches a maximum at some temperature-dependent length scale $L_\mathrm{max}(\beta)$~\cite{Bravyi2013PRL}.

Our objective in this work is to introduce a novel construction of quantum error-correcting codes in three dimensions whose relevant features are enhanced to the extent that they support passive error correction.
In particular, we augment the energy barrier while exponentially suppressing entropic factors. 
Notably, our codes also lack a thermal phase transition; in fact, it is known that while a stable ordered phase is sufficient for self-correction, as observed for example in the toric code in four dimensions~\cite{alicki2010thermal}, it is not necessary.
This result derives from recent studies of models defined on expander graphs~\cite{Hong2025Quantum}.
There the essential feature is a technical property known as ``linear confinement'' which characterizes topological spin glasses~\cite{placke2024topological}, and whose effect is essentially to impose a lower bound on the energy barrier of the code.
While linear confinement is not possible in geometrically local stabilizer models, these results do show that self-correction can arise from a favorable energy barrier with sufficient entropy suppression, independent of finite-temperature order.

We introduce a family of codes, termed \emph{cored product codes}, and argue that these codes exhibit algebraic scaling of energy barriers along with exponential suppression of entropic factors when coupled to a thermal bath.
Our primary mechanism for both features is essentially to eliminate as much spatial symmetry as possible, in order to mimic the glassy aspects of Haah's cubic codes but with the additional contribution of disorder.
We conjecture that this family of codes affirmatively solves the problem of passive error correction in three dimensions. 
These codes are obtained by first selecting from a certain type of classical linear code satisfying some geometric constraints.
These codes are then used as factors in a hypergraph product, after which a ``coring'' operation is performed, reducing the code graph to a form which is embeddable in fewer dimensions but maintains desirable properties like CSS structure and the code distance.
In some sense, the reduction scheme is similar to the lifted or balanced product codes~\cite{Panteleev_ITI2022,Breuckmann_TIT2021}, but instead of exploiting symmetry makes use of the geometric features of the input codes.

The outline of the paper is as follows: in Sec.~\ref{sec:classical_codes} we introduce classical factor codes suitable for our construction, whose properties can be determined from an associated graph we refer to as a ``slead''.
In Sec.~\ref{sec:cored_product_codes} we explain a geometric procedure for reducing the size of the hypergraph product code and apply this procedure to the product of slead codes.
We refer to this application as the coring algorithm because it produces the 2-core of a graph associated with the quantum code.
The cored product code can then be embedded into a manifold with lower spatial dimension.
In Sec.~\ref{sec:finite_temp} we present numerical Monte Carlo simulations of our memories coupled to a bath at finite temperature and discuss the data, which indicates growing lifetimes for codes up to $60\,000$ qubits.

\subsection{Background and prior work}
\label{sec:intro_background}

The area of self-correcting quantum memories has attracted vast attention in the past two decades, and much of the philosophy refined in the body of literature has informed our work. In low dimensions, a range of proposals have been designed to circumvent known no-go theorems~\cite{Bravyi2009NoGo, Haah_PRA2011, Bravyi2011Energy, Yoshida2011Feasibility}, including inducing a linear barrier in the toric code via coupling to a bosonic bath~\cite{Pedrocchi2013Enhanced}, welding patches of topological orders in such a way as to frustrate the logicals~\cite{Michnicki2014}, or embedding codes in a fractal lattice~\cite{Brell2016}. More recent constructions utilize topological defect networks~\cite{Williamson_Layer2023} and sophisticated concatenation~\cite{Lin_arXiv2024} to saturate bounds on the parameters of geometrically local codes, including the distance and energy barrier. However, across this rich theory landscape, no definite proof nor convincing numerical evidence of quantum self-correction has thus far been achieved.

The guiding principle of removing spatial symmetries to reduce entropic contributions in memory dynamics culminates from a line of prior work.
An early attempt studied codes on fractal subsets of four-dimensional lattices with Hausdorff dimension less than three~\cite{Brell2016}.
Unfortunately, effectively string-like logical operators were ultimately found in the codes~\cite{lin2024proposals}, a problem known to be fatal to quantum memories due to the low energy barriers they induce~\cite{Bravyi2009NoGo, Haah_PRA2011}.
The layer code construction breaks spatial symmetry by sewing patches of topological orders (e.g.~surface codes) in an unstructured manner~\cite{Williamson_Layer2023}, but overwhelming entropic contributions persist within each patch~\cite{baspin2025}.
Our work carries this principle to its natural limit by seeking to remove entropic factors throughout the entire code.

A separate regime of work examines relaxed constraints on the problem, which has allowed considerable theoretical progress. For example, recent work in higher dimensions has yielded stable memories from non-Abelian topological orders~\cite{Hsin2024NonAbelian}. A different approach to self-correction is dynamical protocols which protect quantum information in non-stationary states based on cellular automaton decoders rather than Hamiltonian thermodynamics. Recent developments have unveiled protocols capable of stabilizing even the two-dimensional toric code under a phenomenological noise model~\cite{Balasubramanian2024Local}.

\subsection{Intuitive overview of construction}

We give a high-level summary of the code construction before detailing the necessary formalism and technical aspects in the following sections.
As described previously, we adopt the philosophy that spatial symmetries like translation invariance, while convenient for analytic treatment, incur entropic penalties too great to permit self-correction.
These entropy factors include both those which scale with the size of the symmetry group (up to extensively for fully translation-invariant models), as well as more important exponentially scaling factors associated with self-similarity.

Such an approach prevents the use of many of the tools used to construct and study quantum models; however, one that remains applicable is the family of product code constructions, a mechanism for generating CSS stabilizer quantum error-correcting codes from sets of classical linear codes.
Thus, in the spirit of maximally eliminating spatial symmetry, we define a family of quantum product codes based on pairs of two-dimensional classical codes on open boundary conditions, hosting codewords that are non-self-affine, in fact neither translationally invariant nor self-similar, in spatial structure.
We do require certain structures within these codes, chiefly that they have a fundamentally anisotropic character allowing us to associate a special graph, known as a ``slead'', to each code, which constrains the spatial flow of information about the classical spin state.
One way in which this structure aids understanding is in the treatment of codewords, which can be tunably inserted by a local deformation of the code.

The hypergraph product of such codes is a four-dimensional CSS stabilizer model, with the enlarged space consisting of a pair of two-dimensional subspaces associated with each factor, similar to the Cartesian product.
It thus remains to perform a reduction of the four-dimensional hypergraph product code rendering it embeddable in three-dimensional space in a geometrically local manner.
We achieve this reduction by a process we describe as ``coring'', being reminiscent of the algorithm for taking the 2-core of a graph.
The main idea is simply to eliminate qubits and stabilizers with low weight, whose existence is guaranteed by the slead property of the classical factors; we thus find this to be the natural language for formalizing our work.
Moreover, we show that the coring operation on the quantum code preserves logical qubits, code distance, and some properties of the energy landscape, thus enabling a viable route to self-correction in three dimensions.

\section{Classical slead codes}
\label{sec:classical_codes}

To construct suitable quantum product codes requires classical factor codes with certain geometric properties.
Specifically, we study geometrically local linear codes whose numbers of bits $n$ and checks $m$ are equal or nearly equal.
A convenient feature of such codes is that in the absence of redundancy, tuning this imbalance allows for precise control over the dimension of the code space.
In addition, this allows grouping together paired bit and check nodes in the Tanner graph into single vertices in a new directed graph with self-loops which we will refer to as a ``slead''.
By removing a small number of checks (in fact, only one) from the system, one adds nontrivial codewords whose properties can be determined from the slead; for this reason, we call the codes themselves slead codes.

A crucial feature of slead codes is the ability to introduce disorder, and in both this feature and the reliance on imbalance described above, they are like Euclidean variants of proposals realizing topological spin glass~\cite{placke2024topological}.
However, in many ways they resemble or intersect with translation-invariant codes described by the well-known algebraic formalism~\cite{Yoshida_PRB2013,Haah_Commuting2013,Haah_Algebraic2016}; thus, we refer to this framework as well.
The definitions below formalize the notion of a code defined on a graph which is geometrically local when embedded into a finite-dimensional space.

The codes we study are defined on a Delone point set $V$ embedded in $D$ dimensions.
Each vertex $v \in V$ at $\bm x_v \in \R^D$ hosts both a spin variable $\sigma_v = \{0,1\}$ and a check term $c_v = \prod_{v' \in S_v} (1-2\sigma_{v'})$ acting on a subset $S_v \subseteq N_v$, where $N_v$ is the neighborhood of vertices within a $D$-ball of fixed radius $r$ centered on $v$.
In a finite code, both the number of checks $m$ and number of bits $n$ are equal to $|V|$.
In the algebraic formalism, $V$ is a hypercubic lattice of linear size $L$ with the topology of a $D$-torus, and the check term is translation-invariant and specified by a $D$-variate polynomial $f$ (we refer readers to details of the formalism in Ref.~\onlinecite{Yoshida_PRB2013}).
We will precisely define both of the following terms in subsequent sections, but we note here such codes are capable of supporting an algebraic distance $d \sim L^\alpha$, $0 < \alpha \leq D$, and a logarithmic code barrier $\E \sim \log L$~\cite{Newman_SIAM2003}.

\subsection{Boundaries}
\label{sec:boundaries}

It is necessary to break translation invariance in order to obtain classical codes with suitable energetic and entropic properties, and the first step in doing so is to introduce boundaries to the system.
We require fully bounded spatial volumes with trivial topology.
Boundaries are introduced to both translation-invariant and more general codes by restricting to a hypercubic volume $\Lambda = [0,L]^D$, and truncating checks in the neighborhood of the bounding hyperplanes by removing their support on spins lying outside of $\Lambda$.
Because inclusion is determined by the position of vertices, $m=n$ after truncation.

Along with the introduction of boundaries, we impose two technical conditions on the classical codes, which together allow for precise control of the properties of the codewords by precluding redundancies and directing the flow of information through the graph.
First, each check must act nontrivially on its co-located bit.
For algebraic codes, this is equivalent to requiring $1 \in f$.
Second, in order to introduce the structure of a preferred direction to the graph, we require checks to satisfy the following ``half-space'' condition: there exists a vector $\bm t \in \R^D$ such that $|\bm\delta| > 0$ implies $\bm \delta \cdot \bm t > 0$ for all $\bm\delta = \bm x_{v'} - \bm x_v$, $v' \in S_v$, and $v \in V$.
That is, there is a codimension-1 hyperplane with normal vector $\bm t$ dividing the space into positive and negative halves, such that when shifted by $\bm x_v$ the check on $v$ acts only on spins strictly within the positive half, with the exception of its co-located spin.
For example, both conditions are satisfied by the Newman--Moore code $f=1+x+y$ with periodic boundaries, for $\bm t = (1,1)$.
In fact there is a subtlety, which is that the half-space condition can be \emph{locally} satisfied on a finite system with nontrivial topology but cannot be globally consistent.
In contrast, the trivial topology of $\Lambda$ allows the \emph{global} half-space condition, which turns out to be crucial to the following discussion.

\begin{figure}[t]
\centering
\includegraphics[width=0.8\columnwidth]{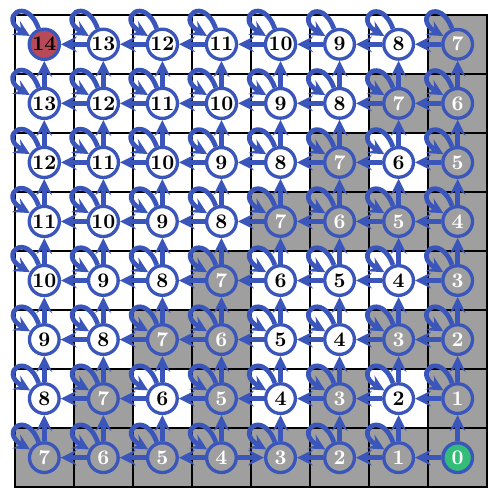}
\caption{\label{fig:slead}
\textbf{Example of a slead code.} The code shown has local Newman--Moore check $f=1+x+y$, up to truncation.
In this figure bits and checks are placed on the faces of the square lattice drawn in black, on the two-dimensional interval $[0,8]^2$.
The equivalent slead is drawn in blue, with labels indicating the level $p$ of each vertex in the topological ordering.
The single source vertex at $p=0$ is shaded green and the single sink vertex at $p=14$ is shaded red.
The self-loop on the source vertex is removed, and the induced codeword is indicated in gray on both the physical system and the slead.}
\end{figure}

All codes satisfying the above conditions have an associated graph we refer to as a \emph{self-loop enriched acyclic digraph} (slead), from which we name them slead codes.
An example is illustrated in Fig.~\ref{fig:slead}.
The slead is a graph $G = (V,E)$ comprising the vertices $V$ as well as a set of directed edges $E$ which are ordered pairs of elements in $V$.
An edge $(v,v')$ is included in $E$ if the check on vertex $v'$ acts nontrivially on the spin on vertex $v$.
(Graphically, an arrow points from $v$ to $v'$ if the check on $v'$ acts on the spin on $v$.)
The requirement that checks act on co-located bits includes all self-loops $(v,v)$ in $E$.
However, the strict inequality in the half-space condition applied locally guarantees that if $(v,v') \in E$ then $(v',v) \not\in E$ for $v' \neq v$.
The global half-space condition implies that there can be no directed cycles beside the self-loops.
(In fact, slead codes are more general than the half-space condition: e.g., one can generalize $\bm t$ to a vector field $\bm t(\bm x)$, apply any invertible deformation on $\Lambda$, and make the same statements.)

The self-loop attached to every vertex distinguishes a slead from a true acyclic digraph, as it contains $|V|$ directed cycles of unit length.
However, important properties of acyclic digraphs are unaffected: in particular, a slead supports the same \emph{topological ordering}.
This refers to a partial order on $V$ defined by reachability, where $v' > v$ if $v'$ is reachable from $v \neq v'$ via edges in $E$.
Any such finite graph necessarily contains \emph{source} vertices, which are not reachable from any other vertex, and \emph{sink} vertices, which do not reach any other vertex.
Further, the topological ordering partitions $V$ by assigning all source vertices to a level $p=0$, and other vertices $v$ to level $p+1$ if $p$ is the maximum level of all vertices reaching $v$.
In the example code in Fig.~\ref{fig:slead}, the level $p$ is indicated on each vertex.
For a geometrically local code the maximum level $p_\text{max} = \Omega(L)$.
We provide more illustrated examples of sleads in App.~\ref{app-sec:examples}.

The slead can immediately be used to show that boundaries trivialize the code space, even if a code protects logical bits under periodic boundaries.
Observe that a codeword is a ``flipped'' subset of vertices $C \subseteq V$ having even-parity overlap with all checks; equivalently, every vertex is directly reachable by an even number of vertices in $C$, including its self-loop.
(We refer to a vertex reachable from another by a single edge as ``directly reachable''.)
As sources are not reachable from any other vertex, the spins located on these vertices cannot be flipped in any codeword.
For the present purpose, we can thus consider the source vertices to be pruned from the slead; however, doing so simply introduces new source vertices---because sources must be present in any such graph---and these again cannot participate in any codeword.
By induction, the only codeword is the trivial one.

\subsection{Codewords}
\label{sec:classical_logical}

To protect a logical bit thus requires modification of a slead code, a process which is greatly simplified by the structures defined in the previous section.
To introduce one codeword, we remove the check on a single vertex while preserving its co-located bit.
Now the imbalance $m=n-1$ ensures that a nontrivial spin configuration involving flipping this bit will be introduced to the codespace, and because information about the spin configuration flows only one way along the slead, only spins on other vertices at higher levels are flipped in the new codeword.
That is, the topological ordering of the slead bounds the codeword, based on the level of the chosen vertex.
The specifics of the codeword, including its weight and its actual spin configuration, are determined by details of the code. 

The scheme of ``check depletion'' for injection of logical bits we employ here was used previously in the construction of fracton models without translation invariance~\cite{Tan2024Fracton}.
Because all checks are originally linearly independent by construction, the linear map $\omega: \F_2^n \to \F_2^m$ from the space of spins (identified with a basis $\sigma_i \equiv \hat e_i$, $i=1,\ldots,n$) to checks (identified with a basis $c_j \equiv \hat e_j$, $j=1,\ldots,m$) is invertible.
There are thus $|V|$ unique preimages $\omega^{-1}(c_j)$ violating only a single check $c_j$; each of these is a set of spins, described by a binary error vector in $\F_2^n$.
These are local minima of the classical Hamiltonian $H = -\sum_{v \in V} c_v$ with excitation energy $2$.
Our strategy for protecting logical information is to upgrade some of these local minima to global minima by removal of the corresponding violated checks.
Doing so introduces nullity into $\omega$, making it noninvertible.

In the slead, check depletion transforms an existing vertex $v$ at level $p$ into a source, which moreover lacks a self-loop so is capable of nucleating a codeword by flipping its spin.
We note that no other vertex at level $p$ or lower participates in the codeword. 
Moreover, the support of the codeword must be contained in the causal cone of $v$\footnote{By the \emph{causal cone} of $v$ we denote all vertices reachable from $v$.}.
Geometric locality implies that this causal cone has volume $O((p_\mathrm{max}-p) L^{D-1})$, as each level must contain at most $L^{D-1}$ vertices on average and a maximum of $p_\mathrm{max}-p$ levels are accessible.
Thus, the choice of depletion level provides a tunable upper bound on the distance, independent of the details of the code itself.

The codeword is constructed by proceeding level by level through the slead, flipping spins as necessary to satisfy the local parity checks.
To begin, all checks on vertices at level $p$ or lower are automatically satisfied.
The violated checks on level $p' \geq p+1$, which is the lowest level of vertices reachable from $v$, can be satisfied simply by flipping the co-located spin of each.
These spin flips---along with the spin flipped at level $p$---violate checks at level $p'' \geq p+2$ which themselves can be satisfied in the same way.
Repeating this process eventually satisfies all checks, completing the codeword.
In this way, the topological ordering controls the way information contained in the spin state is allowed to flow through the graph.
We refer readers to App.~\ref{app-sec:examples/classical_slead_codes} for an illustrated example of this process.

\subsection{Energetic properties of codes}
\label{sec:classical_barrier}

As suggested above by the language of causality, a codeword is naturally interpreted as the worldline of an excitation created at the vertex $v$, whose level $p$ is associated with a time index $t=0$, flowing along the slead until reaching the absorbing sink vertices at $t= p_\mathrm{max}-p$.
It is important to note that $\Z_2$ charge is not generally conserved, and in moving between time steps excitations will branch and collide, creating and annihilating charge, respectively.

The above ``quantization'' notion of time should be compared to a distinct concept of time associated with the thermalizing error channel studied in Sec.~\ref{sec:finite_temp}.
That is, any error, or collection of spin flips, $e \in \F_2^n$ has an associated \emph{energy barrier} $\E(e)$, which is the minimum energy penalty required to implement $e$ via a time-ordered sequence of single spin flips $\bm \sigma = \{\sigma_1,\sigma_2,\ldots\}$, or ``classical walk''.
Explicitly,
\begin{equation}
\E(e) = \min_{\bm \sigma \in \Sigma_e} \max_{\sigma_i \in \bm \sigma}\,|H \cdot w_i(\bm\sigma)|~,
\end{equation}
where $|\cdot|$ denotes Hamming weight, $e_i(\bm\sigma) = \sum_{j=1}^i \hat e_j$ is the spin configuration after observing $i$ steps of $\bm \sigma$, and the set $\Sigma_e$ contains all walks implementing $e$: that is, $\Sigma_e = \{ \bm \sigma \mid e_{|\bm \sigma|}(\bm \sigma) = e\}$.
The minimum energy barrier over codewords is referred to as the energy barrier of the code, and plays an important role in thermal memory: in particular, memory time is naively estimated to increase exponentially in the barrier~\cite{Brown2016RMP}.
However, we note that energy barriers are not fundamental to a linear code in the same sense as its distance or rate, but are instead properties of the classical Hamiltonian, or equivalently the set of checks.
For example, introducing redundant checks can enhance energy barriers without affecting the code space.

One way to guarantee sufficiently large energy barriers is through a property known as \emph{confinement}.
Although definitions vary, we use the following: a code on $n$ bits is confining if every error $e$, $|e| \leq \delta(n)$, satisfies $\gamma(|H\cdot e|) \geq |e|$ for some monotonically increasing functions $\gamma$ and $\delta$~\cite{Quintavalle_PRX2021}.
If $\gamma(x) = O(e^x)$, the code is said to be logarithmically confining, and if $\gamma(x) = O(x^\alpha)$ for some $\alpha \geq 1$, the code is algebraically confining.
Confinement implies \mbox{$\E(e) \geq \gamma^{-1}(\mathrm{min}[|e|,\delta(n)])$}.
The strongest possible case is linear confinement, which implies single-shot error correction~\cite{Bombin_PRX2015} and is closely related to the presence of topological spin-glass order on expander graphs~\cite{placke2024topological}, but cannot be realized in finite-dimensional geometrically local stabilizer codes.

\subsubsection{Translationally invariant checks}
\label{sec:fractal_eb}

\begin{figure}[ht!]
\centering
\includegraphics[width=0.48\columnwidth]{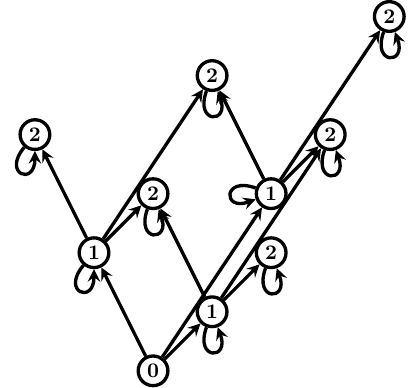}
\includegraphics[width=0.48\columnwidth]{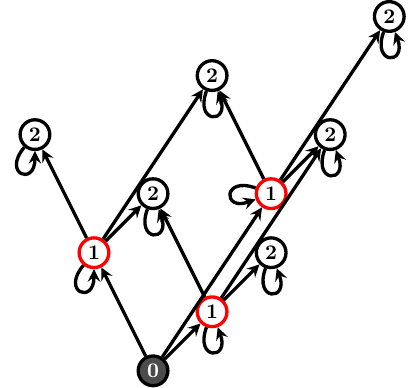}\\
(a)\hspace{4cm}(b)\\
\includegraphics[width=0.48\columnwidth]{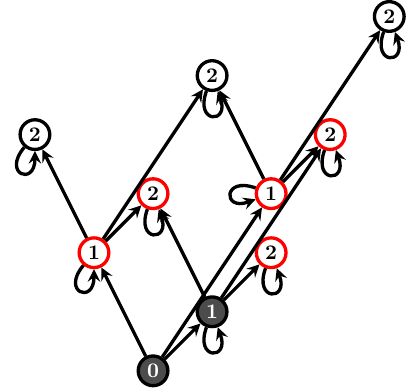}
\includegraphics[width=0.48\columnwidth]{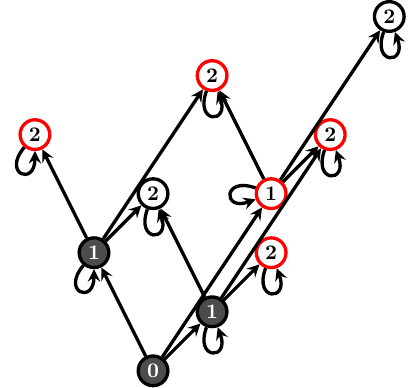}\\
(c)\hspace{4cm}(d)\\
\includegraphics[width=0.48\columnwidth]{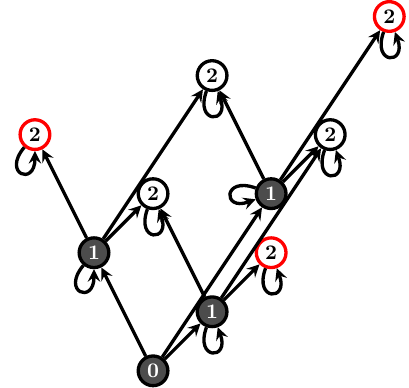}
\includegraphics[width=0.48\columnwidth]{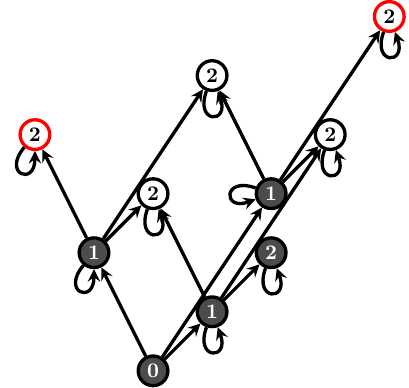}\\
(e)\hspace{4cm}(f)\\
\includegraphics[width=0.48\columnwidth]{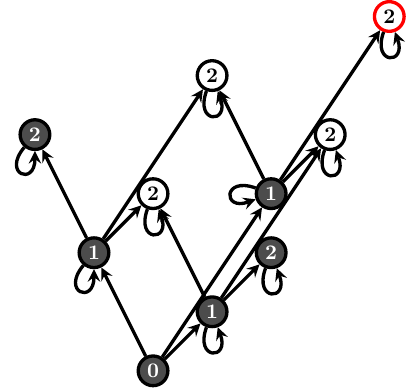}
\includegraphics[width=0.48\columnwidth]{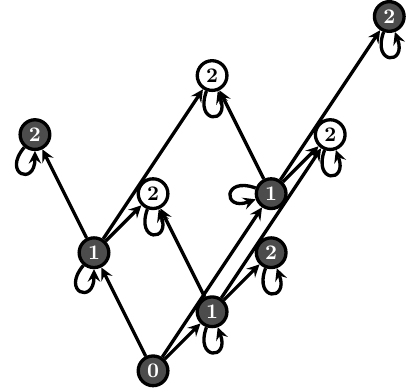}\\
(g)\hspace{4cm}(h)\\
\caption{\label{fig:ti_barrier}
\textbf{Cancellation in algebraic codes.} Panels (a)--(h) show a classical walk implementing the codeword in a small slead code described by a polynomial $f$ in the bulk.
A filled vertex indicates a spin flip, and a red boundary indicates a violated local check.
Cancellation occurs at levels $p=2^a$, $a \in \Z$, and despite the introduction of boundaries the walk achieves a logarithmic barrier.
Note that the number of spin flips at both levels $p=1,2$ in the codeword is only $|f|-1=3$.
}
\end{figure}

Translation-invariant codes described by the algebraic formalism exhibit scale invariance in their codewords, and with this residual symmetry is associated a logarithmic code barrier~\cite{Yoshida_PRB2013}.
We claim that this slow scaling is not generic in slead codes without translation invariance, and to support this argument we first explicitly derive the logarithmic barrier for the case of an algebraic slead code described by a polynomial $f$.
In the following we will again treat the process of sequentially implementing a codeword as propagating a worldline along the time-slices specified by the graph topological ordering.

Suppose that checks are described by an $\eta$-term polynomial $f$, so that a spin flip at vertex $v$ violates $\eta$ checks on vertices separated from $v$ by a monomial in the term-wise inverse $\overline f$, one of which is located on $v$.
By flipping the spins on these vertices, we violate checks on vertices separated from $v$ by a term in $\overline f^2$, which again has only $\eta$ terms over $\F_2$.
We have thus performed a collective transport operation on $\eta$ excitations.
This collective transport process is illustrated in Fig.~\ref{fig:ti_barrier}, which shows an example of the implementation of a codeword.
Sequential applications of this pattern always generate $\eta$ excitations but at larger separations $\overline f^4$, $\overline f^8$, and so on.
The energy cost of the original spin flip is evidently $\eta$, and one can confirm that the energy upon subsequent spin flips $i = 1,\ldots,\eta-1$ is $(i+1)(\eta-i)$.
The barrier for a single collective transport step $\overline f \to \overline f^2$ is thus $\E_1 = \lfloor\frac{\eta+1}{2}\rfloor\lceil\frac{\eta+1}{2}\rceil = O(\eta^2)$.

Now for the step $\overline f^2 \to \overline f^4$ one repeats the above sequence in its entirety for each component of $\overline f^2$, of which there are $\eta-1$, excluding the excitation on vertex $v$.
This amounts to asynchronous evolution of these $\Z_2$ charges.
The energy required to implement component $i = 1,\ldots,\eta-1$ is given by $\E_1 + i(\eta-i-1) + 1$, and the barrier for the complete transport operation is $\E_2 = \E_1 + \lfloor \frac{\eta-1}{2} \rfloor \lceil \frac{\eta-1}{2} \rceil + 1$.
This form applies generally to the step $\overline f^{2^q} \to \overline f^{2^{q+1}}$; thus one sees that the barrier at scale $q$ is $\E_q = O(q \eta^2)$.
If the original vertex $v$ is a source with depleted check at level $p$, $p_\mathrm{max}-p \sim L$, so that $q \sim \log L$, then this process implements a fractal codeword with $\mathrm{poly}(L)$ weight while observing only a $\log(L)$ barrier.

\subsubsection{Disordered checks}
\label{sec:random_barrier}

The preceding section establishes how translation invariance reduces the energy cost of scale-invariant errors in a slead code, including  codewords, by providing sets of $\Z_2$ charges which quickly annihilate within a worldline.
By evolving these charges together via spin flips, one avoids having to maintain many costly intermediate charges simultaneously.
We argue in the present section that such structures are exponentially unlikely without translation invariance.

\begin{figure}[t]
\centering
\includegraphics[width=\columnwidth]{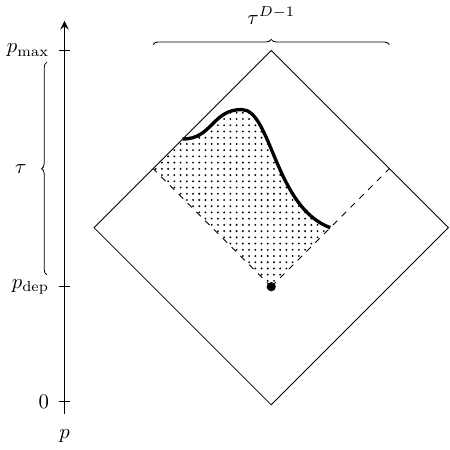}
\caption{\label{fig:typ_barrier}
\textbf{Worldline implementations in unstructured codes.}
A slead code is shown, oriented such that the graph topological ordering labels $p$ correspond to the vertical axis.
We perform check depletion on an indicated vertex at level $p_\mathrm{dep} = p_\mathrm{max} - \tau$, and the resulting lightcone is indicated by the dashed line.
Due to geometric locality, the width of the lightcone scales as $\tau^{D-1}$.
A partial worldline implementation of the codeword is indicated, with excitations generated along the thick boundary between the shaded (implemented spin flips along the worldline) and unshaded (no spin flips) regions.
This is a coarse-grained counterpart to the middle panels of Fig.~\ref{fig:ti_barrier}.
}
\end{figure}

Consider a slead code with no additional structure beyond geometric locality in $D$ dimensions.
A codeword is inserted via check depletion on a vertex $v_\mathrm{dep}$ at level $p_\mathrm{dep}$, leading to a wordline with weight $d$ scaling as $\tau^\alpha$, $0 \leq \alpha \leq D$, where $\tau = p_\mathrm{max}-p_\mathrm{dep}$.
The causal cone of $v_\mathrm{dep}$ defines the region of the graph bounding the support of the codeword and has maximum width scaling as $\tau^{D-1}$.
This geometry is illustrated in Fig.~\ref{fig:typ_barrier}.

A classical walk for a worldline implementation of the codeword begins by flipping the local spin at $v_\mathrm{dep}$ and continues as in Fig.~\ref{fig:ti_barrier}, at each timestep flipping a spin co-located with a parity check violated by the current configuration.
The coarse-grained picture of the spin state at intermediate times consists of two regions, one connected to $v_\mathrm{dep}$ in which the codeword is already implemented, and one region containing no spin flips.
The energy of the configuration arises along the interface between these regions, shown in Fig.~\ref{fig:typ_barrier} as a thick solid line.

In order to completely implement the codeword the interface between regions must achieve a length scaling at least as $\tau^{D-1}$.
Due to the assumption of statistical homogeneity, the number of spin flips connected to checks on the interface scales as $\tau^{\alpha-1}$.
If each check observing a flipped spin has equal probability of being satisfied or violated, in the limit $\tau \gg 1$ the energy $\E$ of the state
is well approximated by a Gaussian of mean $\mu = \frac12 t^{\alpha-1}$ and variance $\sigma^2 = \frac14 t^{\alpha-1}$:
\begin{align}
p(\tau,\E) &= \frac12 \left(1 + \mathrm{erf}\left[\frac{2\E - \tau^{\alpha-1}}{\sqrt 2 \tau^{\frac{\alpha-1}{2}}}\right]\right)\\
&\sim \frac{\tau^{\frac{\alpha-1}{2}}}{\sqrt{2\pi}(\tau^{\alpha-1} - 2\E)} \exp\left[-\left(\frac{\tau^{\alpha-1} - 2\E}{\sqrt 2 \tau^{\frac{\alpha-1}{2}}}\right)^2\right].
\end{align}
That is, the likelihood of any parametric reduction of $\E$ below $\tau^{\alpha-1}$ scales as
\begin{equation}
p(\tau,\E \ll \tau^{\alpha-1}) \sim \tau^{-\frac{\alpha-1}{2}}\exp\left(-\tau^{\alpha-1}\right)~.
\end{equation}

\subsection{Pinwheel slead codes}
\label{sec:pinwheel_slead_codes}

\begin{figure}[t]
\centering
\includegraphics[width=\columnwidth]{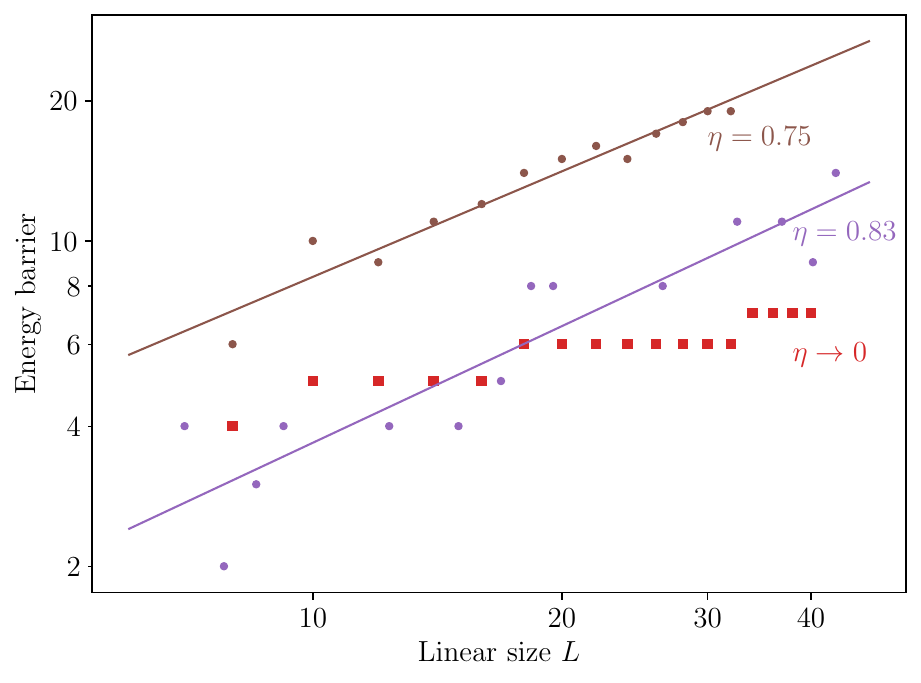}
\caption{\label{fig:eb_classical}
\textbf{Energy barriers in classical fractal codes.} Numerical upper bounds on energy barrier versus code size in two-dimensional classical fractal codes, obtained by greedy search over minimal Pauli walks. As a contrasting example, the Newman-Moore code with periodic boundary conditions (red) exhibits logarithmic barrier with code size. The pinwheel codes (purple) discussed in  Sec.~\ref{sec:pinwheel_slead_codes}, and another family of fractal codes that are translation-invariant in the bulk but with open boundaries (brown), exhibit algebraic scaling of energy barrier $\E \sim L^\eta$ for $\eta > 0$.}
\end{figure}

Based on the arguments in the previous section, we seek to break translation symmetry in slead codes in order to take advantage of the more energetically favorable typical behavior.
We examine two strategies for doing so: first, through a code having a translation-invariant bulk but with boundaries as discussed in Sec.~\ref{sec:boundaries}; and second, by defining the code on a disordered aperiodic lattice.
In both cases, numerics suggest that the barrier to implement a codeword scales algebraically.

A slead code obtained by introducing boundaries to a translation-invariant fractal code (as in, for example, Fig.~\ref{fig:slead}) can exhibit at best only logarithmic confinement, as errors in the bulk are not affected.
However, it may be that an algebraic barrier applies to codewords, if they have nontrivial support along the boundaries.
We provide numerical support for this claim in Fig.~\ref{fig:eb_classical}, and note that a similar effect was observed in the related setting of boundaries for Haah's cubic code~\cite{Aitchison_Strings2023}.
For details of the calculations shown in Fig.~\ref{fig:eb_classical}, see App.~\ref{sec:barrier_bound}.
Nevertheless, the lack of algebraic confinement allows large errors to exist in the bulk with low energy penalty, which is undesirable for the purposes of the memory.
This is because the entropy of walks implementing a codeword with minimal or nearly minimal energy barrier is greatly increased due to low-energy ``haven'' spin configurations providing a partial implementation of the codeword.

A stronger approach is to introduce disorder into the system via a geometrically local graph without spatial symmetry.
For this purpose we utilize the aperiodic ``pinwheel tiling'' of the two-dimensional plane, which lacks any exact symmetries but is statistically homogeneous and isotropic~\cite{Radin_Space1992,Radin_1994Pinwheel,Radin_Aperiodic1997,Frettloh_Substitution2008}.
An aperiodic point set is preferable to pure randomness because of its feature of hyperuniformity, meaning that density fluctuations vanish in spatial regions that are large compared to the typical spacing~\cite{torquato2003local}.
Consequently, no rescaling is required in order to maintain consistent locality; rare regions do not interfere with the code properties; and both degree and locality radius of the resulting graph are controlled.
The cost paid for these properties is the development of finite higher moments in the lattice correlation functions.
It is therefore a conjecture that the arguments about truly unstructured codes made in Sec.~\ref{sec:random_barrier} apply also to disordered codes of this type.

\begin{figure}[ht]
\centering
\includegraphics[width=0.65\columnwidth]{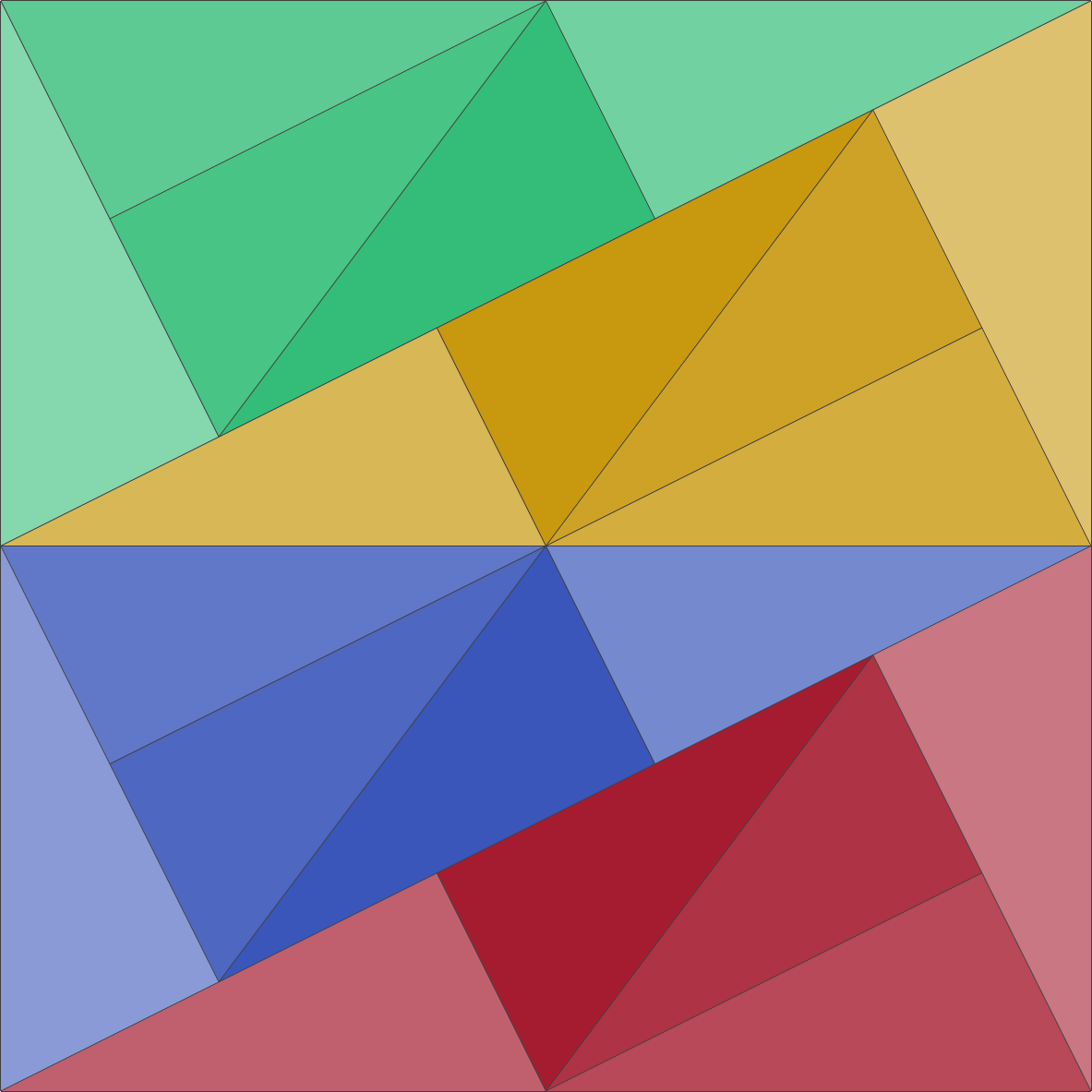}\\
(a)\\
\includegraphics[width=\columnwidth]{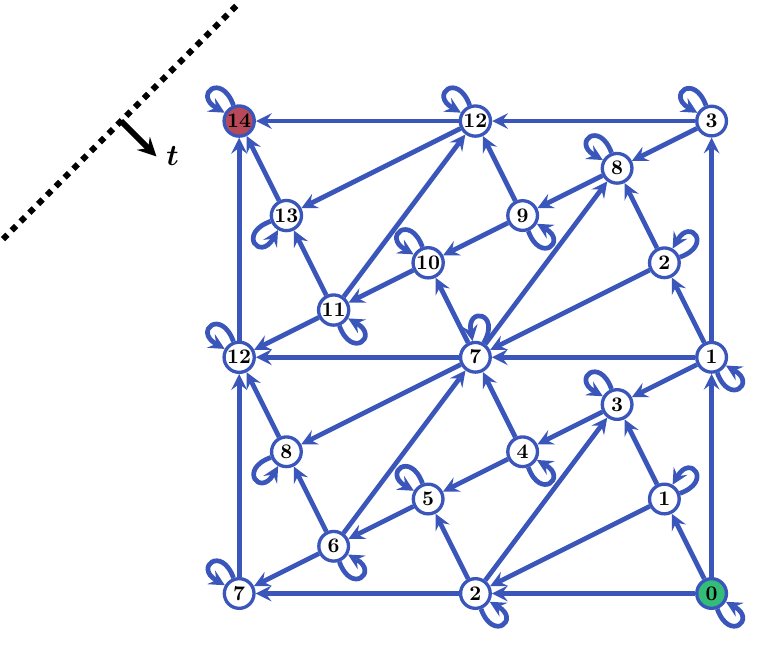}\\
(b)
\caption{\label{fig:pinwheel}
\textbf{Pinwheel codes.}
\textbf{(a)} A pinwheel tiling of a square is pictured, based on four triangular prototiles subject to a single round of substitution.
While the original tiles are oriented along the coordinate axes, descendant triangles appear at an irrational angle $\arctan\left(1/2\right)$, leading to the statistical isotropy of the tiling after many rounds of substitution.
Due to the choice of initial tiles, after any number of rounds of substitutions the tiling has a single symmetry of order two generated by a $\pi$ rotation about the center.
\textbf{(b)} A pinwheel slead code associated with the tiling is shown, obtained by applying the half-space criterion with direction $\bm t$.
Vertex labels indicate level $p$ in the topological ordering, with the single source vertex shaded green and sink vertex shaded red. 
Note that the choice of either $\bm t$ or a check depletion site breaks the rotational symmetry of the underlying pinwheel tiling.
}
\end{figure}

The pinwheel tiling was previously used to define fracton models in three and four dimensions~\cite{Tan2024Fracton}. 
In that case the quantum codes are again based on classical codes, having local check terms determined by the binary reduction of the graph Laplacian.
Such codes do not satisfy the slead property, so we instead again use the half-space condition defined in Sec.~\ref{sec:boundaries} to obtain a suitable code.
See Fig.~\ref{fig:pinwheel} for an example.

The pinwheel tiling supports a substitution rule, in which all prototiles of a given tiling are deterministically replaced by new tiles, followed by rescaling the entire tiling: the result is a new tiling that is locally isomorphic to the original.
Starting with a finite patch, we use the substitution rule to define a family of codes by applying the half-space condition to the graph after one or more rounds of substitution.
In particular, the family studied here begins with an initial set of four triangular prototiles covering a square, as shown shaded in different colors in Fig.~\ref{fig:pinwheel}.
As shown, each prototile divides into five congruent descendant tiles in each round of substitution.
The same substitution pattern (rotated and reflected, as necessary) is then applied to each of these tiles to generate a more fine-grained tiling, and accordingly a larger classical code.
Owing to the arrangement of the prototiles, this tiling does have a single spatial symmetry generated by a $\pi$ rotation about the center of the square, though the code does not share this symmetry.
We refer to Sec.~\ref{sec:finite_temp/codes} for more details of the exact factor codes utilized in simulations.

\section{Cored product codes}
\label{sec:cored_product_codes}

\subsection{Product codes}
\label{sec:cored_product_codes/product_codes}

A quantum product code is generated from a set of classical factors, which are linear codes on $n$ bits with $m$ terms as described in Sec.~\ref{sec:classical_codes}, with parameters $[n,k,d]$, where $k$ is the dimension of the code space and $d$ is the distance~\cite{MacKay_2003}.
These codes need not be either low-density or geometrically local.
Such a classical code can be described by a map $\omega:\F_2^n \to \F_2^m$.
Similarly, a quantum CSS code can be represented by a symplectic stabilizer map 
\begin{equation}
\omega=\left(\begin{array}{@{}c|c@{}}
\omega_X & 0 \\
0 & \omega_Z
\end{array}\right)~,
\end{equation}
with $\omega_X$ and $\omega_Z$ specifying the action of $X$ and $Z$ stabilizers, respectively~\cite{Calderbank_PRA1996,Steane_PRL1996}.
Given two classical factors with maps $\omega_1$ and $\omega_2$, the \emph{hypergraph product} code is defined by~\cite{Tillich_TIT2014}
\begin{equation}
\omega=\left(\begin{array}{@{}cc|cc@{}}
\omega_1 \otimes \I & \I \otimes \overline\omega_2 & 0 & 0 \\
0 & 0 & \I \otimes \omega_2 & \overline \omega_1 \otimes \I
\end{array}\right),
\label{eq:hgp}
\end{equation}
where $\overline\omega_i = \omega_i^{-1}$ is the dual and $\I$ the identity map.
A convenient way to understand a classical linear code is as a length-1 chain complex with boundary map $\omega$, and codewords being elements of $\ker\,\omega$.
Equation \eqref{eq:hgp} is equivalent to the homological product code~\cite{Bravyi_ACM2014}, which reproduces the tensor product of chain complexes, under a change of convention $\omega_2 \leftrightarrow \overline\omega_2$.
We use the present notation for convenience.
The parameters $[[n_\q,k_\q,d_\q]]$ of a hypergraph product code are determined from its classical factors as
\begin{equation}
\begin{gathered}
n_\q = n_1 n_2 + m_1 m_2~,\\
d_\q = \min(d_1,d_2,\overline d_1,\overline d_2)~,\\
k_\q = k_1 k_2 + \overline k_1 \overline k_2~,
\label{eq:hgp_params}
\end{gathered}
\end{equation}
where $\overline k_i$ and $\overline d_i$ denote parameters of the dual code.
Moreover, for a pair of classical factors each defined on a graph, the hypergraph product code is defined on the Cartesian product of the graphs.
There is therefore a natural definition of inclusions $\iota_1$, $\iota_2$ (up to the choice of base points) and projection maps $\pi_1$, $\pi_2$ associated with each factor.
If the classical factors are either low-density or geometrically local, the hypergraph product inherits the same sense of locality.

The hypergraph product has proved to be a highly versatile tool in constructing quantum error correcting codes.
For example, the quest for good quantum LDPC codes with linear distance and dimension eventually succeeded through extensive study of ways to supercede the limitations of the hypergraph product code parameters, in particular the distance bound~\cite{Panteleev_ACM2022,Dinur2023Good}.
A key additional ingredient in the breakthrough constructions was the notion of a lift, which allows a product code to be reduced to its cosets under a symmetry group shared by both classical factors~\cite{Panteleev_ITI2022,Breuckmann_TIT2021}.
In these lifted or balanced product codes, the quotient group is often a product of cyclic groups, each of order $\mathrm{poly}(n)$.

\subsection{Dimensional reduction by coring}
\label{sec:cored_product_codes/coring}

The quotient prescription summarized above cannot be naively applied to product codes lacking symmetries, including those with boundaries.
Unfortunately the slead codes described in Sec.~\ref{sec:classical_codes} with the most favorable properties for use as classical factors are exactly those without symmetry.
To solve the problem of obtaining a quantum code in fewer spatial dimensions, we utilize the geometric rather than topological structure of the product.
As an overview, in this section we first discuss a general measurement-based protocol for qubit deletion in stabilizer codes which does not cause anticommutation or increase the logical dimension (i.e., number of logical qubits).
We then demonstrate that when specialized to act on a hypergraph product of slead codes, this process of \emph{coring} can ultimately produce codes capable of being embedded in fewer spatial dimensions while maintaining geometric locality.

\subsubsection{Measurement protocol for deletion}

We implement qubit and stabilizer deletion in such a way as to maintain commuting generators and avoid introducing additional logical qubits, by following a certain protocol of Pauli measurement.
Generally, consider the mixed state in the code space of a stabilizer code on $n$ qubits generated by $\S$:
\begin{equation}
\rho = \prod_{s \in \S} \Pi_s \rho^\infty_n \Pi_s~,
\quad
\Pi_s = \frac{1+s}{2}~,
\end{equation}
where $\rho^\infty_n = \I_n$ is the infinite-temperature mixed state on $n$ qubits and equalities are up to normalization.
Upon measuring a Pauli operator $q \in \mathcal P_n$ (ignoring the measurement outcome), if $q \in \langle\S\rangle$, the measurement is deterministic and has no effect on the state.
If instead $q \notin \langle\S\rangle$, $q$ may or may not commute with $\S$.
In the former case, $q$ is an element of the logical space; after measurement it is added to $\S$ and the rank of the mixed state is reduced---equivalently the logical dimension decreases by one.
In the latter case, a basis can be chosen such that $q$ anticommutes with only one generator $s' \in \S$.
After the measurement $s'$ is replaced by $q$ in $\S$, with the logical dimension unchanged.
Pauli measurement cannot increase the logical dimension; moreover, if $q$ is local and anticommutes with only a single element of a local stabilizer basis, the stabilizer group remains locally generated.
Our algorithm relies on the ability to perform measurements satisfying these criteria in order to preserve both logical dimension and locality.

We now specialize to CSS codes $\S = \S_X+\S_Z$, which include all examples discussed so far.
In order to maintain locality we do not allow any rotation increasing the support of any stabilizer generator; to this end, we measure only single-qubit Pauli operators, on qubits acted on by a single generator from either the $X$ or $Z$ sector.
For example, if a qubit $i$ is acted on by one $s^Z_j \in \S_Z$, along with some elements of $\S_X$, one can measure $X_i$ with the result that $s^Z_j$ is replaced by $X_i$ in the generating set.
Additionally, any element of $\S_X$ acting on qubit $i$ may be multiplied by $X_i$, so that the only stabilizer generator acting nontrivially on qubit $i$ is $X_i$ itself.
The qubit is thereby decoupled from the code and exists by itself as a trivial paramagnet.
Equivalently, we say that the measurement destroys the coherence of the qubit, leaving only a classical bit, namely the outcome.
Both qubit and stabilizer can thus be removed from the system with no further effect.
Incidentally, this rule was recently identified as a gadget for fault-tolerant chain maps in a different context~\cite{pesah2025}.

\begin{figure*}
{\centering
\begin{minipage}{\linewidth}
    \begin{algorithm}[H]
    \caption{\raggedright Measurement-based deletion for CSS quantum codes}
    \label{alg:css_bd}
    \begin{algorithmic}
    \Require Stabilizer generators $\S^0 = \S^0_X + \S^0_Z$, mixed state in the initial code space $\rho^0 = \prod_{s \in \S^0} \Pi_s \rho^\infty_{Q^0} \Pi_s$ on qubits $Q^0 = \{1,\ldots,n^0\}$, and deletable set $D \subseteq Q^0$.
    \Ensure Stabilizer generators $\S = \S_X + \S_Z$, mixed state in the final code space $\rho = \prod_{s \in \S} \Pi_s \rho^\infty_Q \Pi_s$ on qubits $Q \subseteq Q^0$, $|Q| = n$.
    \algrule
    \State $(\S,\rho,Q,n_\mathrm{del}) \gets (\S',\rho',Q',1)$
    \While{$n_\mathrm{del} > 0$}
        \Comment{Continue with next round so long as previous had nontrivial effect.}
        \State $n_Z \gets~$\Call{Measure}{$Z,\S,Q,D,\rho$}
            \Comment{Measure out $Z$-type stabilizer generators.}
        \State $n_X \gets~$\Call{Measure}{$X,\S,Q,D,\rho$}
            \Comment{Measure out $X$-type stabilizer generators.}
        \State $n_\mathrm{T} \gets~$\Call{RemoveTrivial}{$\S,Q,\rho$}
            \Comment{Remove trivialized stabilizer generators and their support.}
        \State $n_\mathrm{del} \gets n_Z + n_X + n_\mathrm{T}$
            \Comment{Number of deleted qubits in current round.}
    \EndWhile
    \State \Return $(\S,\rho)$
    \algrule
    \Function{Measure}{$A,\S,Q,D,\rho$} \Comment{Measures out stabilizer generators in sector $A \in \{X,Z\}$.}
    \State $n_\text{meas} \gets 0$
    \If{$A = X$} $B \gets Z$ \textbf{else} $B \gets X$
        \Comment{Measurement basis $B$ is the conjugate sector to $A$}
    \EndIf
    \State $M \gets \smash{\{i \in D \mid |\partial^A_i| = 1\}}$
        \Comment{Deletable qubits each involved only in a single $A$-type stabilizer generator.}
    \For{$i \in M$}
        \Comment{Loop over qubits $i$ to be deleted.}
        \State $\S_A \gets \S_A - \partial^A_i$
            \Comment{Remove $A$-type stabilizer generators that contain qubit $i$.}
        \State $\S_B \gets \S_B - \partial^B_i + \{s B_i \}_{s \in \partial^B_i}$
            \Comment{Remove qubit $i$ from $B$-type stabilizer generators.}
        \State $Q \gets Q - \{ i \}$
            \Comment{Remove qubit $i$ from the code.}
        \State $n_\text{meas} \gets n_\text{meas} + 1$
        \State $\rho \gets \Pi_{B_i} \rho \Pi_{B_i} = \left( \prod_{s \in \S} \Pi_s \rho^\infty_Q \Pi_s \right) \otimes \Pi_{B_i}$
            \Comment{Mixed state is a product state after measurement.}
        \State $\rho \gets \sum_{s \in \S} \Pi_s \rho^\infty_Q \Pi_s$
            \Comment{Update mixed state to be on remaining qubits.}
    \EndFor
    \State \Return $n_\text{meas}$
    \EndFunction
    \algrule
    \Function{RemoveTrivial}{$\S,Q,\rho$}
        \Comment{Removes trivialized stabilizer generators.}
    \State $n_\text{triv} \gets 0$
    \For{$s_j \in \S$}
        \Comment{Loop over all stabilizer generators.}
        \If{$|\Sigma_j| = 1$}
            \Comment{Act only on stabilizer generators that involve a single qubit.}
            \State $\S \gets \S - \{s_j\}$
                \Comment{Remove the stabilizer generator.}
            \State $Q \gets Q - \Sigma_j$
                \Comment{Remove the single qubit in the support of the stabilizer generator.}
            \State $n_\mathrm{triv} \gets n_\mathrm{triv}+1$
            \State $\rho \gets \sum_{s \in \S} \Pi_s \rho^\infty_Q \Pi_s$
                \Comment{Update the mixed state.}
        \EndIf
    \EndFor
    \State \Return $n_\text{triv}$
    \EndFunction
    \end{algorithmic}
    \end{algorithm}
\end{minipage}
\par
}
\end{figure*}

Measurement-based deletion of a CSS code is presented in detail in Alg.~\ref{alg:css_bd}, which takes as input the mixed state of an initial code along with tabulated stabilizer generators $\S = \S_X + \S_Z$ and returns the mixed state of the reduced code and a table of its stabilizer generators.
As described above, single-qubit Paulis $X_i$ and $Z_i$ are measured on qubits $i\in D \subseteq \{1,\ldots,n\}$, if and only if these operators fail to commute with exactly one generator, and the measured qubit is removed from the system.
Any trivialized stabilizers, whose support has been reduced to a single qubit, are also removed.
This process is repeated until no further measurements can be made.
In the algorithm, we denote by $\partial^X_i \subset \S_X$ and $\partial^Z_i \subset \S_Z$ the $X$ and $Z$ stabilizer generators, respectively, acting nontrivially on qubit $i$, and by $\Sigma_j$ the support of a stabilizer generator $s_j$, $j=1,\ldots,m$.
We do not explicitly update these quantities but assume that they are always queries to the current state of the stabilizer group.

We make some further remarks on Alg.~\ref{alg:css_bd}.
First, actually implementing the measurement is unnecessary unless one truly wants to act on a code state, and in practice one simply iteratively updates the lists of qubits and stabilizers based on their connectivity.
Second, it is not evident that a code should contain any qubit satisfying, say, $|\partial^Z_i| = 1$; as we discuss in the following section, they arise from sink vertices if the classical factors are slead codes.
Third, we have suppressed the outcomes of the Pauli $X_i$ and $Z_i$ measurements but after a single-body measurement the qubit is conventionally considered to be converted into this classical bit of data.
It is thus belaboring the point to update $\rho$ in two steps, and one can think instead of such a measurement in the usual way, as having destroyed the qubit.

Under this algorithm the logical dimension is reduced if and only if the set of measurements and trivialized stabilizers generates any logical representative.
However, even if measurement does not affect the logical dimension, the code distance may decrease, and we consider this possibility in the following section.

\subsubsection{Coring slead product codes}
\label{sec:cored_product_codes/coring_slead_product_codes}

Having introduced the measurement-based deletion protocol in the previous section, we now apply it to the product of slead codes.
In this specialized context, we refer to the protocol as \emph{coring}, as its action is related to the generalization of the graph $2$-core to the slead.
The interaction of the coring algorithm with the method of check depletion for logical insertion in the factor codes will lead to a variety of favorable properties including preservation of ligical dimension, distance and some barriers, as shown in Sec.~\ref{sec:cored_prop}.

We first recall that a hypergraph product, being a CSS code, contains two sectors with separate linear maps $\omega_X$ and $\omega_Z$.
The structure of these maps is given in \eqref{eq:hgp}, from which one observes that the slead property is inherited by each sector.
Throughout this section, we abuse notation by referring to all of the slead, the classical factor, and the classical vertex set indistinguishably where the meaning is clear from context.
For instance, $\omega_X$ as the product slead now contains two species of qubit per vertex, which we refer to as red and blue.
We accommodate this generalization by coloring the directed edges $(v,v')$ either red or blue, depending on which qubit on vertex $v$ is acted on by the stabilizer on vertex $v'$; equivalently, color distinguishes the edges arising from each classical factor.
We note that one self-loop of each color is attached to each vertex in the quantum slead.
The structure of $\omega_Z$ is similar, using classical maps $\omega_2$ and $\overline\omega_1$ for red and blue qubits, respectively.

Henceforth we make two specializations to the measurement-based deletion protocol applied to quantum slead codes with classical factors $\omega_1$ and $\omega_2$.
First, we introduce a single nontrivial codeword to both $\omega_1$ and $\omega_2$ via check depletion as described in Sec.~\ref{sec:classical_logical}; while $\overline{\omega}_1$ and $\overline{\omega}_2$ have trivial codespace.
Second, we specify the deletable subset $D$ referenced in Alg.~\ref{alg:css_bd} to be the set of blue qubits.
Denoting the classical components in a specific sector as $\omega_R$ and $\omega_B$ for red and blue qubits, respectively, vertices arising from sinks in $\omega_B$ are deleted by the procedure.
We note that while red qubits are not themselves candidates for deletion, they may still be removed if the weight of nearby stabilizers is sufficiently reduced.
In this context, the deletion protocol is similar to the computation of the 2-core of a graph.
We thus refer to the measurement-based deletion process on a slead code as \emph{coring}.

\begin{figure}[t]
\centering
\includegraphics[width=\columnwidth]{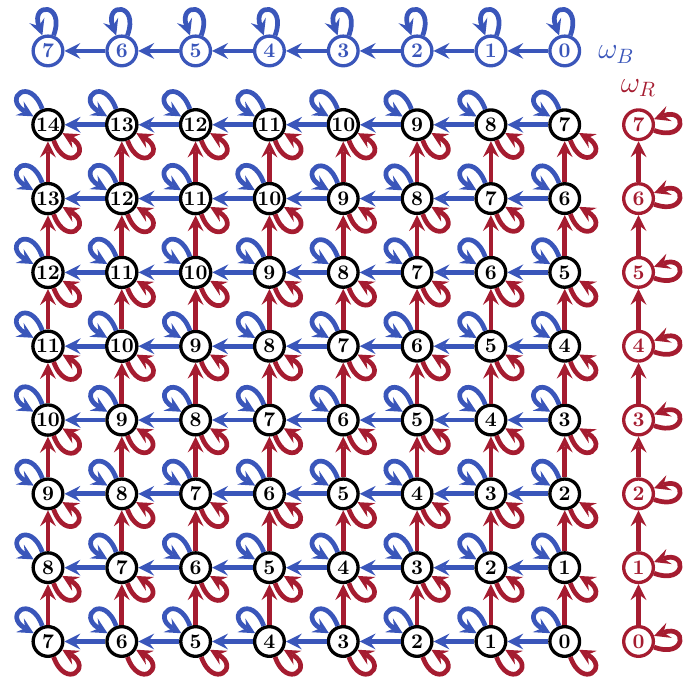}
\caption{\label{fig:quantum_code}
\textbf{Slead representation of an example product code hosting no logical information.}
A slead associated with one sector of a quantum product code is shown.
It is read in the following way: any vertex with arrows into it contains a stabilizer (or classical check), whereas any vertex emanating arrows of one color hosts a qubit of that color (or classical bit); both statements are inclusive of self-loops.
As the classical codes $\omega_R$ and $\omega_B$ are one-dimensional Ising models, the quantum stabilizers in the bulk are those of the two-dimensional toric code.
However, the chosen boundaries do not admit a logical qubit.
Three types of graph topological ordering can be imposed, associated with either red or blue edges, or both.}
\end{figure}

To be more explicit about the action of coring, suppose instead that no check depletion is performed in either factor $\omega_R$ or $\omega_B$, so no logical qubit is encoded.
An example of such a quantum slead without depletion is shown in Fig.~\ref{fig:quantum_code}.
In order to treat both sectors simultaneously, we use the general language of \emph{direct} and \emph{conjugate} sectors, where the direct sector may refer equally well to either $\omega_X$ or $\omega_Z$.
In the first round of coring in the direct sector the blue qubits on vertices $M_1^\mathrm{dir} = \omega_R \times \{p_\mathrm{max}\}_B$, where $\{p_\mathrm{max}\}_B$ denotes the set of sinks of $\omega_B$, are deleted along with their co-located direct stabilizers.
Recall that in the conjugate sector the classical maps are $\overline\omega_B$ and $\overline\omega_R$ for red and blue qubits, respectively.
Thus after the deletion of these blue qubits, the conjugate stabilizers on the vertices in $M_1^\mathrm{dir}$ act only on their co-located red qubits.
Each pair of red qubits and conjugate stabilizers is thus trivial and no longer participates in the code, so is removed.
Thus, in one step all objects hosted on $M_1^\mathrm{dir}$ are eliminated, and these vertices are pruned from the system.
The story proceeds similarly for measurements on blue qubits affecting the conjugate sector, which live on another set of vertices $M_1^\mathrm{conj}$.
It is clear that the entire code is deleted by repeating this process as specified in Alg.~\ref{alg:css_bd}.
This is a direct consequence of the trivial code space: as we demonstrate in the following section, utilizing check depletion to generate nontrivial logical operators also forces the preservation of some portion of the code.

For additional clarity, we provide detailed illustrated examples of the hypergraph product and the coring process as applied to a product code in App.~\ref{app-sec:examples}, in both the Tanner graph and slead representations.

\begin{figure}[t]
\centering
\includegraphics[width=\columnwidth]{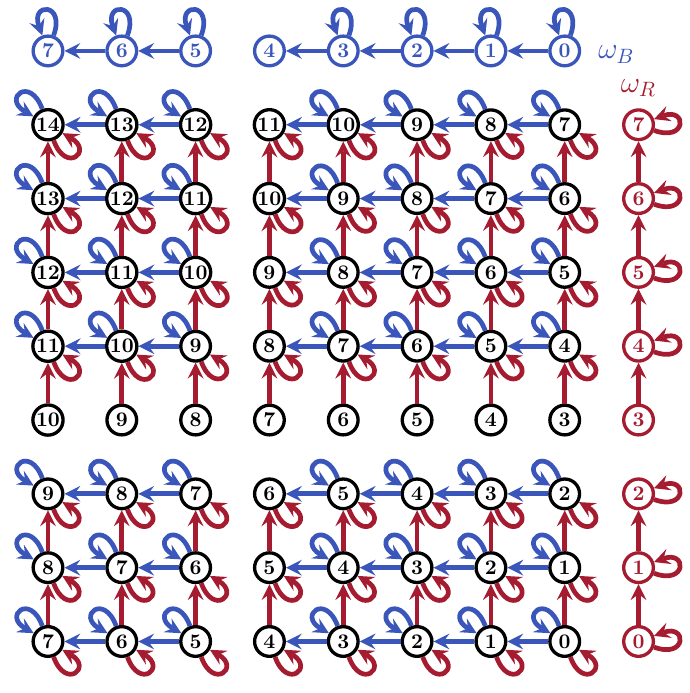}
\caption{\label{fig:quantum_code_2}
\textbf{Slead representation of an example check-depleted product code hosting logical information.}
A slead associated with one sector of a quantum product code is shown.
The classical codes $\omega_R$ and $\omega_B$ are one-dimensional Ising models, with check depletion in $\omega_R$ and $\overline\omega_B$ inserting a single logical.
This depletion is exactly equivalent to choosing appropriate boundaries for the Kitaev surface code.
The resulting slead contains partial vertices lacking one or more qubits or a stabilizer.
The regions on the righthand side of the figure do not permit removing blue qubits in the direct sector, though the bottom righthand patch is pruned by the action on the conjugate sector.}
\end{figure}

\subsection{Properties of cored slead codes}
\label{sec:cored_prop}

By construction, the logical dimension $k_\q$ cannot increase under the measurement and deletion protocol.
The logical dimension may however decrease, which occurs if and only if the measurements generate any representative of a logical operator.
Further, even if $k_\q$ is preserved, it may be that the code distance $d_\q$ decreases.
In this section we show that contrary to the general case, coring decreases neither the distance nor the logical dimension of slead product codes.
Moreover, energy barriers of worldline implementations of Pauli errors are also preserved under coring.

\subsubsection{Qubit support of logicals under stabilizer equivalence}

As a consequence of the check depletion $k_1 = k_2 = 1$ and $\overline k_1 = \overline k_2 = 0$, so $k_\q = k_1 k_2 = 1$.
We denote the codewords of $\omega_1$ and $\omega_2$ as $C_1$ and $C_2$, respectively.
The bare logical operators are immediate from the embedding: for example, in the $X$ sector, these are supported on the red qubits on vertices $\iota_1(C_1,v_2)$, $v_2 \in C_2$, where
\begin{equation}
\iota_1(~\cdot~,v_2):~\omega_1 \mapsto \omega_1\times\{v_2\}
\end{equation}
denotes the inclusion map with base point $v_2$.
Similarly, a minimal logical operator in the $Z$ sector has support on the red qubits on vertices $\iota_2(C_2,v_1)$ for $v_1 \in C_1$.
The union of these sets in either sector is identical, namely $C_1 \times C_2$.

We observe that a classical codeword is a ``stopping set'', meaning that it has even-parity overlap with every check~\cite{di2002finite}.
Remarkably, this basic fact about the factors turns out to imply multiple important features of the coring algorithm.
For instance, the support on red qubits of a dressed logical operator $L$ in the direct sector satisfies 
\begin{equation}
C_R \subseteq \pi_R(\supp(L))~,
\label{eq:L_supp}
\end{equation}
denoting the codeword of $\omega_R$ as $C_R$.
This is evidently true for a bare logical representative $L_\bare$ acting on red qubits on vertices in $\iota_R(C_R,v_B)$ for some $v_B \in C_B$, with $C_B$ the codeword of $\overline\omega_B$.
Now all equivalent logical operators differ from $L_\bare$ by a conjugate stabilizer, whose action on the red qubits is generated by the inclusion $\iota_B(\overline\omega_B,v_R)$, $v_R \in \omega_R$, of the checks of $\overline\omega_B$.
Due to the stopping set property, the odd parity of the overlap of $L_\bare$ with all inclusions $\iota_B(C_B,v_R)$, $v_R \in C_R$, is invariant under multiplication by stabilizers.
That is, while the support of a dressed logical operator $L$ may increase relative to $L_\bare$, it may not decrease and so $\pi_R(\supp(L_\bare)) \subseteq \pi_R(\supp(L))$, proving~\eqref{eq:L_supp}.
Immediately one concludes that all logical representatives have support on at least $d_R = |C_R|$ red qubits and cannot be measured by pruning only blue qubits, as is done in the coring process.

\subsubsection{Logical dimension and distance}

Although pruning blue qubits cannot directly measure a logical, the indirect effect on red qubits could in principle reduce the logical dimension or distance of the code.
In this section we show that this does not occur in the coring algorithm, thus the logical distance and dimension are preserved.

A consequence of the preceding section is that all logical operators of the direct sector are supported on at least $d_R$ red qubits in the vertex set
\begin{equation}
C_R \times C_B = \bigcup_{v_B \in C_B} \iota_R(C_R,v_B) = \bigcup_{v_R \in C_R} \iota_B(C_B,v_R)~.
\end{equation}
We must consider whether by pruning blue qubits one can remove stabilizers acting on red qubits in $C_R\times C_B$.
Recall that a direct stabilizer in the product code arises from a check in $\omega_R$ and a check in $\omega_B$.
Recall that $C_B$ is the codeword of $\overline\omega_B$.
Thus, the stopping set property applied to the dual code implies that every blue qubit in an inclusion $\iota_B(C_B,v_R)$, $v_R \in \omega_R$, is acted on by an even number of direct stabilizers in the same inclusion $\iota_B(C_B,v_R)$.
That is, every stabilizer acting on a red qubit in $\omega_R \times C_B$ is protected from removal under coring in the direct sector.
Applying the same property in the conjugate sector shows that all red qubits within the intersection $C_R \times C_B$ are protected from removal under coring in either sector.
We conclude that the coring algorithm preserves logical dimension and distance in both sectors.

\subsubsection{Energy barriers}

Energy barriers are known to be inherited from the classical codes in the hypergraph product~\cite{Zhao2024Barrier}.
It may be, however, that the coring algorithm reduces the barrier of the product code.
We cannot exclude this possibility in general, but the preceding results imply that the barriers of a certain class of Pauli walk implementing a logical error are preserved.
Namely, it is shown above that no stabilizer acting on red qubits in $C_R \times C_B$ can be pruned.
Consequently, any Pauli walk implementing a bare logical $\iota(C_R,v_B)$, $v_B \in \omega_B$ by acting on qubits in the support of the logical operator will have its energy barrier preserved by coring.
This class contains the quantum inclusion of the worldline implementations of codewords discussed in Sec.~\ref{sec:classical_barrier}, which are known to be minimal for quantum codes with algebraic factors.

\subsubsection{Geometric locality}
\label{sec:cored_locality}

\begin{figure}
    \centering
    \includegraphics[width=0.60\columnwidth]{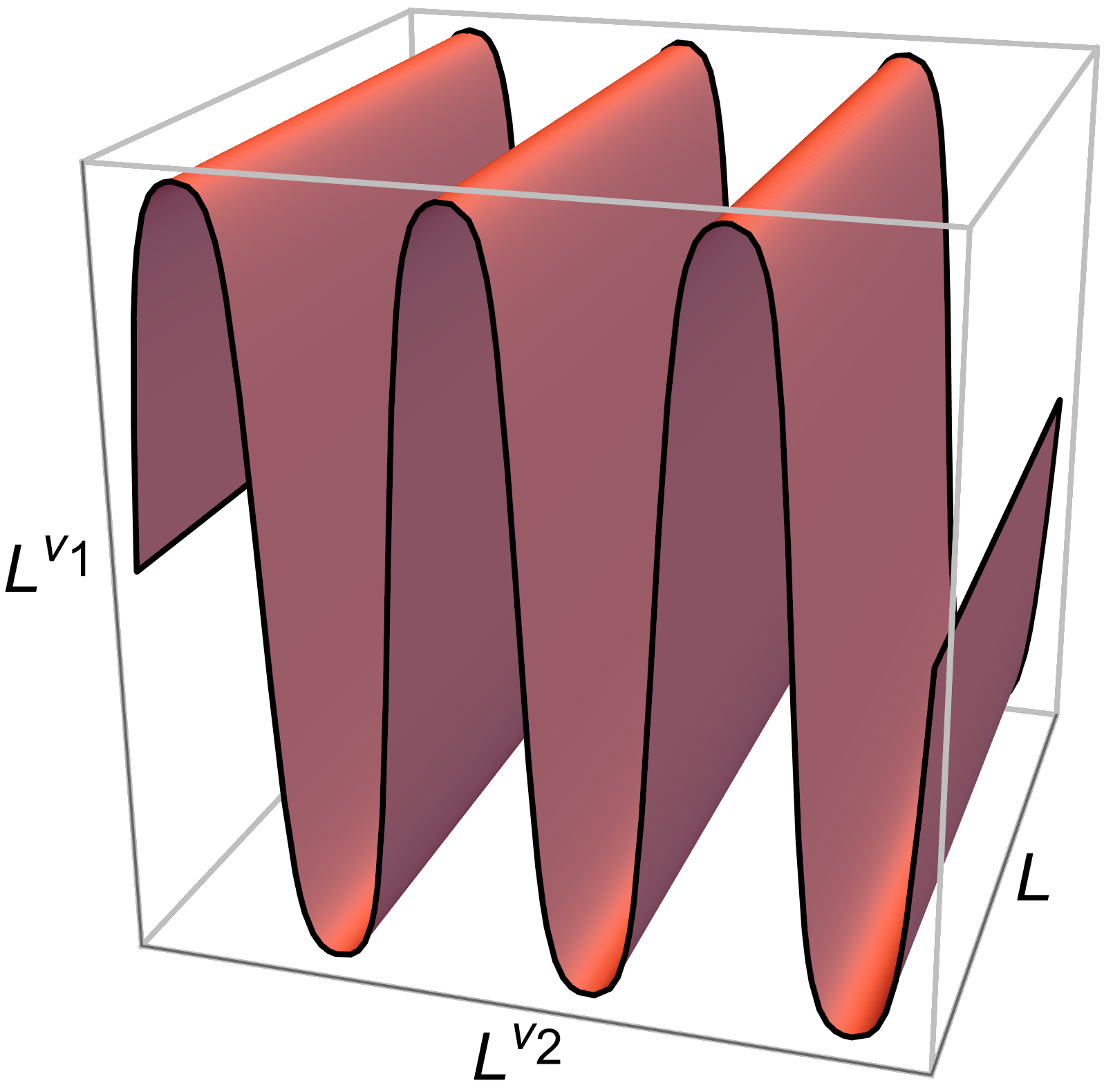}
    \\[2pt](a)
    \\[8pt]
    \includegraphics[width=0.80\columnwidth]{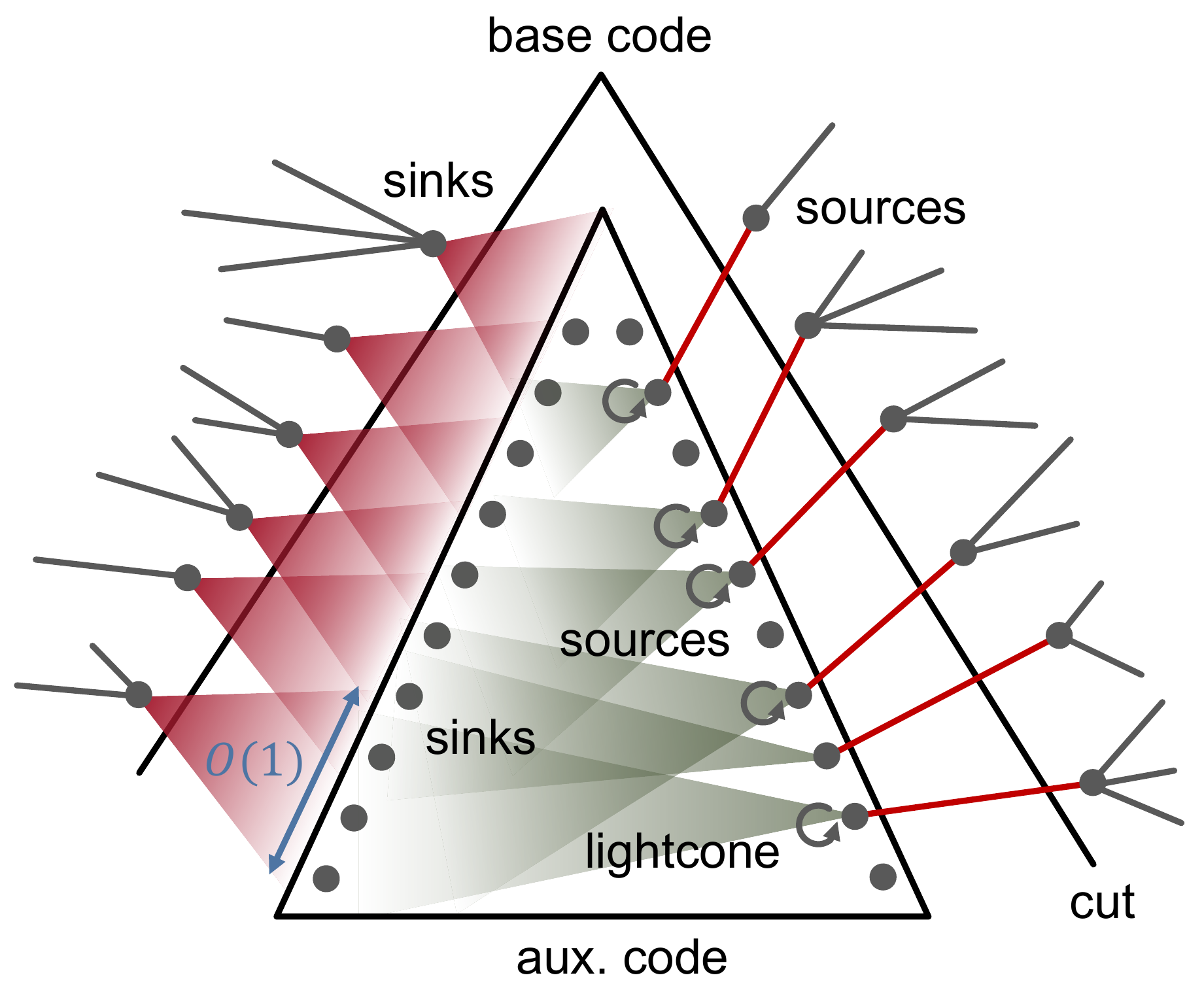}
    \\[2pt](b)
    \\
    \caption{\textbf{Strategy for geometric locality in three dimensions.} \textbf{(a)} Illustration of a space-filling curve, drawn as a sinusoid in this example, winding through the cored quantum code.
    The quantum code is four-dimensional and the curve is a three-dimensional membrane; two linear dimensions of the code scale as $L^{\nu_1}$ and $L^{\nu_2}$ and all remaining linear dimensions scale as $L$, in the notation of Sec.~\ref{sec:cored_locality}.
    \textbf{(b)} An example of mediating all of the long-range parity checks crossing a cut by using an auxiliary slead code.
    At a high level, the information about the spin configuration on one side of the cut is ``encoded'' into the auxiliary code on one side (using the fundamental directionality of slead codes) and ``decoded'' on the other by judiciously modifying the parity checks in the base code.
    We refer to these as source and sink, though they are not the global sources and sinks of the code.
    For specific details of this process, see App.~\ref{app:aux_codes}.
    }
    \label{fig:cored_locality}
\end{figure}

We have shown that a quantum slead code with classical factors $\omega_1$ and $\omega_2$ contains at least the set of vertices $C_1 \times C_2$ after coring.
If $C_1$ is injected by depleting a check at level $p_1$ in $\omega_1$ and $C_2$ at level $p_2$ in $\omega_2$, then the cored code contains a subset of the vertices arising from levels $p_1$ to $p_{1,\mathrm{max}}$ in $\omega_1$ and $p_2$ to $p_{2,\mathrm{max}}$ in $\omega_2$, and does not contain vertices arising from levels lower than $p_1$ in $\omega_1$ or $p_2$ in $\omega_2$.
The code maintains geometric locality relative to the background manifold on the remaining vertices.

Consider a family of factor codes $(\omega_1^{(i)},\omega_2^{(i)})$, $i = 0,1,\ldots$, with quantum code size $n_q^{(i)} = n_1^{(i)} n_2^{(i)} + m_1^{(i)} m_2^{(i)} = O(L_i^D)$, where each $\omega_1^{(i)}$ is embedded in a $D_1$-dimensional manifold with characteristic length $L_i$ and $\omega_2^{(i)}$ a $D_2$-dimensional manifold also with characteristic length $L_i$, and $D = D_1+D_2$.
If fixed depletion levels $p_1^{(i)} = p_1$ and $p_2^{(i)} = p_2$ are chosen for all $i$, coring does not in general alter the asymptotic scaling of $n_q^{(i)}$, as only a vanishing fraction of vertices arise from lower levels in the sleads and are therefore guaranteed to be pruned.
In contrast, a size-dependent choice of depletion level can alter the asymptotic scaling of the cored code.
Recalling that $p_\mathrm{max} = \Omega(L)$ for geometrically local codes of length scale $L$, one sees that if $p^{(i)}_{1,\mathrm{max}} - p_1^{(i)} \sim L^{\nu_1}$, then $n_1^{(i)} = O(L_i^{D_1-1+{\nu_1}})$.
That is, a vanishing fraction of the system is involved in the encoding in the asymptotic limit.
In this way, by choice of $p_1^{(i)}$ and $p_2^{(i)}$ one obtains a cored code with volume scaling as $L^{D-2+\nu_1+\nu_2}$ for $0 < \nu_1,\nu_2 \leq 1$.

The tunability of the volume of the cored code does not affect its geometric locality, which remains inherently $D$-dimensional.
A geometrically local code in some lower target number of dimensions $D' < D$ thus requires further modification.
For example, it may be that a fractal structure inherent to the logical operators allows an effective reduction of the locality~\cite{Brell2016}; here we instead apply a general proposal of projecting to a $D'$-dimensional space-filling curve within the $D$-dimensional background manifold, with both $D$ and $D'$ integers.

Suppose the cored code has spatial extent $L$ along basis directions $\hat e_j$, $j=3,\ldots,D$ and $L^{\nu_1}$ and $L^{\nu_2}$ along the basis directions $\hat e_1$ and $\hat e_2$, respectively; that is, it becomes increasingly anisotropic.
We will project from this region onto a $D' = (D-1)$-dimensional ``folded'' surface.
Let $\nu_1 \leq \nu_2$; then we choose $\hat e_1$ to become the folding axis of the space-filling curve, and $\hat e_2$ the direction along which to perform the fold.
For example, if the folding function is taken to be a sinusoid, the amplitude direction is along $\hat e_1$ and the phase direction $\hat e_2$.
An illustration of the folding geometry in a three-dimensional volume is shown in Fig.~\ref{fig:cored_locality}(a).
Then the spatial extent of the code projected to the $D'$-dimensional surface is $L^{\nu_1+\nu_2}$ along $\hat e_1'$ and $L$ along $\hat e_j'$, $j = 2,\ldots,D-1$.

The projection to the folded surface results in a finite density of long-range interactions in the direction $\hat e_1'$.
The interaction range of these scales as $L^{\nu_1}$, and in order to restore geometric locality we mediate these connections by introducing auxiliary error-correcting codes on additional qubits, as shown in Fig.~\ref{fig:cored_locality}(b).
Observe that a long-range edge has length scaling as $\ell = O(L^{\nu_1})$, and we introduce $O(\ell)$ additional qubits per edge in order to restore geometric locality.
The total scaling of the volume is thus $L^{D-2+2\nu_1+\nu_2}$ for the geometrically local code embedded in the $D'$-dimensional surface.
We arrange the embedding of the quantum code so that the deformation is performed at the level of the classical factors, and allows freedom in the actual implementation.
For a discussion specific to the codes studied in numerics, see Sec.~\ref{sec:finite_temp/codes} and App.~\ref{app:aux_codes}.

\section{Finite temperature simulations}
\label{sec:finite_temp}

\subsection{Modeling memory coherence time}
\label{sec:finite_temp/monte_carlo}

To study the behavior of a many-body system at finite temperature, it is often sufficient to couple locally to a Markovian bath, under an assumption that all of the relevant scales are well separated.
This is the approach we employ to simulating the coherence time of the quantum memory.
The code thus weakly coupled to a reservoir undergoes open system dynamics described by the Lindbladian
\begin{equation}
\dot \rho = -i [H,\rho] + \mathcal L(\rho)~,
\end{equation}
which is known as the Davies master equation~\cite{davies1974}.
The generator
\begin{equation}
\mathcal L(\rho) = \sum_\alpha \left(\hat L_\alpha \rho \hat L_\alpha^\dagger - \frac12 \{\hat L_\alpha^\dagger \hat L_\alpha,\rho\}\right)
\label{eq:lindblad}
\end{equation}
captures dissipative coupling to the environment through a set of Lindblad operators $\hat L_\alpha$.
The magnitudes of these Lindblad operators are chosen so that the equilibrium state of the stabilizer Hamiltonian $H = -\sum_{s \in \mathcal S} s$ at a fixed temperature $\beta$ is a steady state of the dynamics.
Due to the CSS structure of the codes studied here, under an error model of independent local Pauli noise $\hat L_\alpha \sim X_i,Z_i$, the $X$ and $Z$ sectors of the code can be simulated independently as classical Hamiltonians.

We emphasize that thermalizing noise of this type is distinct from other error channels like Pauli noise, despite utilizing the same quantum jump operators.
This is because the probability of an error is not only spatially inhomogeneous but also dependent on the current state of the system and the energy functional.
Rather than applying an i.i.d.~local noise channel to all qubits, the bath is constantly updating its probabilities, aware of how costly any particular error is and preferentially applying low-energy noise based on the Boltzmann factor.
Self-correction without measurement and feedback is not possible within the i.i.d.~framework, which is effectively an infinite-temperature bath~\cite{fernando2009how}.

As the CSS codes studied here have distinct $X$ and $Z$ sectors, we study these independently, performing classical time evolution through kinetic Monte Carlo but with quantum decoding identifying logical errors.
More concretely, we use the following protocol to estimate the coherence time of the memory~\cite{Bravyi2013PRL}:
\begin{enumerate}
\item Initialize the memory at time $t=0$ in the trivial ground state.
\item Stochastically evolve the system in one sector using quantum jump operators at inverse temperature $\beta$ as specified in the Lindbladian~\eqref{eq:lindblad}.
\item After a set $T_\mathrm{ec}$ time interval, attempt error correction using a standard QEC decoder:
\begin{itemize}
\item If decoding succeeds, the recovery operation is not performed and the simulation continues.
\item If decoding fails, the simulation ends and the failure time is taken as the memory lifetime.
\end{itemize}
\item Repeat prior steps until encountering decoding failure (i.e., a logical error).
\end{enumerate}
Performing many trials of the above we obtain ensembles of coherence times parameterized by $\beta$, and we take the mean of each distribution as the lifetime of the memory at $\beta$. 
We discuss the concrete cored product codes we use in our simulations in the next subsection and memory lifetime results in Sec.~\ref{sec:finite_temp/lifetime_results}; technical details of the kinetic Monte Carlo implementation and decoder are described in Apps.~\ref{app-sec:kinetic_monte_carlo} and \ref{app-sec:decoding}.

\begin{figure}[t]
    \centering
    \includegraphics[width=\linewidth]{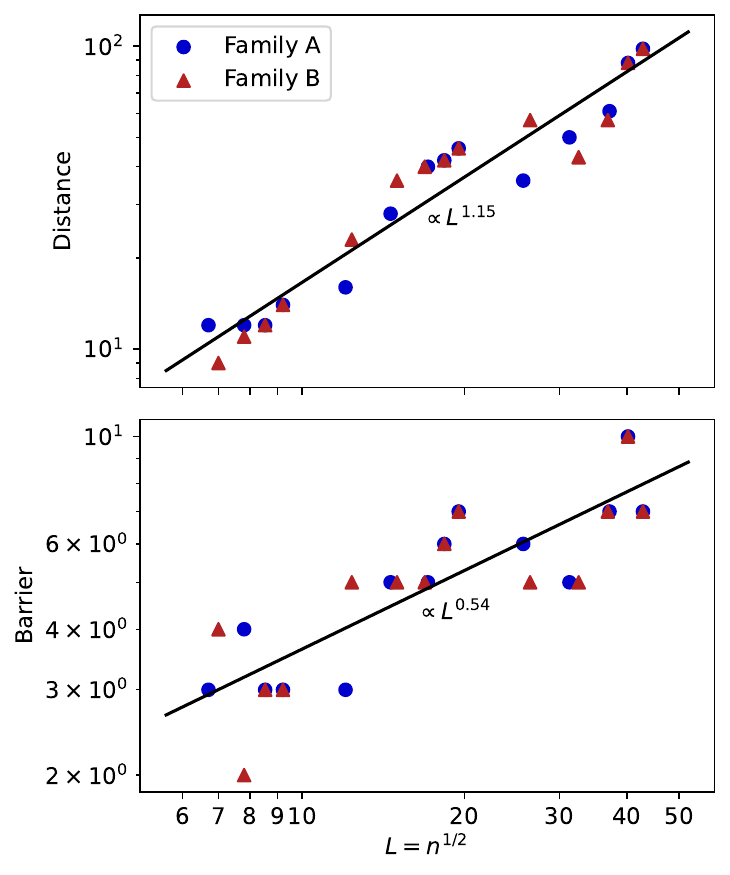}
    \caption{\textbf{Characteristics of two pinwheel factor code families used in numerical experiments.} Exact distances found through integer programming and barrier upper bounds found by greedy search are plotted against the characteristic side length of the two-dimensional classical pinwheel codes $L = n^{1/2}$. The two families of pinwheel codes differ in the permutations used to recursively construct generations of the codes (see Sec.~\ref{sec:finite_temp/codes}). Both the distance and barrier of the codes scale algebraically with code size.}
    \label{fig:pinwheel_props}
\end{figure}

\subsection{Cored product codes for quantum memories}
\label{sec:finite_temp/codes}

\begin{figure}
\centering
\includegraphics[width=0.9\columnwidth]{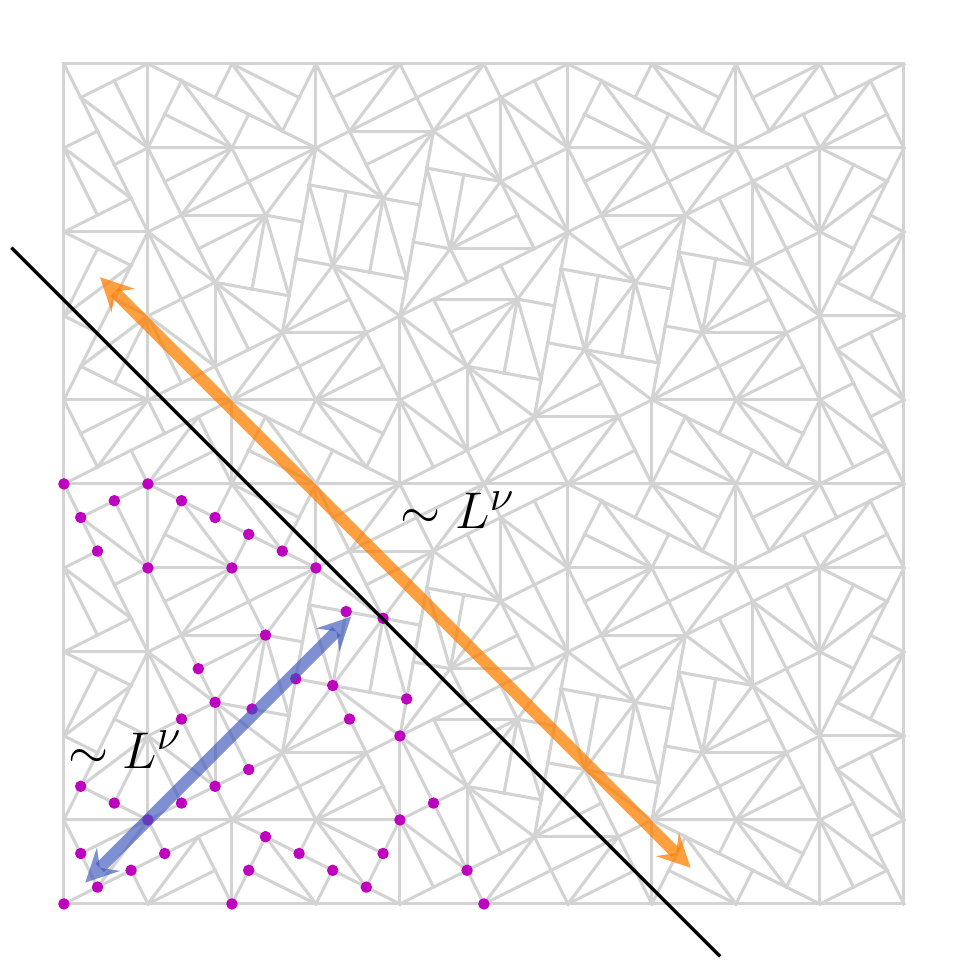}
\caption{\textbf{Schematic of classical factor geometry.} We illustrate the underlying lattice of the pinwheel factor code (gray lines with qubits at vertices), a nontrivial codeword induced by check depletion (purple vertices), and spatial region with particular $L^\nu$ scaling, which bounds the extent of the code after coring.
The choice $\nu_1 = \nu_2 = \nu$ here is not the optimal choice as described in Sec.~\ref{sec:cored_locality}, but happens to better fit the shape of typical codewords, increasing distance.}
\label{fig:pinwheel_patch}
\end{figure}

\begin{figure}
    \centering
    \includegraphics[width=1\linewidth]{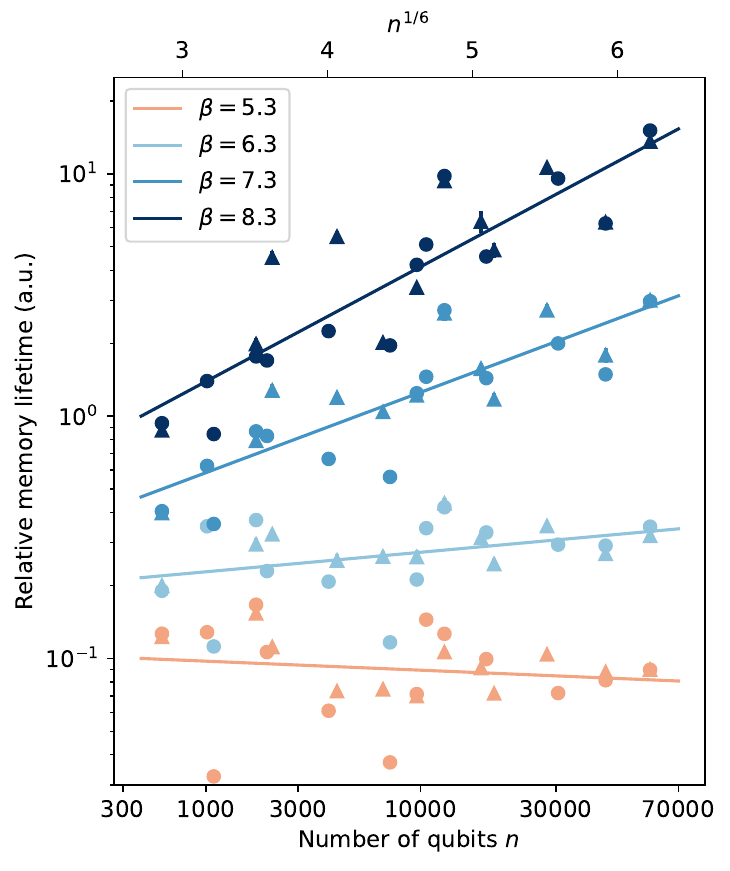}
    \caption{\textbf{Relative quantum memory lifetimes versus code size.} Passive quantum memory lifetimes 
    for two families (circles and triangles) of cored product codes arising from pinwheel factor codes, found through kinetic Monte Carlo numerics at different inverse temperatures $\beta$.
    Trendlines are stretched exponentials in code size, of the form $a_\beta \exp(\smash{b_\beta n^{1/6}})$ for temperature-dependent $a_\beta, b_\beta$, motivated by the Arrhenius equation.
    For compactness, lifetimes have been normalized for each $\beta$ so that the trendlines are equally spread at the smallest code size; this entails only uniform vertical offsets in datapoints for each $\beta$ and has no effect on relative trends in memory lifetimes.
    Each lifetime data point is averaged over $256$ independent shots; error bars are $1 \sigma$ standard errors (most are smaller than the markers).
    Scatter in lifetimes at different finite code sizes is expected due to the nature of the codes, which are inhomogeneous and possess no spatial symmetry.}
    \label{fig:memory_lifetimes_single_panel_relative}
\end{figure}

Direct simulation of coherence times as described above requires selecting a suitable family of memory codes based on classical factor codes satisfying all of the properties described in the preceding sections.
Our concrete construction is as follows: we begin with the specific pinwheel tiling of a square in two dimensions shown in Fig.~\ref{fig:pinwheel}, with the first substitution of the four initial triangular prototiles labeled as generation $g=(1,0)$.
In the figure, the half-space vector is chosen to be $\bm t = (-1,1)$.
For the simulated codes, we utilize instead $\bm t = (-1,-1)$, which selects another corner of the square, inequivalent under the $\Z_2$ rotational symmetry of the prototiles.

Subsequent generations $g$ of the code are produced based on a partial substitution of the initial code.
In the substitution rule each tile is replaced by five subtiles, so in every generation beginning with $g=(1,0)$ corresponding labels $\{1,2,3,4,5\}$ may be attached to all tiles. 
Then a partial implementation of the substitution defined by an ordering of the labels allows for intermediate codes between generations.
For example, having chosen the permutation $\sigma = (5,4,3,2,1) \in S_5$ one may define a code at generation $g=(1,1)$ starting from $g=(1,0)$ and applying the substitution rule to all tiles labeled by $\sigma(1) = 5$.
The code $g=(1,2)$ is obtained from $g=(1,0)$ by applying the substitution rule to all tiles labeled by $\sigma(1) = 5$ or $\sigma(2) = 4$.
Consequently, $g=(1,5)=(2,0)$, as all tiles from the previous generation have been substituted.
Evidently the block length of such a code family is increasing, and this pattern has the advantage of implementing spatially homogeneous incremental substitution, in contrast to a random partial substitution which would introduce strong density fluctuations.

The choice $\bm t = (-1,-1)$ places source vertices at the upper righthand corner of Fig.~\ref{fig:pinwheel} and the sink vertices at the lower lefthand corner.
We observe that for such a triangular geometry bounded by the corner of the square and the diagonal hyperplane, both the base and the height scale as $L^\nu$; that is, with the same exponent, which determines the appropriate location of the depleted check for each system size\footnote{The linear size $L$ is determined as $L = \sqrt N$ for each classical code on a two-dimensional square.}, as described in Sec.~\ref{sec:cored_locality}.
So the volume scaling of the cored product code using two such factor codes is $O(L^{4\nu})$, and accounting for the additional factor of $\nu$ arising from folding along any direction we find that $\nu = 3/5$ is sufficient for embedding in three spatial dimensions.

We utilize two permutations $\sigma_A = (1,2,3,4,5)$ and $\sigma_B = (2,4,3,1,5)$ to define families of codes with otherwise identical geometry.
In order to avoid occasional short logical operators, we select an approximate point in $\R^2$ for depletion and query the six nearest vertices, choosing to perform the check depletion on the site producing the highest-weight codeword.

One of the factor codes must finally be deformed by projecting onto a space-filling curve and inserting auxiliary codes as necessary, as described in Sec.~\ref{sec:cored_locality}.
This step must be specialized to individual members of a family of codes, and in particular requires choosing a locality radius $r = O(1)$ as well as a shape for the space-filling curve.
While these details are unimportant asymptotically, they may alter the apparent behavior of the system at small sizes by requiring an optimization specific to each member of the family.
We therefore perform the simulations without introducing the auxiliary codes to the classical factors.
As a result, the volume scaling of the codes presented is not $O(L^3)$ but rather $O(L^{12/5})$.
The effects caused by the auxiliary codes in the specific case studied here are discussed in more detail in App.~\ref{app:aux_codes}.

\subsection{Memory lifetime results}
\label{sec:finite_temp/lifetime_results}

We present memory lifetime results for the two example families of slead codes (detailed in Sec.~\ref{sec:finite_temp/codes}) in Fig.~\ref{fig:memory_lifetimes_single_panel_relative} for a range of temperatures. 
We observe a critical temperature of $5.3 \leq \beta^* \leq 6.3$, above which the quantum memory is expectedly unstable and memory lifetime does not increase with system size $n$. 
Below the critical temperature, we observe a hallmark stretched exponential increase in memory lifetime with system size. 
We do not observe evidence of a plateau or downturn of memory lifetime with increasing $n$, which would indicate partial self-correction, on code instances of up to $60\,000$ qubits.

In these numerical simulations, we use belief propagation with ordered statistics post-processing (BP-OSD)~\cite{panteleev2021degenerate, roffe2020decoding, roffe2022github}, a general purpose polynomial-time decoding algorithm for qLDPC codes of arbitrary geometry, to decode our quantum memory at readout (see App.~\ref{app-sec:decoding} for technical details of the decoder). 
For proper functioning of the decoders, it is necessary to provide appropriate physical error probability priors; however, this is not straightforward for thermal noise. 
To review, in the standard treatment of (active) quantum error correction with Markovian independent, phenomenological or circuit-level noise models, a set of probabilities $\{p_\ell \ll 1\}$ characterizes the physical error channels $\ell$, and the probability priors for decoding are directly computable from $\{p_\ell\}$. 
In contrast, under thermal noise, physical errors ($X$ and $Z$ flips) occur with rates determined by the energetics of the system and are not described by stationary $\{p_\ell\}$ known beforehand. 

We thus adopted a calibration procedure to obtain suitable priors for decoding. 
We sampled the thermal dynamics of our codes through kinetic Monte Carlo (see Sec.~\ref{sec:finite_temp/monte_carlo}) and fitted the averaged qubit flip probabilities to the simple relaxation form
\begin{equation}
p_q(t) = A_q \left( 1 - e^{-t / \tau_q} \right),
\label{eq:decoder_relaxation_form}
\end{equation}
for $q \in [n]$. 
Note that \eqref{eq:decoder_relaxation_form} is expected from a minimal (linearized) mean-field model of spin relaxation in a homogeneous environment, and we fixed $A_q = 1/2$ uniformly as we do not anticipate globally (non-glassy) ordered thermal phases in our codes in general. For illustration, we provide a plot of the averaged qubit flip probabilities used for calibration on one of our codes in Fig.~\ref{app-fig:extra_results/physical_qubit_flip_probabilities}. We refer readers to App.~\ref{app-sec:kinetic_monte_carlo} for details of the kinetic Monte Carlo implementation, which were heavily optimized at the algorithmic and low-level programming levels, and App.~\ref{app-sec:decoding} for details on the decoder.

We remark on an interesting observation in our finite temperature simulations: namely that a quantum memory can survive for long periods of time despite the physical error density (i.e., weight of Pauli errors divided by code size $n$) on the code quickly exceeding $40\%$.
This phenomenon is generic across all code instances investigated in our numerics. Superficially, this observation appears to be in contention with the error-correcting ``thresholds'' of the codes, which are expected to be \mbox{$\lesssim 10\%$,} beyond which effective decoding is not expected to be possible. The resolution is that such code capacity thresholds are conventionally defined with respect to depolarizing or independent $X$ and $Z$ Pauli noise, whereas the noise model considered for self-correcting quantum memories is one of thermal noise; the prospects of self-correcting quantum memories are known to be severely limited in the settings of the former noise models~\cite{fernando2009how, fernando2010limitations}. That is, decoders appear to be stronger, in the sense of being able to correct higher-density errors, under thermal noise than depolarizing or independent Pauli noise.

\section{Discussion}
\label{sec:discussion}

In this paper we have presented a construction of a geometrically local quantum code in three spatial dimensions and argued that it exhibits self-correction, a conclusion supported by numerics at finite temperature.
The key feature of the code is positional disorder, realized by an underlying aperiodic tiling, whose effect is to strongly suppress entropic factors in the free energy.
The philosophy implicit in this construction is that self-correction should be generic for quantum codes even in three dimensions under an appropriate measure.
In this sense, we expect the technical details of our construction are of limited importance, serving mostly to facilitate a specific definition of a code lacking spatial symmetries.

The specific code presented here offers many opportunities for improvement: it protects only a single logical qubit, independent of the block length; it is not single-shot and does not exhibit a finite-temperature phase transition; and the connectivity graph of the classical pinwheel factors is not optimized for performance.
Nevertheless, the fundamental ingredients of the slead framework and coring algorithm provide a natural setting based on the familiar hypergraph product for writing down tractable codes with tunable parameters in reduced dimensions.

In fact, the technical details of the construction presented here can be understood generally as an effort to advance our central goal of studying disordered codes that are LDPC and geometrically local in three dimensions.
While entirely random code constructions can be used to this end, aperiodic ones confer certain advantages afforded by hyperuniformity.
Namely, fluctuations in density and connectivity are reduced, and one need not worry about the effects of rare regions in the thermodynamic limit.

We conjecture that, if one could sample uniformly over LDPC stabilizer codes with growing distance which respect geometric locality in three dimensions---of which translation-invariant or highly symmetric codes are a vanishing subset---the property of self-correction would be typical.
This is due to the dramatic reduction in disordered codes of entropic factors, which have proved very difficult to overcome otherwise~\cite{baspin2025}.
We have attempted here to systematically realize a more typical family of codes drawn form this measure which avoids the specific type of fine-tuning arising from spatial symmetries.
It will be worthwhile to investigate other ways of realizing this goal, perhaps even developing disordered codes whose self-correction properties can be rigorously analyzed and proved.

\begin{acknowledgments}
We gratefully acknowledge insightful discussions with Philip Crowley, Arpit Dua, Jeongwan Haah, Helia Kamal, Ting-Chun Lin, Harry Putterman, Thomas Schuster, and Beni Yoshida.
We especially thank Ethan Lake for helpful feedback on an earlier version of this manuscript.
This work was supported in part by NSF via the STAQ program and the QLCI program (grant no. OMA-2016245).
B.~R.~acknowledges support from the Harvard Quantum Initiative Postdoctoral Fellowship in Science \& Engineering.
J.~M.~K.~acknowledges support from the A*STAR Graduate Academy, Singapore.
N.~Y.~Y.~acknowledges support from a Simons Investigator award.
\end{acknowledgments}

\bibliography{refs}

\clearpage
\newpage

\appendix
\onecolumngrid

\section{Illustrative examples}
\label{app-sec:examples}

\subsection{Correspondence between Tanner graphs and sleads of classical codes}
\label{app-sec:examples/classical_tanner_slead}

\begin{figure}[ht]
    \includegraphics[width=0.9\linewidth]{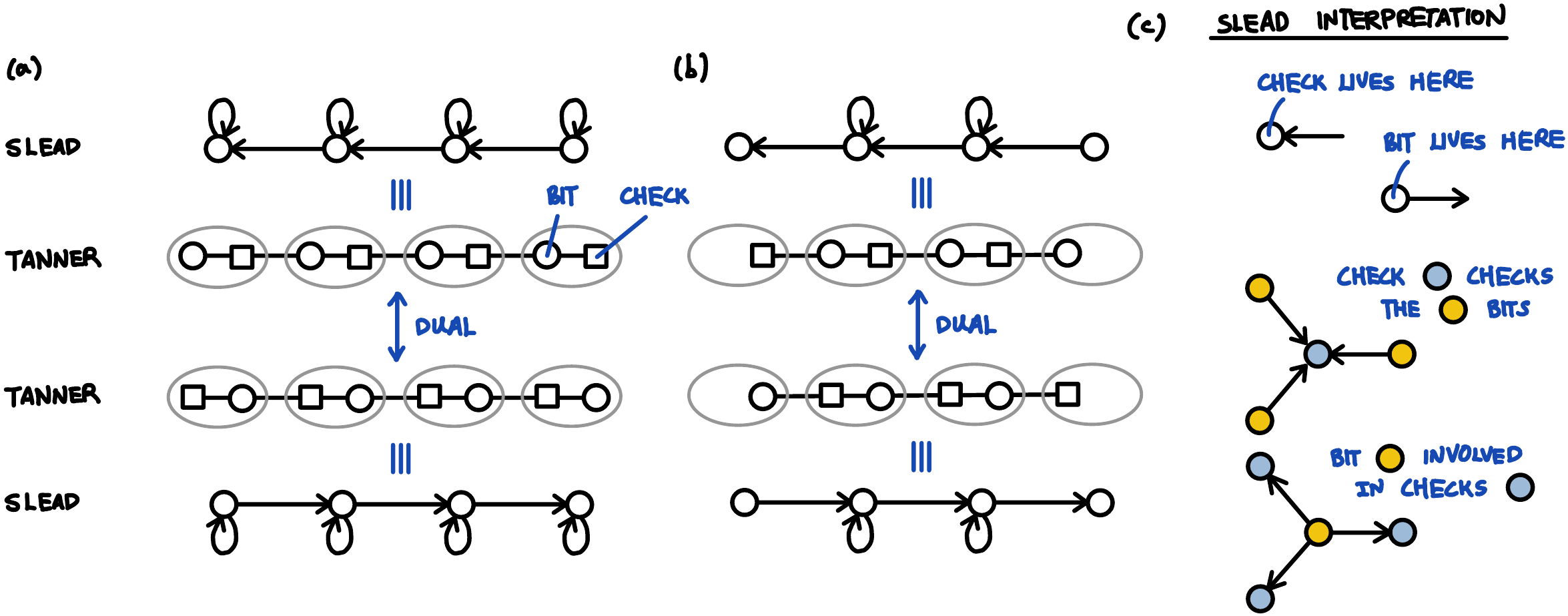}
    \caption{\textbf{Tanner graph and slead representations of classical codes and their duals.} \textbf{(a)} A classical repetition code and its dual in both Tanner graph and slead representations. Squares represent checks and circles represent bits on the Tanner graph. We group a check and a bit together in cells, demarcated as gray ovals, in going to the slead representation. The dual code is the primal code but with checks and bits interchanged. On the Tanner graph, this corresponds to interchanging squares and circles. On the slead, this corresponds to reversing the direction of all edges (arrows). \textbf{(b)} Analogous illustration for the classical repetition code but with different boundary conditions. All comments above apply. \textbf{(c)} Summary of the interpretation of elements of the slead representation of classical codes. Here, colors are only to highlight the groups of vertices referred to and carry no semantic meaning.}
    \label{app-fig:examples/classical_tanner_slead}
    \phantomsubfloat{\label{app-fig:examples/classical_tanner_slead/rep_code_a}}
    \phantomsubfloat{\label{app-fig:examples/classical_tanner_slead/rep_code_b}}
    \phantomsubfloat{\label{app-fig:examples/classical_tanner_slead/slead_interpretation}}
    \vspace{-36pt}
\end{figure}

First, we review the slead representation of classical error correcting codes as introduced in Sec.~\ref{sec:classical_codes} by establishing a correspondence to the conventional Tanner graph representation of codes, which may be familiar to readers versed in error correction. This correspondence is straightforward, as we illustrate and explain in Fig.~\ref{app-fig:examples/classical_tanner_slead}. The slead representation can be thought of as a compressed and minimalistic version of Tanner graphs, where a bit and a check are grouped together to form cells, each cell being represented by a vertex. Going to the dual of a classical code entails simply reversing the direction of all edges in its slead.

\subsection{Classical slead code examples}
\label{app-sec:examples/classical_slead_codes}

\begin{figure}[h]
    \includegraphics[width=0.85\linewidth]{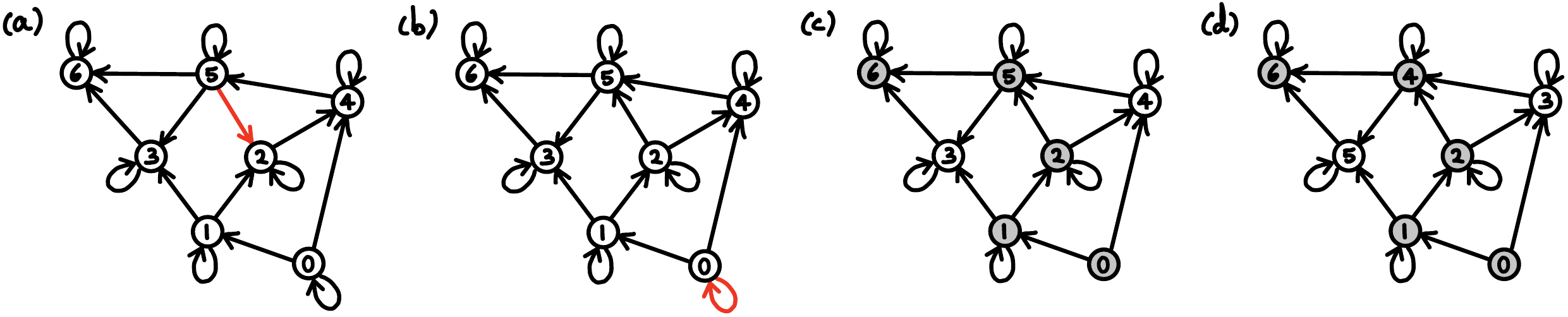}
    \caption{\textbf{Classical slead codes.} \textbf{(a)} A non-example of a classical slead code, as the graph contains a cycle of length ${>} \, 1$, namely the length-$3$ cycle on vertices $\{2, 4, 5\}$. Here, for explicit referencing we have labeled each vertex with an arbitrary unique integer, as opposed to the main text where we have labeled vertices of slead by their levels. \textbf{(b)} The slead code produced by reversing the direction of the red-highlighted edge in (a). This code does not host any nontrivial codewords as there is an equal number of linearly independent checks and bits. \textbf{(c)} Deleting the check at vertex $0$, namely, the red-highlighted edge in (b), produces a slead code that hosts a nontrivial codeword. The support of this codeword is shaded in gray, which can be found by starting at vertex $0$ and traversing the slead, flipping bits as necessary to satisfy the encountered checks. \textbf{(d)} A relabeling of the vertices of (c) according to their levels (i.e., their topological ordering).}
    \label{app-fig:examples/classical_slead_codes}
    \phantomsubfloat{\label{app-fig:examples/classical_slead_codes/non_example}}
    \phantomsubfloat{\label{app-fig:examples/classical_slead_codes/example_no_codeword}}
    \phantomsubfloat{\label{app-fig:examples/classical_slead_codes/example_check_depleted}}
    \phantomsubfloat{\label{app-fig:examples/classical_slead_codes/example_levels}}
    \vspace{-20pt}
\end{figure}

For clarity, we next discuss an illustrative example of the slead representation of classical codes. We first show a non-example in Fig.~\ref{app-fig:examples/classical_slead_codes/non_example}, which is not a slead code as it contains a cycle of length ${>} \, 1$. Note that, to allow explicit referencing in this basic discussion, we have identified each vertex in the slead with a unique integer label---this is in contrast to the main text (e.g.~in Figs.~\ref{fig:slead}, \ref{fig:ti_barrier}, and \ref{fig:pinwheel}), where vertices are labeled by their topological ordering. Removing the cycle in Fig.~\ref{app-fig:examples/classical_slead_codes/non_example} by reversing the direction of an edge produces Fig.~\ref{app-fig:examples/classical_slead_codes/example_no_codeword}, which is a valid slead code. However, this slead code hosts only the trivial (all-zero) codeword, as there are as many linearly independent checks as there are bits. To be explicit, the checks present on this slead code are:
\begin{table}[H]
    \centering
    \begin{tabular}{p{1.5cm} p{3cm}}
        \toprule
        Check & Bits Involved \\
        \midrule
        $0$ & $\{0\}$ \\
        $1$ & $\{0, 1\}$ \\
        $2$ & $\{1, 2\}$ \\
        $3$ & $\{1, 3, 5\}$ \\
        $4$ & $\{0, 2, 4\}$ \\
        $5$ & $\{2, 4, 5\}$ \\
        $6$ & $\{3, 5, 6\}$ \\
        \bottomrule
    \end{tabular}
\end{table}

The parity check matrix of this code can be written
\begin{equation}
    H = \left(
        \begin{array}{@{} c c c c c c c @{}}
            1 & 0 & 0 & 0 & 0 & 0 & 0 \\
            1 & 1 & 0 & 0 & 0 & 0 & 0 \\
            0 & 1 & 1 & 0 & 0 & 0 & 0 \\
            0 & 1 & 0 & 1 & 0 & 1 & 0 \\
            1 & 0 & 1 & 0 & 1 & 0 & 0 \\
            0 & 0 & 1 & 0 & 1 & 1 & 0 \\
            0 & 0 & 0 & 1 & 0 & 1 & 1
        \end{array}
    \right),
\end{equation}
where the rows lists the checks and the columns correspond to the bits of the codes in order. This matrix is full-rank and has a trivial kernel; accordingly the code does not host any nontrivial codeword. 

Check depletion introduces a nontrivial codeword into the code. This is illustrated in Fig.~\ref{app-fig:examples/classical_slead_codes/example_check_depleted}, where the check at vertex $0$ is deleted. The parity check matrix associated with the check-depleted code is exactly $H$ above but with the first row removed. The kernel of the parity check matrix is $1$-dimensional and is spanned by the nontrivial codeword $\left(\begin{array}{@{} c c c c c c c @{}} 1 & 1 & 1 & 0 & 0 & 1 & 1 \end{array}\right)$, which is shaded in Fig.~\ref{app-fig:examples/classical_slead_codes/example_check_depleted}.

Lastly, we show in Fig.~\ref{app-fig:examples/classical_slead_codes/example_levels} a relabeling of the vertices of Fig.~\ref{app-fig:examples/classical_slead_codes/example_check_depleted} according to their topological order. This ordering of vertices, also referred to as levels, is used throughout the main text. We first discussed the topological ordering of vertices on a slead in Sec.~\ref{sec:boundaries}, and, for instance, the illustrations of sleads in Figs.~\ref{fig:slead}, \ref{fig:ti_barrier}, and \ref{fig:pinwheel} use this convention for labeling of vertices.

\subsection{Correspondence between Tanner graphs and sleads of quantum hypergraph product codes}
\label{app-sec:examples/hgp_tanner_slead}

Just as there is a straightforward correspondence between the Tanner graph and slead representations of classical codes, which we discussed in App.~\ref{app-sec:examples/classical_tanner_slead}, there is an analogous correspondence for quantum codes. We illustrate this correspondence explicitly here. We take the hypergraph product of two check-depleted classical repetition codes as an example, which are also drawn in Fig.~\ref{fig:quantum_code_2} of the main text. 

\begin{figure}[ht]
    \includegraphics[width=0.45\linewidth]{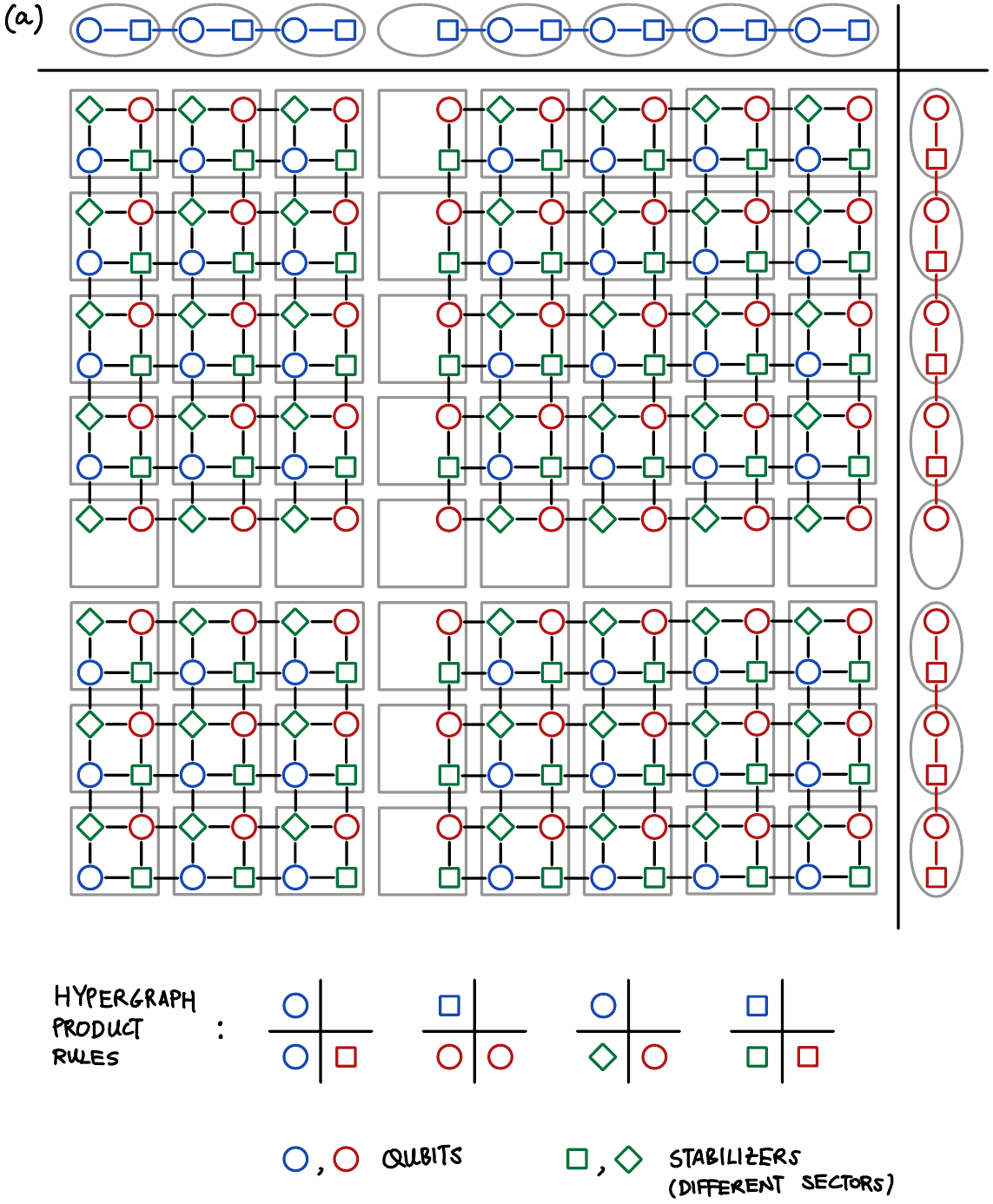}
    \\ \vspace{12pt}
    \includegraphics[width=0.45\linewidth]{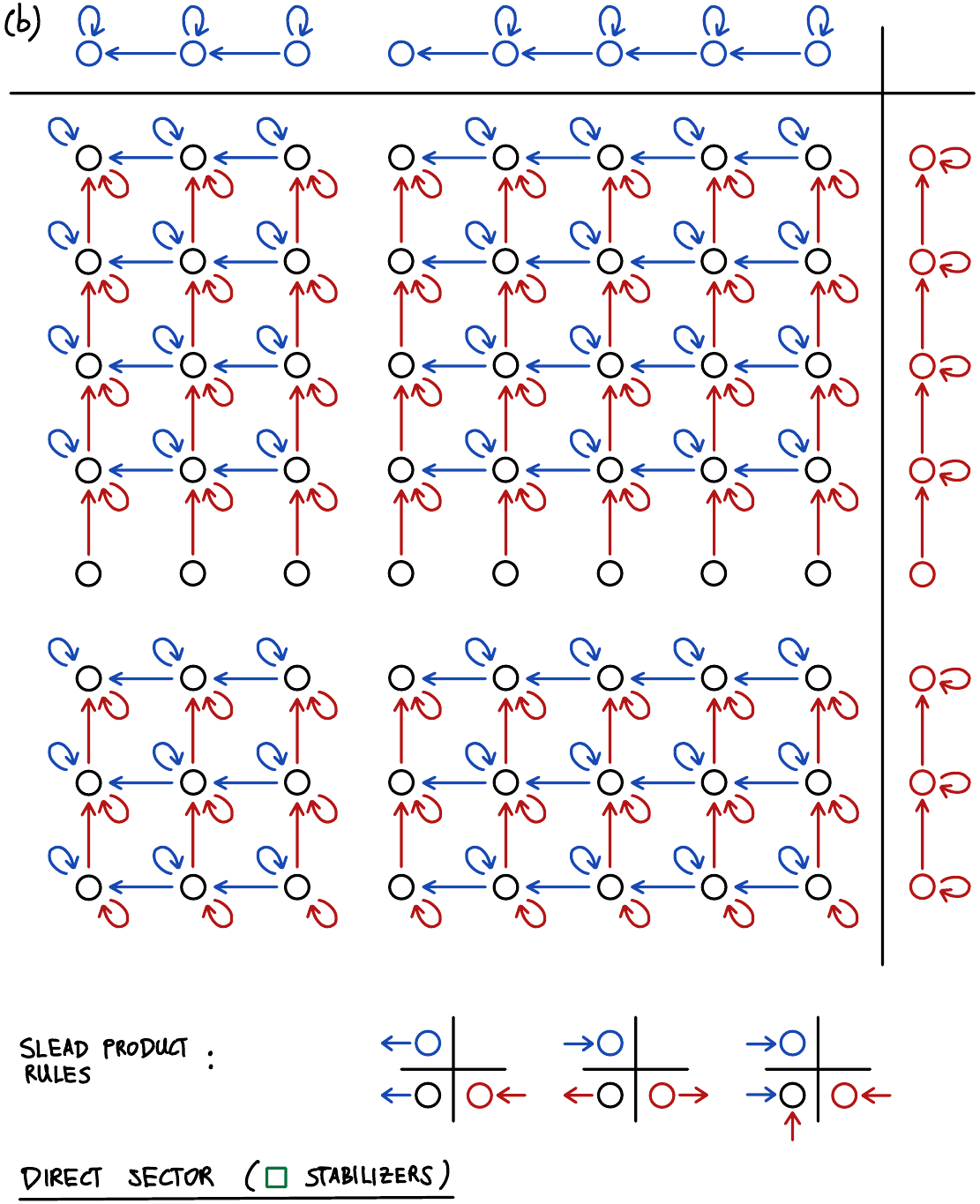}
    \hspace{16pt}
    \includegraphics[width=0.45\linewidth]{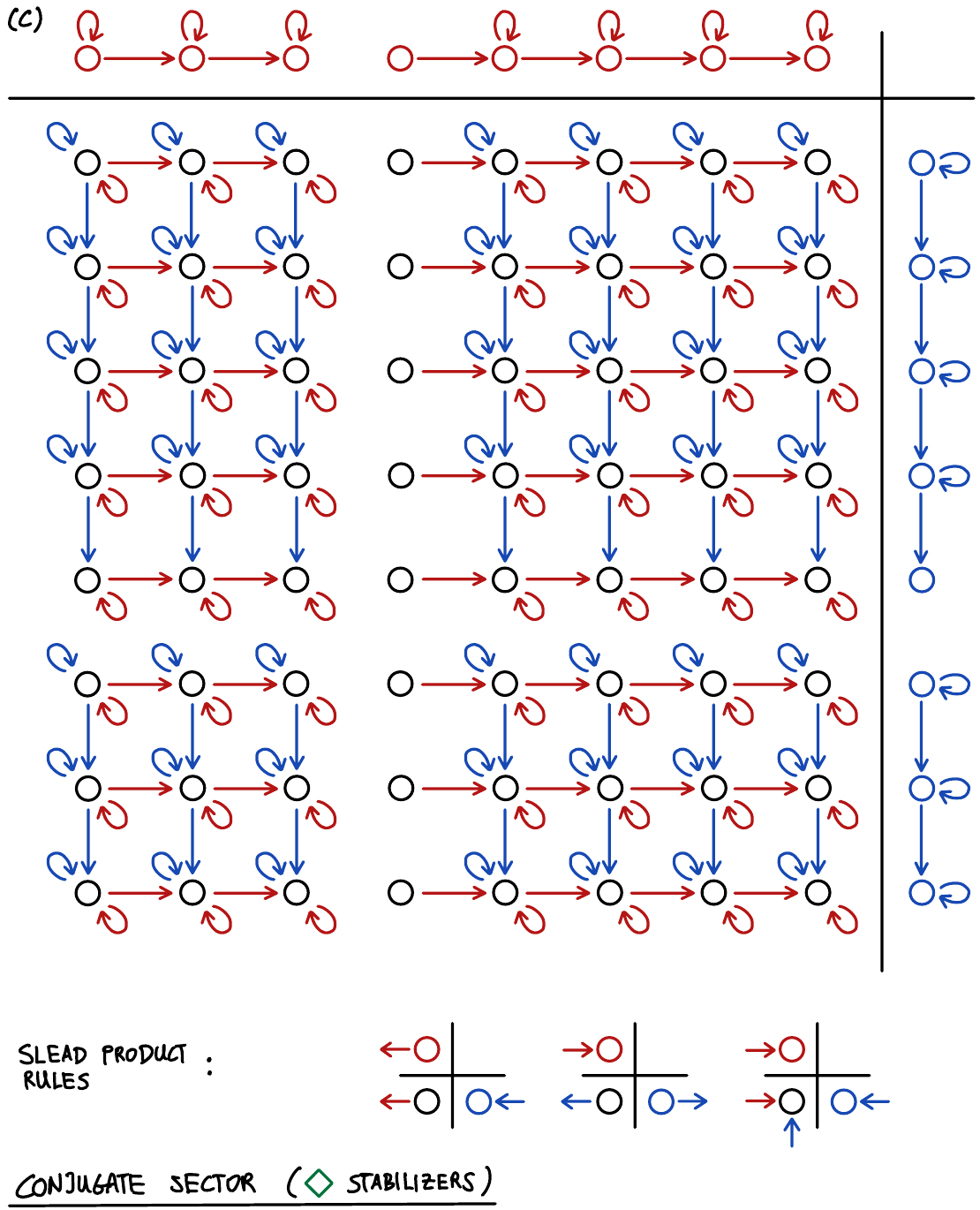}
    \\ \vspace{12pt}
    \caption{\textbf{Tanner graph and sleads of a product code.} \textbf{(a)} Tanner graph of the hypergraph product of two check-depleted classical repetition codes, drawn in blue and red at the top and right sides of the panel. Two types of qubits, colored blue and red, and stabilizers of $X$ and $Z$ types, drawn as green squares and diamonds, are present in the quantum code. A summary of the hypergraph product association rules for the qubits and stabilizers is provided at the bottom of the panel. Gray ovals and squares in the background demarcate cells on the classical and quantum codes. \textbf{(b)} Slead representation of the product code in the direct sector, arbitrarily defined to be capturing the square stabilizers. \textbf{(c)} Slead representation of the product code in the conjugate sector capturing the diamond stabilizers. Corresponding hypergraph product association rules in the slead representation are given at the bottom of the panels.}
    \label{app-fig:examples/hgp_tanner_slead}
    \phantomsubfloat{\label{app-fig:examples/hgp_tanner_slead/tanner}}
    \phantomsubfloat{\label{app-fig:examples/hgp_tanner_slead/slead_direct}}
    \phantomsubfloat{\label{app-fig:examples/hgp_tanner_slead/slead_conjugate}}
\end{figure}

We first show in Fig.~\ref{app-fig:examples/hgp_tanner_slead/tanner} the Tanner graph associated with this product code. Following the main text (namely Sec.~\ref{sec:cored_product_codes/coring_slead_product_codes}, or Figs.~\ref{fig:quantum_code} and \ref{fig:quantum_code_2}), we color the two repetition code classical factors blue and red, and the qubits on the product code accordingly are divided into blue and red species. Here, we denote $X$- and $Z$-type stabilizers as green squares and diamonds. To review, the hypergraph product entails taking the Cartesian product of the Tanner graphs of the two classical factors, and applying a set of association rules to associate products of bits and checks to qubits and stabilizers; we provide a summary of these association rules at the bottom of the panel.

We illustrate the corresponding slead representations of the product code in Figs.~\ref{app-fig:examples/hgp_tanner_slead/slead_direct} and \ref{app-fig:examples/hgp_tanner_slead/slead_conjugate}, in the direct and conjugate sectors, respectively. Here, we arbitrarily select the direct (conjugate) sector as that capturing the $X$-type ($Z$-type) stabilizers. Similar to the case of classical codes, we group qubits and stabilizers into cells in going from the Tanner graph to the slead representation, each of which is represented by a vertex on the slead; this cell structure on the product code is inherited from the cell structure of the classical factor codes. In the slead representation, the hypergraph product entails taking the Cartesian product of the sleads of the classical code factors except at source vertices, and applying corresponding sets of association rules, which we illustrate at the bottom of the panels.

\subsection{Coring process represented on Tanner graphs and sleads}
\label{app-sec:examples/hgp_coring}

We now continue our illustrated example in Fig.~\ref{app-fig:examples/hgp_tanner_slead}, which we remind is a reproduction of the check-depleted product code first drawn in Fig.~\ref{fig:quantum_code_2} of the main text, by depicting the coring process---as described in Sec.~\ref{sec:cored_product_codes/coring}---in a step-by-step fashion. For ease of discussion, as before, we arbitrarily associate squares (diamonds) with $X$-type ($Z$-type) stabilizers on the Tanner graph.

First, as discussed in the main text, we note that the top-right quadrant of the product code is exactly a surface code patch and encodes logical information. The coring procedure is guaranteed to preserve the logical dimension of the code, and in fact leaves this quadrant entirely untouched, as there is no blue qubit available that is involved only in a single stabilizer of either $X$- or $Z$-type for the coring algorithm to start deletion from.

\begin{figure}[ht]
    \includegraphics[width=0.95\linewidth]{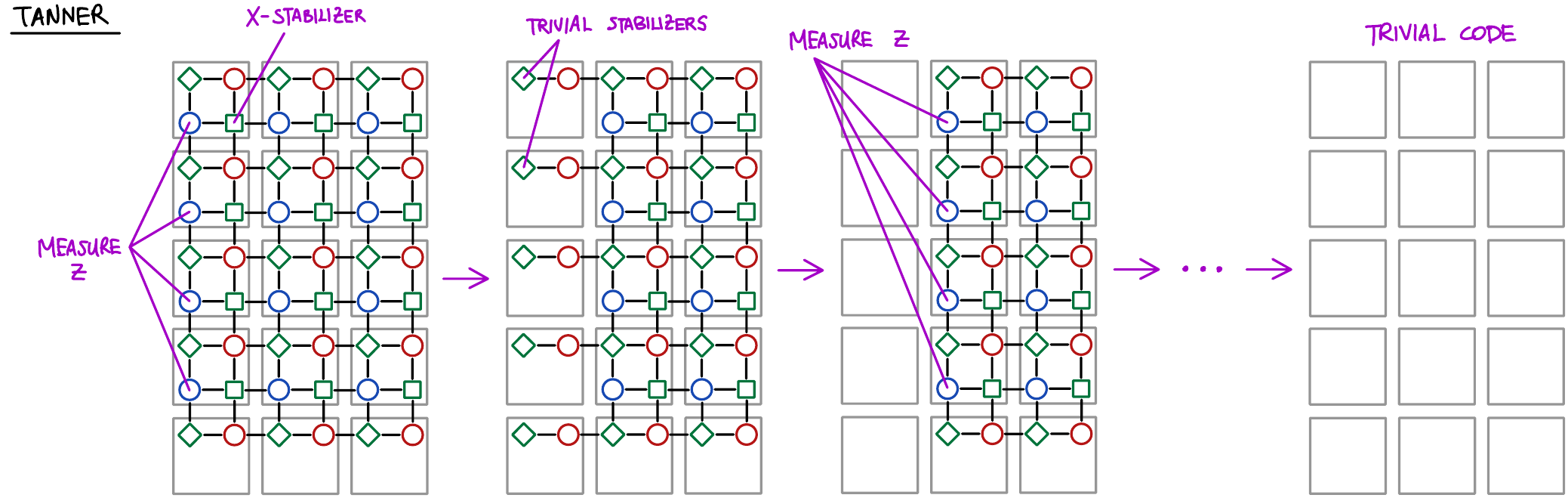}
    \\ \vspace{10pt}
    \includegraphics[width=0.95\linewidth]{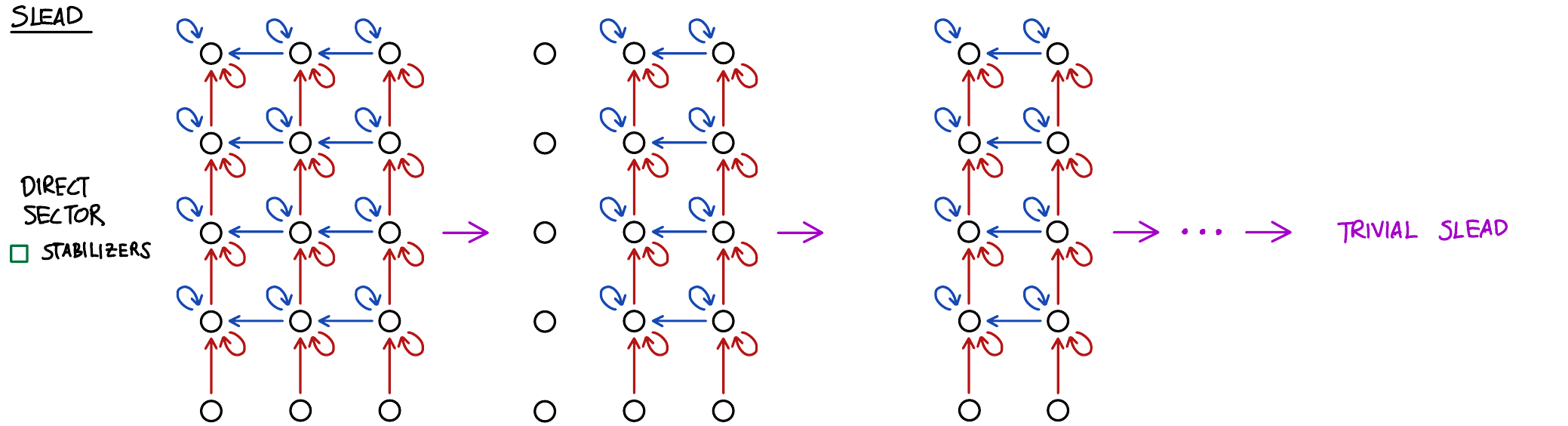}
    \\ \vspace{10pt}
    \includegraphics[width=0.95\linewidth]{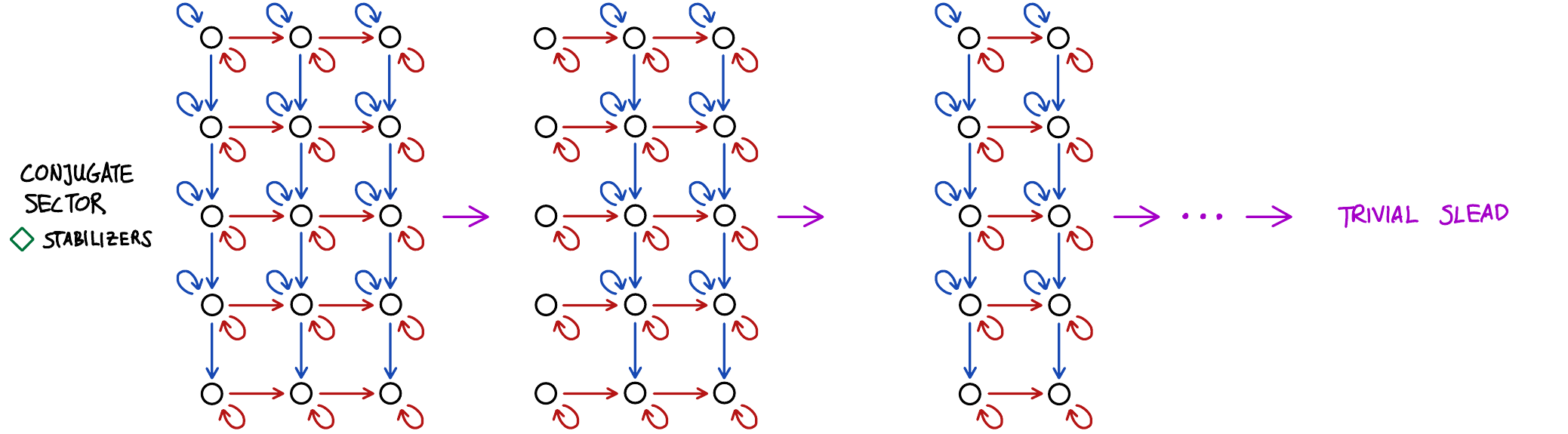}
    \\ \vspace{10pt}
    \caption{\textbf{Coring process on the top-left quadrant of example product code.} We illustrate a step-by-step run-through of the coring algorithm as applied to the top-left quadrant of the product code of Fig.~\ref{app-fig:examples/hgp_tanner_slead}, in both Tanner graph and slead representations. For the slead representation, the direct (conjugate) sector has been arbitrarily chosen to capture the $X$-type ($Z$-type) stabilizers.}
    \label{app-fig:examples/coring_top_left}
\end{figure}

The top-left quadrant, however, does not encode logical information, and indeed has blue qubits on the leftmost boundary that are each involved only in a single $X$-type stabilizer. The coring procedure starts by measuring these qubits in the $Z$-basis, as depicted in Fig.~\ref{app-fig:examples/coring_top_left}, thereby enabling their removal together with the neighboring $X$-type stabilizers. Now, the remaining $Z$-type stabilizers on the leftmost boundary have become trivialized (i.e., they act only on a single qubit each), and are removed by the coring procedure together with their supports. This exposes a new leftmost boundary of structure identical to the initial lattice, and the process repeats, ultimately resulting in the entire quadrant being deleted. In Fig.~\ref{app-fig:examples/coring_top_left} we illustrate this process using both Tanner graph and slead representations of the code for clarity.

\begin{figure}[ht]
    \includegraphics[width=0.95\linewidth]{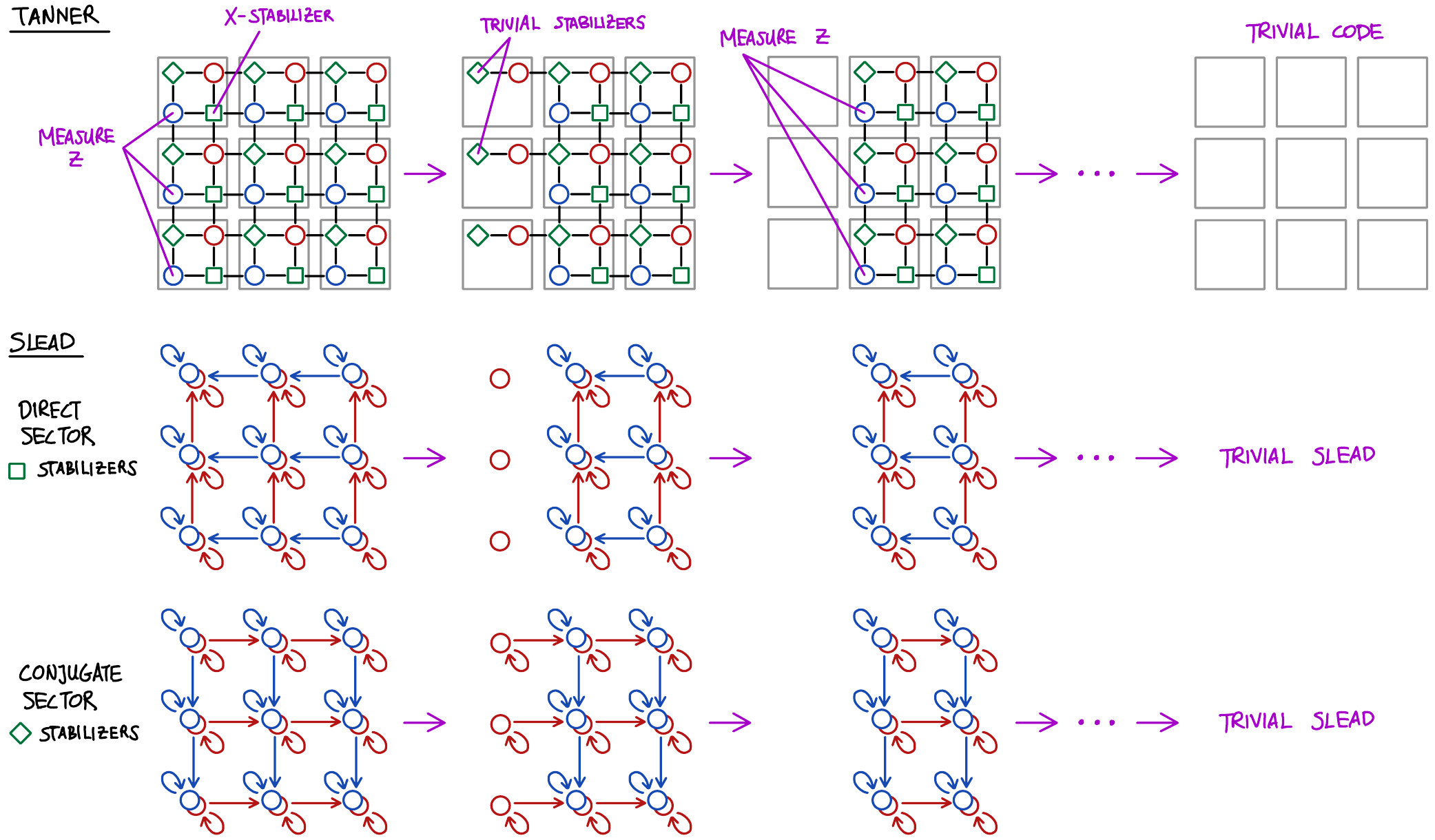}
    \\ \vspace{10pt}
    \caption{\textbf{Coring process on the bottom-left quadrant of example product code.} We illustrate a step-by-step run-through of the coring algorithm as applied to the bottom-left quadrant of the product code of Fig.~\ref{app-fig:examples/hgp_tanner_slead}, in both Tanner graph and slead representations. For the slead representation, the direct (conjugate) sector has been arbitrarily chosen to capture the $X$-type ($Z$-type) stabilizers.}
    \label{app-fig:examples/coring_bot_left}
\end{figure}

\begin{figure}[ht]
    \includegraphics[width=0.95\linewidth]{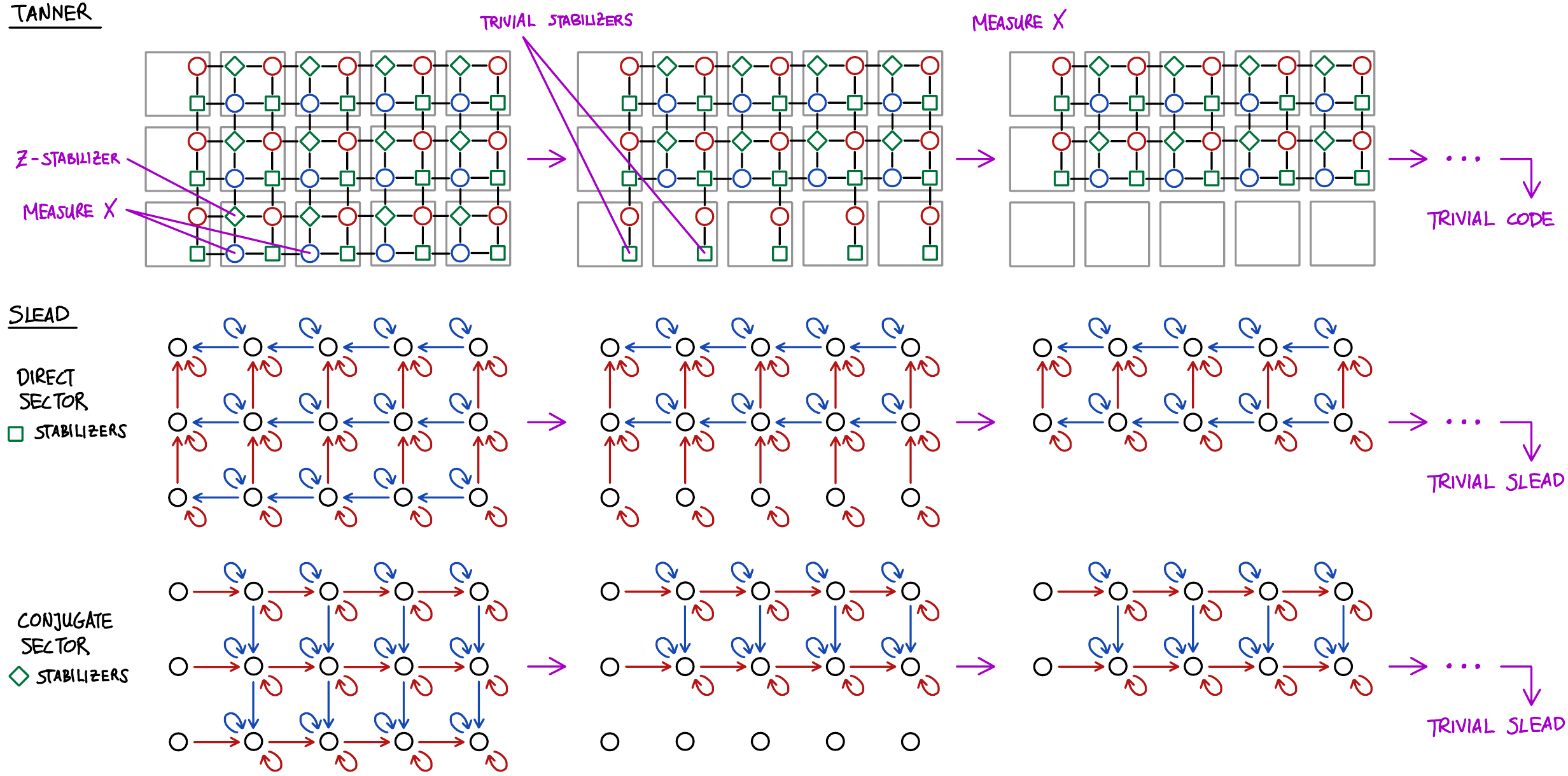}
    \\ \vspace{10pt}
    \caption{\textbf{Coring process on the bottom-right quadrant of example product code.} We illustrate a step-by-step run-through of the coring algorithm as applied to the bottom-right quadrant of the product code of Fig.~\ref{app-fig:examples/hgp_tanner_slead}, in both Tanner graph and slead representations. For the slead representation, the direct (conjugate) sector has been arbitrarily chosen to capture the $X$-type ($Z$-type) stabilizers.}
    \label{app-fig:examples/coring_bot_right}
\end{figure}

Likewise, we illustrate the coring process for the bottom-left and bottom-right quadrants in Figs.~\ref{app-fig:examples/coring_bot_left} and \ref{app-fig:examples/coring_bot_right}, respectively. These quadrants similarly do not encode logical information and are entirely deleted by the coring process. The coring algorithm begins by measuring in the $Z$ basis on the left boundary for the bottom-left quadrant in Fig.~\ref{app-fig:examples/coring_bot_left}, and in the $X$ basis on the bottom boundary for the bottom-right quadrant in Fig.~\ref{app-fig:examples/coring_bot_right}. 

We remark that in reality, the coring of these quadrants occur at the same time as the algorithm proceeds; we have only separated their illustrations here for the sake of visual clarity. In general, the coring procedure removes parts of the code that do not encode logical information whilst preserving parts that do.

\clearpage
\pagebreak

\section{Numerical upper bound on energy barrier}
\label{sec:barrier_bound}

In the main text we present numerical upper bounds on energy barriers based on parity check matrices for linear classical and CSS quantum codes.
In this appendix we summarize the numerical method utilized to compute these upper bounds, which is in essence a greedy search over paths between antipodal points $(0,0,0,\ldots)$ and $(1,1,1,\ldots)$ on the unit hypercube.

In principle, a path $\bm\sigma$ implementing a specified error $e$ may be arbitrarily long, and the number of distinct configurations is $2^n$ for block length $n$.
This greatly complicates the task of searching over such paths, whose only condition is to reach the spin state $e$ at the endpoint.
To make the problem more tractable we rely on understanding from Sec.~\ref{sec:classical_barrier}, noting that for a codeword $C$ of a slead code, the worldline implementation requires flipping spins only within the support of $C$.
Moreover, worldline implementations are minimal in the sense that the length of the path is the minimum, $|C|$, as each spin in the support of $C$ is flipped exactly once.

Even within the class of minimal implementations of a codeword $C$ (suppose $C$ is the unique nonzero codeword, so $|C| = d$) there remain $d!$ possible orderings, and it is over this set that we perform a numerical search.
At any point in the walk the spin state may be represented by a binary vector $v \in \F_2^d$ and the energy barrier by a path-dependent value $E \in \R$.
Given this tuple $(v,E)$ and the classical Hamiltonian $H: \F_2^d \to \R$, we generate the set $\tilde V = \{v + \hat e_i \}$ for all positions $\{i \mid v_i = 0\}$, from which we compute $\tilde E = \min_{\tilde v \in \tilde V} H \cdot \tilde v$.
The process is repeated for every pair $(v',E')$, $v' \in \tilde V$ and $H \cdot v' = \tilde E$, and $E' = \max\,\{E,\tilde E\}$.

The protocol above can be initialized either with the trivial codeword $(\bm 0,0)$, or with a full initial set $\{(\hat e_i , H \cdot \hat e_i)\}$, $i = 1,\ldots,d$ for a slightly broader search.
Note that we do not make use of the topological ordering, so the paths considered form a superset of all worldline implementations. 
Various optimizations are possible, including using bit-packing and bitwise operations to save both memory and time required to compute energies.
Moreover, at any step many paths may produce the same word, and since all subsequent steps will be identical, one many combine redundant histories.
One interesting point is that this algorithm provides a heuristic for the entropy of codewords via the multiplicity of degenerate paths.
We observe exactly the expected exponential scaling of the number of paths implementing translation-invariant fractal codewords by neglecting to combine duplicates, and also see that for disordered codes the growth of duplicates is very strongly suppressed even without a combination step.

\section{Code deformation for geometrically local embedding}
\label{app:aux_codes}

In Sec.~\ref{sec:cored_locality} of the main text, a general process is sketched for deforming cored product codes to achieve a local embedding in fewer dimensions.
The basic scheme is to consider a folded curve of one lower dimension and finite thickness embedded into the higher-dimensional space of the product code, as illustrated for a three-dimensional example in Fig.~\ref{fig:cored_locality}.
The sites are then orthogonally projected onto the nearest point of the lower-dimensional manifold.
This process necessarily generates a finite density of long-range interactions that violate locality in the lower dimension and so must be treated.
We emphasize again that because the treatment of the long-range connections must be specialized to each instance of a code, in order to avoid interfering with the analysis of codes with finite block length we do not implement it for the numerical simulations.
We contend that this does not affect our arguments for the growing memory lifetime.

\subsection{Mediation by auxiliary classical codes}

\begin{figure}[ht!]
    \centering
    \includegraphics[width=0.90\textwidth]{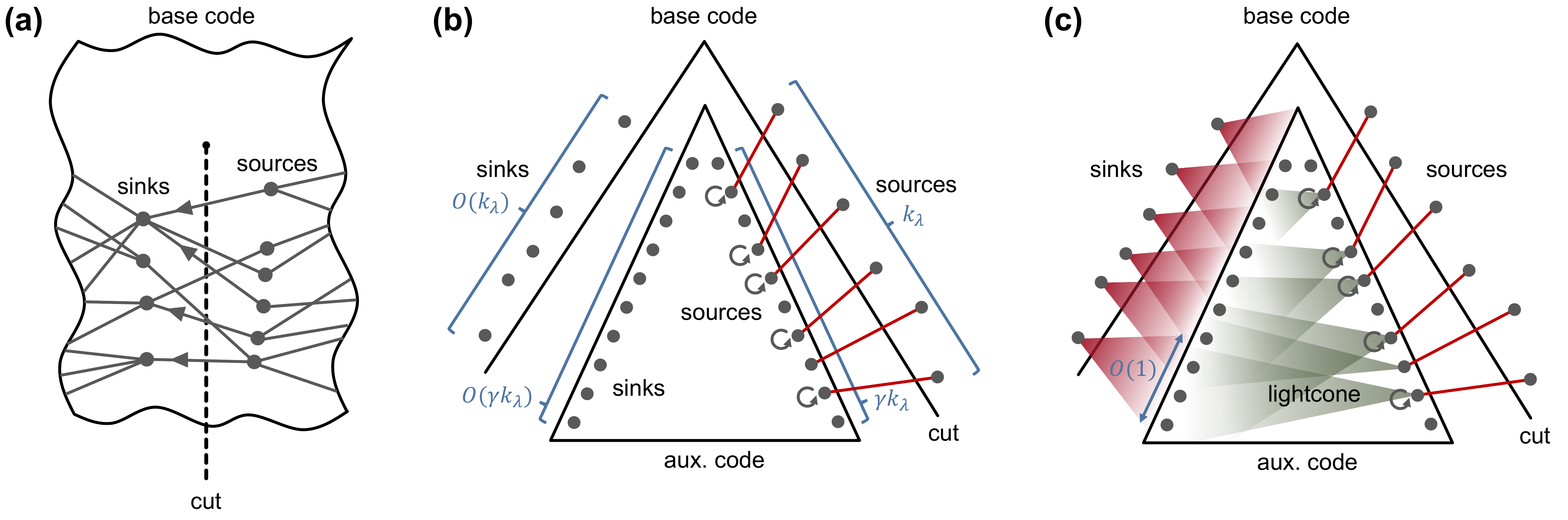}
    \caption{\textbf{Mediation of parity check connections across a cut.} \textbf{(a)} After selecting a lower-dimensional manifold, cuts are made to the base code between vertices whose projected positions become nonlocal.
    Because of the precise alignment of the embedding described here, all edges crossing the cut are similarly aligned and one can unambiguously designate ``source'' and ``sink'' sides.
    \textbf{(b)} The auxiliary slead code is first coupled to the base code by drawing geometrically local edges between vertices on the two source sides.
    Every source vertex in the base code is connected to a single source vertex in the auxiliary code, so that flipping its spin induces a logical bit of the auxiliary code, with some associated spin configuration throughout the bulk and in particular on the sink vertices within its lightcone.
    \textbf{(c)} In the final step, the sink vertices of the auxiliary code are coupled to the sink vertices of the base code in a geometrically local way, ensuring that the codeword of the base code does not acquire an energy penalty from the modified parity checks.
    }
    \label{fig:cored_locality_appendix}
\end{figure}

Our proposal is to ``mediate'' long-range stabilizer connections by replacing them with auxiliary error-correcting codes, as mentioned in the main text.
The scaling of the block length of the code, as specified in Secs.~\ref{sec:cored_locality} and \ref{sec:finite_temp/codes}, takes into account these additional qubits.
Because naively introducing auxiliary qubits to the quantum code in this way does not preserve stabilizer commutation, we propose a concrete protocol for instead modifying one of the classical factors to enable a local embedding of the product code.

The chief decision that must be made is that of the orientation of the three-dimensional folded manifold.
Restricting ourselves to modifying the factor codes constrains its possible orientations in that two of its basis directions must generate one complete subspace, and the associated classical code is not modified.
The manifold projects to a folded one-dimensional curve in the orthogonal subspace, which we are free to align in such a way as to minimally interfere with the memory, so long as its shape does not generate nonlocal connections of a longer range than can be tolerated by the block length, as described in Sec.~\ref{sec:cored_locality}.

Recall that in the instances described in Sec.~\ref{sec:finite_temp/codes}, the slead code is generated by selecting a vector $\bm t \in \R^2$ for the half-space condition.
For concreteness, suppose the space-filling curve contains the entire subspace associated with classical pinwheel code $\omega_2$, and takes the form of a sinusoid in the subspace associated with classical pinwheel code $\omega_1$, as illustrated in Fig.~\ref{fig:cored_locality} with phase direction parallel to $\bm t$.
Now the projection effectively introduces ``cuts'' in $\omega_1$ in the orthogonal amplitude direction $\bm t^\perp$, and any parity checks involving qubits across these cuts become long-ranged.

Observe now that associated with any single cut in $\omega_1$ is a quantity $O(L^\nu)$ of long-range connections, each having length $\ell = O(L^\nu)$.
We mediate all of the long-range connections across a single cut collectively, modifying $\omega_1$ by coupling to auxiliary codes as follows.
For a particular cut, we first disconnect the long-ranged edges in the slead, noting that because the manifold is aligned to $\bm t$, all long-ranged connections are consistently oriented: thus, one side of the cut can be regarded as the ``source'' and the other the ``sink'', as illustrated in Fig.~\ref{fig:cored_locality_appendix}(a).
We refer to these as source and sink, though they are not the global sources and sinks of the code.
For the auxiliary codes we choose pinwheel slead codes, now derived from tiling an isosceles triangle formed from two triangular prototiles.
The tiling generation $g$ used for the auxiliary code is chosen such that the number of boundary vertices is greater than (but within a constant factor of) the number of long-range edges crossing the cut.
Consequently the block length of the auxiliary code is $O(L^{2\nu})$, as desired.
To produce the slead code, a half-space vector $\bm t$ is chosen such that the boundaries of the auxiliary code match the ``source'' and ``sink'' boundaries of the cut.

Given a suitable auxiliary pinwheel code for every cut introduced by the choice of the lower-dimensional manifold, we next couple these codes to the base code $\omega_1$ in such a way as to preserve its relevant properties.
Specifically, we preserve the LDPC structure and do not change the support of $C_1$, the codeword of $\omega_1$; additionally, by typicality arguments similar to those used in the main text we show that the energy penalty of words outside of the code space does not decrease.

The introduction of auxiliary slead codes is done independently for each cut in $\omega_1$.
Consequently we focus on a single cut, and denote the associated auxiliary slead code by $\lambda$.
Recalling that the cut has well-defined ``source'' and ``sink'' sides, we denote the corresponding subsets of vertices as $V_{\omega_1}^\mathrm{src} \subset V_{\omega_1}$ and $V_{\omega_1}^\mathrm{sink} \subset V_{\omega_1}$, respectively.
Similarly, in the auxiliary code we denote the sets of source and sink vertices as $V_\lambda^\mathrm{src} \subset V_\lambda$ and $V_\lambda^\mathrm{sink} \subset V_\lambda$, respectively.
We align the boundaries of $\lambda$ with the cut in the base code, so that the vertices $V_{\omega_1}^\mathrm{src}$ are proximate to the vertices $V_\lambda^\mathrm{src}$, and $V_{\omega_1}^\mathrm{sink}$ are proximate to the vertices $V_\lambda^\mathrm{sink}$.

By construction, $\lambda$ has $O(L^{2\nu})$ vertices, and has $n_\lambda^\mathrm{src} = |V_\lambda^\mathrm{src}| = O(L^{\nu}) = O(\ell)$ source vertices.
We choose $n^\lambda_\mathrm{src}$ to be larger than $n_{\omega_1}^\mathrm{src} = |V_{\omega_1}^\mathrm{src}|$ by a multiplicative constant $\gamma > 1$.
For each of the vertices $v \in V_{\omega_1}^\mathrm{src}$, a nearby vertex $w \in V_\lambda^\mathrm{src}$ is modified by attaching a directed edge from $v$ to $w$.
This includes the spin on $v$ in the parity check on $w$, and is illustrated in Fig.~\ref{fig:cored_locality_appendix}(b).
As in the main text, in order to draw a specific connection we sample from the auxiliary vertices within a ball of fixed radius around $v$, choosing as $w$ the vertex leading to the highest-weight logical of $\lambda$ after check depletion, subject to the condition that no two vertices in $V_{\omega_1}^\mathrm{src}$ can connect to the same vertex in $V_\lambda^\mathrm{src}$.
Now the logical dimension of $\lambda$ is $k_\lambda = n_{\omega_1}^\mathrm{src}$, and flipping a spin in the base code induces a logical bit of the auxiliary code, which can be found by following the worldline prescription.

Under the coupling of source vertices above, the spin configuration on $V_{\omega_1}^\mathrm{src}$ induces a codeword of $\lambda$, with the trivial codeword producing the trivial spin configuration.
By the next step of locally coupling sink vertices $V_\lambda^\mathrm{sink}$ to $V_{\omega_1}^\mathrm{sink}$ we enforce that the codeword $C_1$ is preserved and no additional codewords are introduced.
First, the spin configuration given by the restriction of $C_1$ to $V_{\omega_1}^\mathrm{src}$ is propagated through $\lambda$ to its sink vertices following the worldline prescription to produce a codeword.
Now each of the $n_{\omega_1}^\mathrm{sink}$ parity checks is updated by drawing directed edges to its vertex from a subset of $V_\lambda^\mathrm{sink}$ within a ball of fixed radius.
These connections are drawn in such a way that the parity checks on $V_{\omega_1}^\mathrm{sink}$ are satisfied when the spin configuration on $\lambda$ is the one specified by $C_1$ in this way and the spins in $V_{\omega_1}^\mathrm{sink}$ take the spin configuration given by the restriction of $C_1$.
After this process of coupling $\lambda$ to $\omega_1$, the restriction of the new codeword to the vertices of $\omega_1$ is again $C_1$, and the restriction to the vertices of $\lambda$ is one of its codewords, with the others obtaining an energy penalty.
Thus the mediated code has again only a single logical bit, maintains the slead structure, remains LDPC, and is geometrically local after projection to the lower-dimensional manifold.
A fully coupled auxiliary code is illustrated in Fig.~\ref{fig:cored_locality_appendix}(c).

\subsection{Effects on quantum memory}

It is necessary to show that the mediation procedure described above is possible and, crucially, that it preserves energy barriers.
Before the cut is introduced to the base code $\omega_1$, the checks are satisfied if the spin configuration matches the codeword $C_1$, and perturbations lead to an energy penalty from violated checks.
By removing the connections along the cut, which become long-ranged after projection to the lower-dimensional manifold, the energy penalty is reduced, ultimately eliminating the barrier.
The purpose of the auxiliary codes is to reintroduce the energy penalty in a way that maintains geometric locality on the manifold.
Instead of trying to use a distinct code to replicate the action of each parity check, the auxiliary codes are coupled to the base in such a way as to simultaneously restore the penalties associated with all checks crossing the cut without necessarily reproducing any one check in particular.

The auxiliary code $\lambda$ has logical dimension $k_\lambda = n_{\omega_1}^\mathrm{src}$, and once coupled to the base code every spin configuration on $V_{\omega_1}^\mathrm{src}$ induces a valid codeword of $\lambda$ via the worldline prescription.
Thus the checks within the auxiliary code need not be violated, regardless of the word in $\omega_1$.
Instead, an energy penalty arises from coupling $V_\lambda^\mathrm{sink}$ to the adjacent vertices $V_{\omega_1}^\mathrm{sink}$.
By construction, the parity checks on $V_{\omega_1}^\mathrm{sink}$ are not violated when the spin configuration on $V_{\omega_1}^\mathrm{src} \cup V_{\omega_1}^\mathrm{sink}$ is the restriction of the codeword $C_1$.
As in the main text, we use typicality to argue that not only is such a construction possible but also that having been chosen, local perturbations away from the codeword incur an energy penalty.
For the former, observe that the maximum level in the topological ordering of $\lambda$ is $p_{\max} = O(L^\nu) = O(n_{\omega_1}^\mathrm{src})$.
In the absence of spatial structure fluctuations around typicality are suppressed, so the lightcone of a nontrivial codeword includes $O(L^\nu)$ sink vertices, evenly distributed between flipped and un-flipped spins.
We consider the case that the local spin configuration is the restriction of $C_1$: then the vertices in $V_{\omega_1}^\mathrm{sink}$ adjacent to the intersection of the lightcone and $V_\lambda^\mathrm{sink}$ admit drawing geometrically local directed edges in such a way as to satisfy all parity checks on $V_{\omega_1}^\mathrm{sink}$.
By drawing only a finite number of edges for each vertex, the LDPC property of the code is maintained. 

It remains to consider the case that the spin configuration on $V_{\omega_1}^\mathrm{src} \cup V_{\omega_1}^\mathrm{sink}$ is not the restriction of $C_1$.
We first consider the likelihood of collision, meaning that the spin configuration on $V_\lambda^\mathrm{sink}$ is the same as that induced by $C_1$: this could introduce an undesired codeword.
Note that $n_\lambda^\mathrm{sink} = \gamma k_\lambda$ with $\gamma > 1$, so while there are $2^{k_\lambda}$ codewords, the number of configurations of the sink vertices is $2^{\gamma k_\lambda}$.
Thus, assuming typicality, the probability of a collision with the codeword is exponentially suppressed in $\gamma$ and is independent of block length.
We conclude that collision does not affect the code.

Although collision is suppressed, the energy penalty is a relatively coarser measure, and this does not guarantee that other spin configurations cannot be found which have low energy along the cut.
If the spin configuration on $\lambda$ is found from $V_{\omega_1}^\mathrm{src}$ by the worldline prescription, then the penalty arises entirely from the parity checks on $V_{\omega_1}^\mathrm{sink}$ coupled to the spins on $V_\lambda^\mathrm{sink}$.
As these $n_{\omega_1}^\mathrm{sink}$ checks are by construction fewer than the $n_\lambda^\mathrm{sink}$ spins by a factor of $\gamma$, it may seem inevitable that additional configurations with low or no energy cost can be found.
However, in order to satisfy all checks in the bulk of $\lambda$, its spin configuration must be the worldline of some configuration on $V_{\omega_1}^\mathrm{src}$.
The parity checks then need to distinguish only between configurations generated by the $n_{\omega_1}^\mathrm{src} = O(n_{\omega_1}^\mathrm{sink})$ images on $V_\lambda^\mathrm{sink}$ caused by flipping spins on $V_{\omega_1}^\mathrm{src}$, rather than every possible configuration.
This reinstates the situation prior to mediation, in which $n_{\omega_1}^\mathrm{sink} = O(n_{\omega_1}^\mathrm{src})$ parity checks acted on $O(n_{\omega_1}^\mathrm{src})$ spins across the cut.
It is in fact improved somewhat, as prior to mediation, one can locally nucleate a codeword on the cut and incur only a finite energy penalty (arising only from the edges crossing the cut) for an arbitrarily large perturbation due to geometric locality.
After mediation, the image of a spin flip on $V_{\omega_1}^\mathrm{src}$ is given by the intersection of its lightcone with $V_\lambda^\mathrm{sink}$, which is $O(L^\nu)$, and locally nucleating a codeword on $V_{\omega_1}^\mathrm{src}$ does not guarantee that the energy penalty is confined to the boundary of the perturbation.

As a final observation, one may wonder whether introducing auxiliary codes, each with a large number of codewords, introduces undesirable entropy into the code.
It can be seen that this is not the case from the preceding point, namely that the energy of the codewords of $\lambda$ not associated with the restriction of $C_1$ is typically large, and that low-energy codewords are not necessarily close in Hamming distance, so do not inject entropy in the low-energy landscape near the desired codeword.

\clearpage
\pagebreak

\section{Kinetic Monte Carlo simulations to assess memory lifetimes}
\label{app-sec:kinetic_monte_carlo}

We provide details of our numerical simulations of memory lifetimes in this section. As explained in the main text, we simulated time-evolution of our quantum memory models under thermal noise, formally described by the Davies master equation (see Sec.~\ref{sec:finite_temp}), by rejection-free kinetic Monte Carlo implemented through a variant of the Bortz-Kalos-Lebowitz (BKL) algorithm~\cite{bortz1975new}, also called the residence-time or $n$-fold way algorithm. The rejection-free algorithm is computationally more efficient than direct Monte Carlo with rejection sampling.

\textbf{Preliminaries.} We first give an outline of BKL rejection-free kinetic Monte Carlo applied to our setting. Recall that as our quantum codes are CSS and we consider thermally distributed single-qubit $\{X_j\}_{j = 1}^n$ and $\{Z_j\}_{j = 1}^n$ noise, the $X$ and $Z$ sectors of the code are decoupled and we need only consider a single sector $\mu \in \{X, Z\}$ at a time for simulating dynamics. We denote the Hamiltonian $H_\mu$ in the $\mu$ sector, which comprises the checks (i.e., stabilizer generators) of type $\mu$,
\begin{equation}
    H_\mu = -\sum_{s \in \mathcal{S}_\mu} s,
\end{equation}
and in sector $\mu$ we consider noise of the opposite type $\overline{\mu}$ which anticommute with and would flip checks in $H_\mu$.

The BKL algorithm works by keeping track of the rates of possible transitions from the current state of the system, in the present context the rates of $\{\mu_j\}_{j = 1}^n$ thermal noise occurring on the $n$ qubits. In each time step, the system is evolved by sampling a transition to undertake according to their rates, time is advanced by a Poisson-distributed amount with mean the inverse of the rate of the undertaken transition, and the rates of transitions are updated to prepare for the next time step. For clarity, we illustrate a basic implementation of the BKL algorithm for code dynamics in Algorithm~\ref{alg:BKL-algorithm-naive}; more advanced variants of the algorithm of superior run-time are then described later. 

Algorithm~\ref{alg:BKL-algorithm-naive} is backed by a data structure for tracking transition rates, which supports the functions $\Call{Init}{}$ for initialization, $\Call{Sample}{}$ for sampling a transition, $\Call{TotalRate}{}$ for obtaining the total transition rate, and $\Call{Update}{}$ for updating a transition rate. We illustrate a simple implementation of this data structure in Data Structure~\ref{alg:BKL-data-structure-array}, which stores transition rates in an array and performs linear traversal over the array to tally cumulative rates for sampling.

\begin{algorithm}[H]
\caption{Naïve rejection-free kinetic Monte Carlo}
\label{alg:BKL-algorithm-naive}
\begin{algorithmic}
    \Require Qubit labeling $[n]$, stabilizer labeling $[m]$ in sector $\mu \in \{X, Z\}$, code Hamiltonian matrix $H_\mu \in \mathbb{F}_2^{m \times n}$ in sector $\mu$, inverse temperature $\beta \in \mathbb{R}_+$, time $T \in \mathbb{R}_+$ to evolve to.
    \Ensure State of qubits $\bm{w} \in \mathbb{F}_2^n$, syndromes $\bm{e} \in \mathbb{F}_2^m$, energy $E \in \mathbb{N}$ of state, at smallest sampled time $t \geq T$. 
    \algrule
    \State $(\bm{w}, \bm{e}, E, t) \gets (\bm{0}_n, \bm{0}_m, 0, 0)$
        \Comment{Start from $\ket{0}^{\otimes n}$ qubit state in $\mu$-basis.}
    \State $\Call{Init}{\mathbf{1}_m \cdot H_\mu}$
        \Comment{Column sums ($\mathbf{1}_m \cdot H_\mu$) of $H_\mu$ are state energies after flipping each qubit.}
    \While{$t < T$}
        \Comment{Take time steps until $T$ is reached.}
        \State $t \gets t + \ln(1 / \Call{Uniform}{}) / \Call{TotalRate}{}$
            \Comment{Advance time by a Poisson-distributed amount.}
        \State $q \gets \Call{Sample}{}$
            \Comment{Sample a qubit $q$ to be flipped.}
        \State $w_q \gets \neg w_q$
            \Comment{Flip qubit $q$}.
        \State $\bm{e} \gets H_\mu \odot \bm{w}$
            \Comment{Compute syndromes of current state.}
        \State $E \gets \mathbf{1}_m \cdot \bm{e}$
            \Comment{Compute energy of current state.}
        \For{$p = 1, 2, \ldots, n$}
            \Comment{Update transitions.}
            \State $\bm{w}' \gets \bm{w}$
                \Comment{Temporary copy of qubit states.}
            \State $w'_p \gets \neg w'_p$
                \Comment{Flip qubit $p$.}
            \State $\bm{e}' \gets H_\mu \odot \bm{w}'$
                \Comment{Compute syndromes of transition state.}
            \State $E' \gets \mathbf{1}_m \cdot \bm{e}'$
                \Comment{Compute energy of transition state.}
            \State $\Call{Update}{p, E' - E}$
                \Comment{Update transition energy difference.}
            \EndFor
        \EndWhile
    \State \Return $(\bm{w}, \bm{e}, E, t)$
\end{algorithmic}
\algrule
\noindent\textit{Remarks:} Matrix and vector multiplication are denoted $\cdot$ over the integers and $\odot$ over $\mathbb{F}_2$. No distinction is made between row and column vectors; transpositions when needed are implied. $\neg x = 1 \oplus x$ denotes negation of the binary variable $x$. $\mathbf{0}_m$ and $\mathbf{1}_m$ denote a length-$m$ vector with all zero and unit entries respectively. $\Call{Uniform}{}$ returns a uniformly distributed real number in the range $[0, 1)$. The algorithm is backed by an underlying data structure for tracking transition rates, which supports $\Call{Init}{}$ for initialization, $\Call{Sample}{}$ for sampling a transition, $\Call{TotalRate}{}$ for obtaining the total transition rate, and $\Call{Update}{}$ for updating a transition rate. A simple implementation of this data structure is given in Data Structure~\ref{alg:BKL-data-structure-array}.
\end{algorithm}

\begin{algorithm}[H]
\makeatletter
\renewcommand{\ALG@name}{Data Structure}
\makeatother
\caption{Naive linear array data structure for tracking transition rates}
\label{alg:BKL-data-structure-array}
\renewcommand{\algorithmicrequire}{\textbf{Parameters:}}
\renewcommand{\algorithmicensure}{\textbf{Data:}}
\begin{algorithmic}
    \Require Qubit labeling $[n]$, inverse temperature $\beta$.
    \Ensure Array $\bm{r} \in \mathbb{R}^n_+$ storing transition rates, float $R \in \mathbb{R}_+$ storing total transition rate.
    \algrule
    \Function{Init}{$\bm{\Delta}$}
        \Comment{Initialize transition rates given transition energy differences $\bm{\Delta}$.}
        \State $(\bm{r}, R) \gets (\bm{0}_n, 0)$
        \For{$q = 1, 2, \ldots, n$}
            \State $\Call{Update}{q, \Delta_q}$
            \EndFor
        \EndFunction
    \algrule
    \Function{TotalRate}{}
        \Comment{Yields total transition rate $R$.}
        \State \Return $R$
        \EndFunction
    \algrule
    \Function{Sample}{}
        \Comment{Samples a transition $q$ with probability $r_q / R$.}
        \State $R^{\mathrm{targ}} \gets \Call{Uniform}{} \cdot \Call{TotalRate}{}$
        \State $R^{\mathrm{curr}} \gets 0$
        \For{$q = 1, 2, \ldots, n$}
            \Comment{Find smallest index $q$ such that $\sum_{k = 1}^{q - 1} r_q < R^{\mathrm{targ}} \leq \sum_{k = 1}^q r_q$.}
            \State $R^{\mathrm{curr}} \gets R^{\mathrm{curr}} + r_q$
            \If{$R^{\mathrm{targ}} \leq R^{\mathrm{curr}}$}
                \State \textbf{break}
                \EndIf
            \EndFor
        \State \Return $q$
        \EndFunction
    \algrule
    \Function{Update}{$q, \Delta$}
        \Comment{Updates transition rate for qubit $q$ with energy difference $\Delta$.}
        \State $R \gets R - r_q$
        \State $r_q \gets p(\Delta_q)]$
            \Comment{Transition rate subject to detailed balance.}
        \State $R \gets R + r_q$
        \EndFunction
\end{algorithmic}
\algrule
\noindent\textit{Remarks:} All arrays are $1$-indexed to be consistent with mathematical convention. $\Call{Uniform}{}$ returns a uniformly distributed real number in the range $[0, 1)$. $\mathbf{0}_m$ denotes a length-$m$ vector with all zero entries. $p(\Delta)$ is a transition rate function satisfying detailed balance (see Eqs.~\ref{app-eq:transition-rate-metropolis} and \ref{app-eq:transition-rate-glauber}).
\end{algorithm}

At inverse temperature $\beta$, the rate of a transition that changes the energy of the system by $\Delta$ under Metropolis dynamics is
\begin{equation}
    p(\Delta) = \min[1, \exp(-\beta \Delta)],
    \label{app-eq:transition-rate-metropolis}
\end{equation}
and under Glauber dynamics is
\begin{equation}
    p(\Delta) = \frac{1}{2} \left[ 1 - \tanh{\left(\frac{\beta \Delta}{2} \right)} \right],
    \label{app-eq:transition-rate-glauber}
\end{equation}
both of which satisfy detailed balance. We used Metropolis dynamics for our simulations in this study to be consistent with seminal Refs.~\onlinecite{Haah_PRL2013, Haah_arxiv2011}.

\begin{table*}[!t]
    \centering
    \begin{tabular}{@{} 
        p{1.8cm} p{4.5cm} 
        x{1.6cm} x{1.6cm} x{1.8cm} x{1.6cm}
        x{2.0cm} x{2.0cm} @{}}
        \toprule 
        Algorithm
            & Data structure
            & \multicolumn{4}{c}{Time complexity of data structure operations}
            & \multicolumn{2}{c}{Algorithm complexity} 
        \\
        \cmidrule(lr){3-6}
        \cmidrule(lr){7-8}
        {} 
            & {}
            & $\Call{Init}{}$
            & $\Call{Sample}{}$
            & $\Call{TotalRate}{}$
            & $\Call{Update}{}$
            & Space $\mathcal{M}$
            & Time $\mathcal{T}_{\mathrm{step}}$
        \\
        \midrule 
        Dense
            & Array
            & $\O{n}$
            & $\O{n}$
            & $\O{1}$
            & $\O{1}$
            & $\O{n}$
            & $\O{n^3}$
        \\
        Sparse
            & Array
            & $\O{n}$
            & $\O{n}$
            & $\O{1}$
            & $\O{1}$
            & $\O{n}$
            & $\O{s^3 + n}$
        \\
        Sparse
            & Fenwick tree
            & $\O{n}$
            & $\O{\log n}$
            & $\O{1}$
            & $\O{\log n}$
            & $\O{n}$
            & $\O{s^3 \log n}$
        \\
        Sparse
            & Binning (array-backed)
            & $\O{n}$
            & $\O{s}$
            & $\O{1}$
            & $\O{1}$
            & $\O{n}$
            & $\O{s^3}$
        \\
        Sparse
            & Binning (Fenwick tree-backed)
            & $\O{n}$
            & $\O{\log s}$
            & $\O{1}$
            & $\O{\log s}$
            & $\O{n}$
            & $\O{s^3 \log s}$
        \\
        \bottomrule
    \end{tabular}
    \caption{\textbf{Space and time complexities of rejection-free kinetic Monte Carlo}. Implementations of rejection-free Monte Carlo using dense operations (\cref{alg:BKL-algorithm-naive}) and exploiting sparsity of the error correcting code (\cref{alg:BKL-algorithm-optimized}) can be paired with different underlying data structures for tracking transition rates. We summarize time complexities for the data structure operations used in the Monte Carlo algorithm, as well as the overall algorithm space requirement $\mathcal{M}$ and the time complexity per time step $\mathcal{T}_{\mathrm{step}}$. Here, $n$ is the number of qubits in the code, and $s$ is the sparsity of the code such that each qubit is involved in $\leq s$ stabilizers and each stabilizer contains $\leq s$ qubits. Note that $\mathcal{M} = \Omega(n)$ is a trivial lower bound for any algorithm as $n$ bits are required to store the (generically non-sparse) state of the qubits.}
    \label{tab:BKL-complexities}
\end{table*}

\textbf{Exploiting sparsity.} The simple BKL implementation in Algorithm~\ref{alg:BKL-algorithm-naive} with Data Structure~\ref{alg:BKL-data-structure-array} is straightforward but not efficient. In particular, it requires $O(n^3)$ time per Monte Carlo time step for a code with $n$ qubits (see first row of Table~\ref{tab:BKL-complexities}), which is prohibitively expensive for large codes (recall $n \sim 60\,000$ in our numerical study). Major improvements in run-time can be obtained by exploiting sparsity in the code---i.e., that each check involves at most a small number $s$ of qubits, and each qubit is involved in at most a small number $s$ of qubits. Sparsity enables quicker Monte Carlo time steps on multiple fronts. First, in updating the energy of the system upon selecting a transition (i.e., $X_q$ or $Z_q$ for a qubit $q$) to undertake, it suffices to consider only the $\leq s$ checks involving qubit $q$ that become flipped; all other checks remain with the same signs. Second, in updating the transition rates from the evolved system state, only transitions (i.e., $X_p$ or $Z_p$ for a qubit $p$) that is connected\footnote{Given a stabilizer generator set $\mathcal{S}$, two qubits $q$ and $q'$ are stabilizer-connected if there is an $s \in S$ involving both $q$ and $q'$, that is, the support of $s$ contains both $q$ and $q'$.} to qubit $q$ by a check have their energy deltas $\Delta_p$ and hence rate modified; transitions on all other qubits have unchanged energy differences and rates. There are $\leq s^2$ qubits $\{p\}$ that are connected to $q$ to be considered. Third, the first trick again applies in computing $\Delta_p$ for each of these qubits---flipping qubit $p$ flips only the $\leq s$ checks containing qubit $p$. Putting these tricks together, we present an optimized implementation of the BKL algorithm in Algorithm~\ref{alg:BKL-algorithm-optimized} that exploits code sparsity. The second row of Table~\ref{tab:BKL-complexities} describes the improved $O(s^3 + n)$ run-time per Monte Carlo time step when paired with Data Structure~\ref{alg:BKL-data-structure-array}.

\begin{algorithm}[H]
\caption{Fast rejection-free kinetic Monte Carlo exploiting sparsity}
\label{alg:BKL-algorithm-optimized}
\begin{algorithmic}
    \Require Qubit labeling $[n]$, stabilizer labeling $[m]$ in sector $\mu \in \{X, Z\}$, stabilizer sets $\{ \mathcal{I}_q \}_{q = 1}^n$ where $\mathcal{I}_q \subseteq [m]$ are the stabilizers of sector $\mu$ containing qubit $q$, qubit degrees $\bm{d}$ where $d_q = |\mathcal{I}_q|$, stabilizer-connected qubit neighbor sets $\{ \mathcal{N}_q \}_{q = 1}^n$ where $\mathcal{N}_q \subseteq [n]$, inverse temperature $\beta \in \mathbb{R}_+$, time $T \in \mathbb{R}_+$ to evolve to.
    \Ensure State of qubits $\bm{w} \in \mathbb{F}_2^n$, syndromes $\bm{e} \in \mathbb{F}_2^m$, energy $E \in \mathbb{N}$ of state, at smallest sampled time $t \geq T$. 
    \algrule
    \State $(\bm{w}, \bm{e}, E, t) \gets (\bm{0}_n, \bm{0}_m, 0, 0)$
        \Comment{Start from $\ket{0}^{\otimes n}$ qubit state in $\mu$-basis.}
    \State $\bm{\Delta}^{\mathrm{tr}} \gets d_q$
        \Comment{$\bm{\Delta}^{\mathrm{tr}} \in \mathbb{Z}^n$ are the changes in energy after flipping qubit $q = 1, 2, \ldots, n$.}
    \State $\Call{InitTransitions}{\beta, \bm{\Delta}^{\mathrm{tr}}}$
        \Comment{Initialize data structure keeping track of transition rates.}
    \While{$t < T$}
        \Comment{Take time steps until $T$ is reached.}
        \State $t \gets t + \ln(1 / \Call{Uniform}{}) / \Call{TotalRate}{}$
            \Comment{Advance time by a Poisson-distributed amount.}
        \State $q \gets \Call{SampleTransitions}{}$
            \Comment{Sample a qubit $q$ to be flipped using data structure.}
        \State $w_q \gets \neg w_q$
            \Comment{Flip qubit $q$}.
        \For{$u \in \mathcal{I}_q$}
            \Comment{Update syndromes and energy due to flip.}
            \State $(e_u, E) \gets (\neg e_u, E + 1 - 2 e_u)$
            \EndFor
        \For{$p \in \mathcal{N}_q$}
            \Comment{Update transitions.}
            \State $\Delta_p \gets 0$
            \For{$v \in \mathcal{I}_p$}
                \Comment{Compute change in energy for a hypothetical flip of qubit $p$.}
                \State $\Delta_p \gets \Delta_p + 1 - 2 e_v$
                \EndFor
            \State $\Call{UpdateTransition}{p, \Delta_p}$
                \Comment{Push update to data structure.}
            \EndFor
        \EndWhile
    \State \Return $(\bm{w}, \bm{e}, E, t)$
\end{algorithmic}
\algrule
\noindent\textit{Remarks:} $\Call{Uniform}{}$ returns a uniformly distributed real number in the range $[0, 1)$. An underlying data structure is used to track transition rates, which supports the operations $\Call{InitTransitions}{}$, $\Call{SampleTransitions}{}$, $\Call{TotalRate}{}$, and $\Call{UpdateTransition}{}$. Different choices of data structures lead to different complexities. Good candidates supporting efficient operations are a Fenwick tree~\cite{fenwick1994new, ryabko2002fast} or a binning data structure (see \cref{alg:BKL-binning-data-structure}.)
\end{algorithm}

\textbf{Data structures.} More sophisticated data structures enable further reductions in run-time. To break the $O(n)$ run-time per time step barrier, the simple Data Structure~\ref{alg:BKL-data-structure-array} can be replaced with a Fenwick tree~\cite{fenwick1994new, ryabko2002fast}, also called a binary indexed tree, to store the transition rates. On a code of $n$ qubits, a Fenwick tree allows $O(1)$-time retrieval of the total transition rate ($\Call{TotalRate}{}$), $O(\log n)$-time prefix-sum search\footnote{A prefix-sum search on a list of elements finds the lowest index $j$ such that the cumulative sum up to the $j^{\text{th}}$ element in the list is at least a given value.} and therefore sampling of transitions ($\Call{Sample}{}$), and $O(\log n)$-time updating of any single transition rate ($\Call{Update}{}$). The third row of Table~\ref{tab:BKL-complexities} describes the further reduced $O(s^3 \log n)$ run-time per Monte Carlo time step of Algorithm~\ref{alg:BKL-algorithm-optimized} exploiting sparsity when paired with a Fenwick tree. (Incidentally, replacing the linear search in $\Call{Sample}{}$ of the simple array-based Data Structure~\ref{alg:BKL-data-structure-array} with a binary search does not suffice to reduce the time complexity to below $O(n)$, as linear time is still needed to compute cumulative rates on the array.)

To remove dependence of run-time per time step on $n$ entirely, we illustrate an alternative ``binning'' data structure in Data Structure~\ref{alg:BKL-binning-data-structure}. The idea is to exploit the integral nature of energy changes caused by transitions; in fact the energy changes lie in the integer interval $[-s, +s]$ for an $s$-sparse $H_\mu$ as each qubit is involved in at most $s$ checks. All transitions of degenerate energy change occur with identical probability; thus storing degenerate transitions in the same bin, sampling transitions to undertake boils down to first sampling over the bins followed by trivial (constant-time) uniform sampling in the selected bin. The implementation details of the data structure, which use an array to store energies of transitions and a combination of dynamic arrays to store forward- and back-pointers for book-keeping of bins, allow $O(1)$-time updating of any single transition rate without the overhead\footnote{While hash tables are parametrically efficient, their time overhead and poor cache locality leads to poor performance in practice. The performance requirements of our numerics are strict at the $\lesssim \SI{10}{\nano\second}$ level per data structure call---see our discussion.} of hash tables. Pairing Algorithm~\ref{alg:BKL-algorithm-optimized} with this data structure enables $O(s^3)$ run-time per Monte Carlo time step independent of $n$ (see fourth row of Table~\ref{tab:BKL-complexities}), which is the best parametric complexity we present in this work. A natural further consideration is to use a Fenwick tree instead of an array to store the subtotal transition rates of the bins, but this leads to an inferior $O(s^3 \log s)$ run-time per time step (see fifth row of Table~\ref{tab:BKL-complexities}). 

\begin{algorithm}[H]
\caption{Binning data structure for tracking transition rates}
\label{alg:BKL-binning-data-structure}
\renewcommand{\algorithmicrequire}{\textbf{Input:}}
\renewcommand{\algorithmicensure}{\textbf{Data:}}
\begin{algorithmic}
    \Require Qubit labeling $[n]$, sparsity parameter $s$ of code Hamiltonian $H_\mu$.
    \algrule
    \Ensure Length-$n$ array $\texttt{energies}$ to contain integers in $[-s, +s]$, length-$n$ array $\texttt{locations}$ to contain integers in $[n]$, length-$(2s + 1)$ array $\texttt{bins}$ of initially empty dynamic arrays each to contain integers in $[n]$, array $\bm{r} \in \mathbb{R}_+^{2s + 1}$ storing subtotal transition rate of each bin, float $R \in \mathbb{R}_+$ storing total transition rate.
    \algrule
    \Function{Init}{$\bm{\Delta}$}
        \Comment{Initialize transition rates given transition energy differences $\bm{\Delta}$.}
        \State $(\bm{r}, R) \gets (\bm{0}_{2s + 1}, 0)$
        \For{$q = 1, 2, \ldots, n$}
            \State $\Call{*Add}{q, \Delta_q}$
            \EndFor
        \EndFunction
    \algrule
    \Function{*Add}{$q, \Delta$}
        \Comment{Internal helper to add transition indexed at $q$ with energy difference $\Delta$.}
        \State $\texttt{energies}[q] \gets \Delta$
            \Comment{Record energy.}
        \State $\texttt{locations}[q] \gets \texttt{bins}[\Delta].\Call{Len}{}$
            \Comment{Record a pointer to upcoming entry in $\texttt{bins}$ dynamic array.}
        \State $\texttt{bins}[\Delta].\Call{Push}{q}$
            \Comment{Add entry at the back of $\texttt{bins}$ dynamic array.}
        \State $r_{\Delta} \gets r_{\Delta} + p(\Delta)$
            \Comment{Add to subtotal transition rate of bin.}
        \State $R \gets R + p(\Delta)$
            \Comment{Add to total transition rate.}
        \EndFunction
    \algrule
    \Function{*Remove}{$q$}
        \Comment{Internal helper to remove transition indexed at $q$.}
        \State $\Delta \gets \texttt{energies}[q]$
            \Comment{Look up energy of targeted transition.}
        \If{$\texttt{bins}[\Delta].\Call{Len}{} > 1$}
            \Comment{Need to swap targeted transition to the back of $\texttt{bins}$ dynamic array.}
            \State $q' \gets \texttt{bins}[\Delta].\Call{Back}{}$    
                \Comment{Index of transition currently occupying the back of dynamic array.}
            \State $\ell \gets \texttt{locations}[q]$
                \Comment{Location of targeted transition in dynamic array.}
            \State $\texttt{locations}[q'] \gets \ell$
                \Comment{Move $q'$-indexed transition up to take the place of the targeted transition.}
            \State $\texttt{bins}[\Delta][\ell] \gets q'$
        \EndIf
        \State $\texttt{bins}[\Delta].\Call{Pop}{}$
            \Comment{Remove the targeted transition at the back of dynamic array.}
        \State $r_{\Delta} \gets r_{\Delta} - p(\Delta)$
            \Comment{Subtract from subtotal transition rate of bin.}
        \State $R \gets R - p(\Delta)$
            \Comment{Subtract from total transition rate.}
        \EndFunction
    \algrule
    \Function{TotalRate}{}
        \Comment{Yields total transition rate $R$.}
        \State \Return $R$
        \EndFunction
    \algrule
    \Function{Sample}{}
        \Comment{Samples a transition $q$ with probability $r_q / R$.}
        \State $R^{\mathrm{targ}} \gets \Call{Uniform}{} \cdot \Call{TotalRate}{}$
        \State $R^{\mathrm{curr}} \gets 0$
        \For{$\Delta = -s, -s + 1, \ldots, s$}
            \Comment{Find smallest bin $\Delta$ such that $\sum_{\delta = 1}^{\Delta - 1} r_\delta < R^{\mathrm{targ}} \leq \sum_{\delta = 1}^\Delta r_\delta$.}
            \State $R^{\mathrm{curr}} \gets R^{\mathrm{curr}} + r_\Delta$
            \If{$R^{\mathrm{targ}} \leq R^{\mathrm{curr}}$}
                \State \textbf{break}
                \EndIf
            \EndFor
        \State $\ell \gets \Call{Uniform}{}$
            \Comment{Random transition to be sampled from bin $\Delta$.}
        \State \Return $\texttt{bins}[\Delta][\ell]$
        \EndFunction
    \algrule
    \Function{Update}{$q, \Delta$}
        \Comment{Updates transition rate for qubit $q$ with energy difference $\Delta$.}
        \State $\Call{*Remove}{q}$
        \State $\Call{*Add}{q, \Delta}$
        \EndFunction
\end{algorithmic}
\algrule
\noindent\textit{Remarks:} For notational ease, the arrays $\texttt{bins}$ and $\bm{r}$ are indexed by integers in $[-s, +s]$; all other arrays are $1$-indexed to be consistent with mathematical convention. Dynamic arrays are arrays that support amortized $O(1)$-time growth in length---for instance the $\texttt{vector}$ container in C++, $\texttt{ArrayList}$ in Java, and $\texttt{list}$ in Python. $\texttt{A}.\Call{Len}{}$ obtains the size of a dynamic array $\texttt{A}$, $\texttt{A}.\Call{Push}{x}$ appends an element $x$ to the back of the dynamic array, $\texttt{A}.\Call{Back}{}$ peeks at the back-most element in the dynamic array, and $\texttt{A}.\Call{Pop}{}$ discards the back-most element; these operations are $O(1)$-time. $\Call{Uniform}{}$ returns a uniformly distributed real number in the range $[0, 1)$. $\mathbf{0}_m$ denotes a length-$m$ vector with all zero entries. $p(\Delta)$ is a transition rate function satisfying detailed balance (see Eqs.~\ref{app-eq:transition-rate-metropolis} and \ref{app-eq:transition-rate-glauber}).
\end{algorithm}

\textbf{Further optimizations and lower-level considerations.} We comment that under Metropolis dynamics (see Eq.~\ref{app-eq:transition-rate-metropolis}) the rate of all transitions of energy change $\Delta \leq 0$ is capped at $1$; therefore these transitions can all be considered degenerate in Data Structure~\ref{alg:BKL-binning-data-structure} and stored in the same, say $\Delta = 0$, bin. The $[-s, -1]$-indexed parts of arrays $\texttt{energies}$, $\texttt{bins}$, and $\bm{r}$ are then no longer needed and can be removed.

We remark also on numerical stability over many time steps: in particular, Data Structures~\ref{alg:BKL-data-structure-array} and \ref{alg:BKL-binning-data-structure} involve incremental updates to subtotal and total transition rates, as expressed in the $R \gets R \pm r_q$, $R \gets R \pm p(\Delta)$ and $r_\Delta \gets r_\Delta \pm p(\Delta)$ statements. As transition rates can be exponentially small, performing such updates naïvely runs the risk of floating-point catastrophic cancellation, which is disastrous for long-time simulations. This issue can be mitigated by using compensated (i.e., Kahan) summation~\cite{kahan1965pracniques, higham1993accuracy}, or more straightforwardly, to recompute the total transition rates from scratch every short interval, with the recomputation cost amortized over time steps. 

Several lower-level optimizations can further reduce (constant factors in) run-time. First, in Data Structures~\ref{alg:BKL-data-structure-array} and \ref{alg:BKL-binning-data-structure}, the values of $p(\Delta)$ for integer $\Delta = -s, \ldots, +s$ can be precomputed and stored in an array for look-up, thus avoiding recomputation of $p(\Delta)$ in each $\Call{Update}{}$ call (which involves expensive floating-point multiplication and exponentiation). Second, in the for-loops of Algorithm~\ref{alg:BKL-algorithm-optimized}, namely the $(e_u, E) \gets (\neg e_u, E + 1 - 2 e_u)$ and $\Delta_p \gets \Delta_p + 1 - 2 e_v$ updates, the $+1$ increments can be batched as $|\mathcal{I}_q|$ and $|\mathcal{N}_q|$ are known and can be done in a single operation outside of the loops. Lastly, all negations and multiply-by-twos can be done through bit-wise operations.

\textbf{Implementation.} We implemented Algorithm~\ref{alg:BKL-algorithm-optimized} with Data Structure~\ref{alg:BKL-binning-data-structure} in C++ for performance. Our implementation operated at $\lesssim \SI{100}{\nano\second}$ per Monte Carlo time step at $n \sim 60000$; but as our quantum memories survive to $\gtrsim 10^{11}$ time steps, the simulations are ultimately resource-intensive.

\section{Details on decoding}
\label{app-sec:decoding}

\textbf{Decoder.} We used belief propagation with ordered statistics post-processing (BP-OSD)~\cite{panteleev2021degenerate, roffe2020decoding, roffe2022github} for decoding in our numerical experiments. Without specific fine-tuning and so that decoding is reasonably fast, we adopted minimum-sum parallel-schedule belief propagation with a decaying scaling factor~\cite{roffe2020decoding} performed for a constant $100$ iterations independent of code size, and used the combination-sweep strategy~\cite{roffe2020decoding, fossorier2002soft} at order $10$ for the ordered statistics post-processor. 

\textbf{Calibration.} As summarized in Sec.~\ref{sec:finite_temp/lifetime_results} of the main text, we performed calibration of the qubit flip probabilities, which characterize the error channels and are taken as input parameters to the BP-OSD decoder, by sampling the dynamics of our codes through kinetic Monte Carlo beforehand and fitting the averaged empirical qubit flip probabilities to the simple exponential relaxation form in \eqref{eq:decoder_relaxation_form}. We employed $256$ shots for each code, at each $\beta$, for this calibration procedure.

\textbf{Decoding interval.} It is infeasibly expensive to perform decoding after every Monte Carlo time step (see App.~\ref{app-sec:kinetic_monte_carlo}) to check for memory failure, as BP-OSD takes seconds to run at $n \sim 1000$ and a few minutes at $n \sim 60\,000$. We therefore run the decoder only after a larger time interval $T_{\text{ec}}$, following the same practice as Refs.~\onlinecite{Haah_PRL2013, Haah_arxiv2011}. At each inverse temperature $\beta$, the same $T_{\text{ec}}$ is used for all code sizes for fairness; the $T_{\text{ec}}$ is selected so that all codes survive for $\gtrsim 100$ intervals.

\clearpage
\pagebreak

\begin{figure}
    \centering
    \includegraphics[width=1\linewidth]{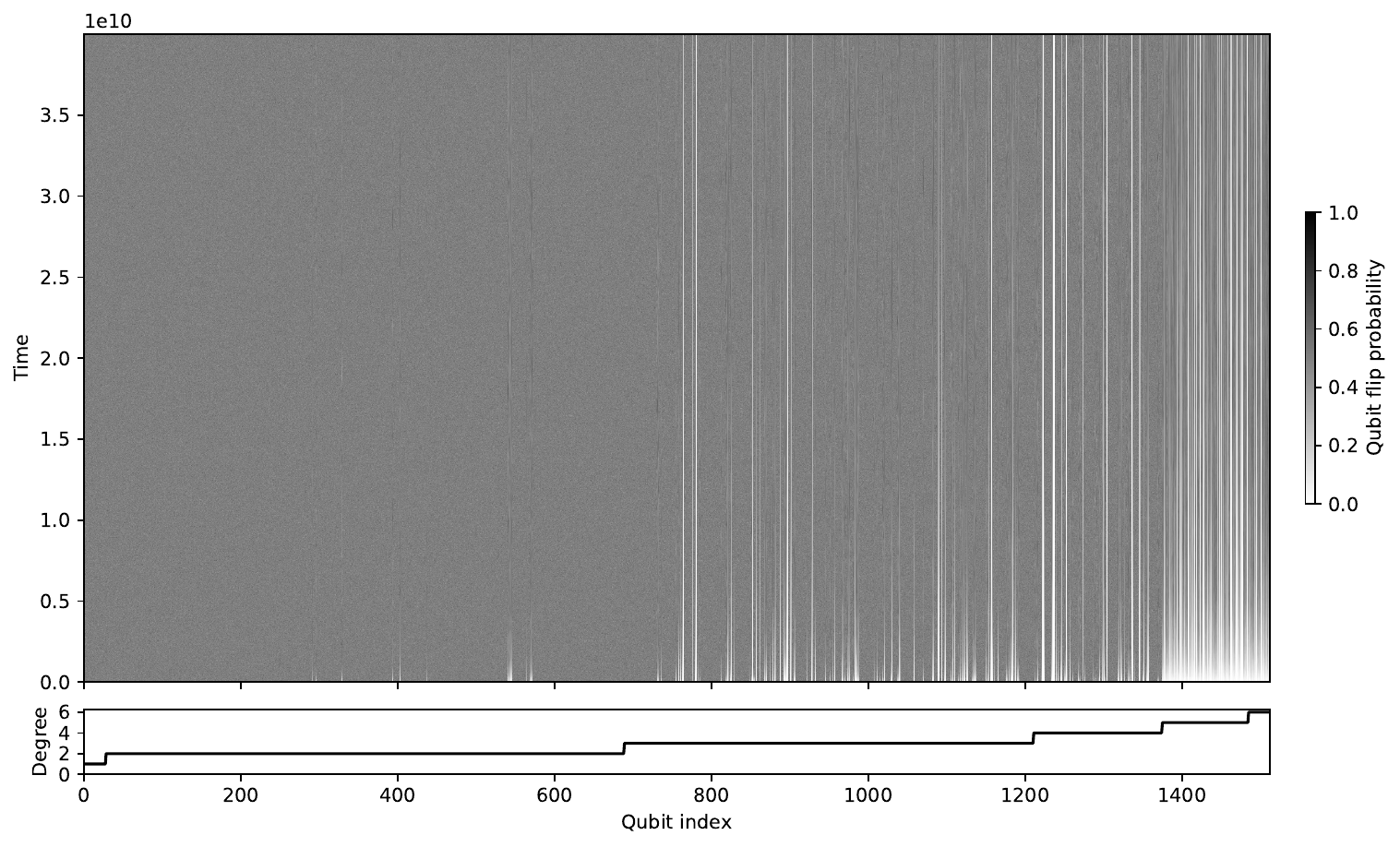}
    \caption{\textbf{Physical qubit lifetimes in a cored pinwheel quantum code.} (Top) Qubit flip probabilities averaged over $256$ shots of thermal dynamics simulated through kinetic Monte Carlo, on a cored quantum pinwheel code formed from the product of two classical pinwheel codes. The particular code used here is one of the codes used in our numerical study (see Fig.~\ref{fig:memory_lifetimes_single_panel_relative}), with construction described in Sec.~\ref{sec:finite_temp/codes} of the main text, at $\beta = 8.3$. The total duration simulated here exceeds the memory lifetime of the code; at the memory lifetime, the error density averaged across qubits in the code is ${\sim} \, 45\%$. (Bottom) The degree of the qubits in the code, defined as the number of stabilizers in the memory basis the qubits are involved in. There is an observable correlation between physical qubit lifetimes and their degrees, which is explored further in Fig.~\ref{app-fig:extra_results/physical_qubit_lifetime_vs_degree}.}
    \label{app-fig:extra_results/physical_qubit_flip_probabilities}
\end{figure}

\begin{figure}
    \centering
    \includegraphics[width=0.5\linewidth]{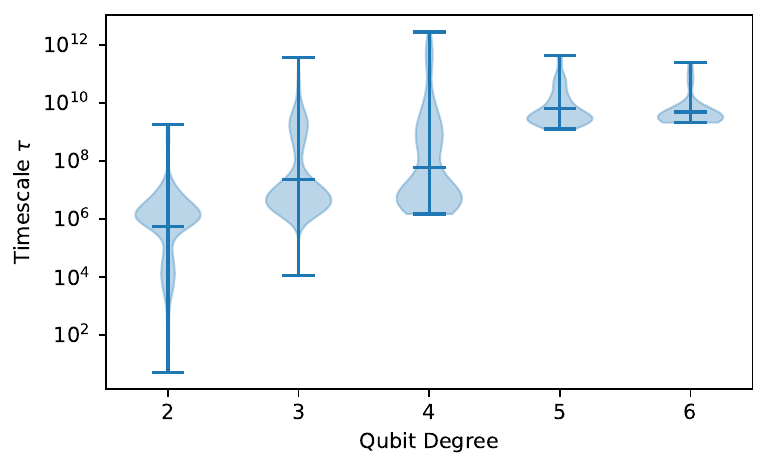}
    \caption{\textbf{Physical qubit lifetimes versus qubit degree in a cored pinwheel quantum code.} Violin plots of qubit flip timescales $\tau$ obtained when the averaged flip probabilities in Fig.~\ref{app-fig:extra_results/physical_qubit_flip_probabilities} are fitted to a simple exponential relaxation form \eqref{eq:decoder_relaxation_form} plotted against qubit degrees. The horizontal bars on the violin plots demarcate the minimums, means, and maximums. The lifetimes of physical qubits span many orders of magnitude.}
    \label{app-fig:extra_results/physical_qubit_lifetime_vs_degree}
\end{figure}

\end{document}